\RequirePackage{fix-cm}
\documentclass{article}
\usepackage{iclr2027_conference_numeric,times}
\makeatletter
\renewcommand\section{\@startsection{section}{1}{\z@}%
  {-1.1ex plus -.3ex minus -.2ex}{.1ex plus .1ex}{\large\sc\raggedright}}
\renewcommand\subsection{\@startsection{subsection}{2}{\z@}%
  {-1.1ex plus -.3ex minus -.2ex}{.1ex plus .1ex}{\normalsize\sc\raggedright}}
\makeatother

\usepackage{amsmath,amsfonts,bm}

\def\eqref#1{equation~\ref{#1}}

\def\1{\bm{1}}

\DeclareMathAlphabet{\mathsfit}{\encodingdefault}{\sfdefault}{m}{sl}
\SetMathAlphabet{\mathsfit}{bold}{\encodingdefault}{\sfdefault}{bx}{n}

\usepackage[T1]{fontenc}
\usepackage{graphicx,booktabs,longtable,array,xcolor}
\usepackage{subcaption}
\usepackage{placeins}
\usepackage{hyperref}
\usepackage{etoc}
\usepackage{url}
\usepackage{tikz}
\usetikzlibrary{arrows.meta,fit,backgrounds}
\usepackage[breakable]{tcolorbox}

\usepackage{enumitem}  
\usepackage{pmboxdraw} 
\newtcolorbox{contributions}{
  colback=blue!5!white,
  colframe=blue!65!black,
  boxrule=0.5pt,
  arc=2pt,
  boxsep=0pt,
  left=4pt,
  right=4pt,
  top=4pt,
  bottom=2pt,
  before skip=4pt,
  after skip=4pt,
}

\usepackage{wrapfig}
\hypersetup{pdfauthor={},pdftitle={LAION-Audio-XL: A Web-Scale Corpus and Scaling Studies for Contrastive Language-Audio Pretraining}}
\graphicspath{{figures/}}
\title{LAION-Audio-XL: A Web-Scale Corpus and Scaling Studies for Contrastive Language-Audio Pretraining}

\author{Gijs Wijngaard$^{1,2,*}$ \quad Felix F\"orster$^{3,*}$ \quad Ross Wightman$^{2}$ \quad Marianna Nezhurina$^{1,2}$ \\
\bf Christoph Schuhmann$^{2}$ \quad Jenia Jitsev$^{1,2,\diamond}$ \quad A. Sophia Koepke$^{3,4,\diamond}$ \\[0.5em]
$^{1}$J\"ulich Supercomputing Center (JSC), Forschungszentrum J\"ulich (FZJ) \quad
$^{2}$LAION \\
$^{3}$Technical University of Munich, MCML \quad
$^{4}$University of T\"ubingen, T\"ubingen AI Center \\[1.3em]
\textbf{\url{https://projects.laion.ai/clap2}}}
\iclrfinalcopy
\begin{document}

\etocdepthtag.toc{mainbody}
\maketitle
\fancyhead{} 
\ificlrfinal
\begingroup
\let\thefootnote\relax
\begin{NoHyper}\footnotetext{$^{*}$Equal contribution, decided by coin flip.\quad $^{\diamond}$Equal supervision.}\end{NoHyper}
\endgroup
\fi

\begin{abstract}
Recent progress on contrastive audio-language models can be attributed mainly to the scale and quality of their training data. Larger training datasets expose models to a broader range of sounds and corresponding audio captions. To exploit this, we collect 1.48 billion audio samples from academic datasets, internet datasets, web collections, Common Crawl, and YouTube, and process them with a multi-stage curation pipeline. 
Furthermore, we generate short and long captions for samples in our corpus that lack them. 
We use our dataset to study the scaling behavior of contrastive audio-language models based on the NaFlex architecture for the audio encoder. Our experiments identify trade-offs between model size and training data for a given compute budget. Moreover, our resulting models, CLAP2, outperform prior audio-text models of comparable size on the MAEB and HEAR benchmarks. We release our dataset, captions, model checkpoints, and code.

\end{abstract}

\vspace{0.2em}
\section{Introduction}
\vspace{0.2em}

\label{sec:intro}

Recent developments in audio-language learning have shown that large-scale contrastive pretraining results in strong downstream performance~\citep{radfordLearningTransferableVisual2021,wuLargeScaleContrastiveLanguageAudio2023,elizaldeCLAPLearningAudio2023}. In contrastive language-audio pretraining (CLAP), models learn to map audio recordings and their corresponding text descriptions into a shared embedding space. This facilitates tasks such as text-audio retrieval~\citep{oncescuAudioRetrievalNatural2021,koepkeAudioRetrievalNatural2023} or audio classification. In addition to that, CLAP models support text-to-audio generation~\citep{liuAudioLDMTexttoAudioGeneration2023,hungTangoFluxSuperFast2026} and audio source separation~\citep{maCLAPSepLeveragingContrastive2024}. CLAP embeddings have also proven useful for data curation, including for detecting dataset overlap~\citep{wijngaardAudioLanguageDatasetsScenes2025}, filtering audio-text pairs~\citep{haji-aliTamingDataTransformers2026,valleFugatto1Foundational2025}, selecting synthetic training examples~\citep{ghoshSynthioAugmentingSmallScale2025}, and refining audio captions \citep{baiAudioSetCapsEnrichedAudioCaption2025}.

As datasets grow, CLAP models are also improving. For instance, LAION-CLAP used 633,526 audio-text pairs \citep{wuLargeScaleContrastiveLanguageAudio2023}, M2D-CLAP used 2.6 million pairs~\citep{niizumiM2DCLAPMaskedModeling2024}, MS-CLAP 2023 increased this to 4.6 million pairs~\citep{elizaldeNaturalLanguageSupervision2024}, and SLAP was trained with 109 million pairs~\citep{meiSLAPScalableLanguageAudio2026}. These models report increasingly strong classification and retrieval results, while also differing in model size. Choosing an appropriate model size is important for making the best use of available data and compute.

Scaling laws address this question by showing that increasing both training data and model size generally improves downstream performance~\citep{kaplanScalingLawsNeural2020, hoffmannTrainingComputeOptimalLarge2022}. In contrastive language-image pretraining (CLIP), scaling training data and model size has been a major source of performance gains~\citep{chertiReproducibleScalingLaws2023}. Recent work measured this in greater detail to compare models and datasets and estimate how many training samples to use for a given compute budget \citep{nezhurinaScalingLawsRobust2026}. Moreover, data selection has been shown to be crucial~\citep{gadreDataCompSearchNext2023}.

The availability of large-scale audio-text data remains a bottleneck for studying the scaling of audio models. While vision and language pretraining have scaled to open, billion-scale web-scraped datasets with transparent curation pipelines~\citep{schuhmannLaion5bOpenLargeScale2022,gadreDataCompSearchNext2023}, public audio-text collections remain smaller and less documented. Open caption datasets for general sound reach about 50 million text-audio pairs~\citep{haji-aliTamingDataTransformers2026}, and the largest open release, Bagpiper's pre-training collection, pairs 392 million mostly speech clips (933K hours) with captions from a single captioning model~\citep{tianBagpiperSolvingOpenEnded2026}. AF-Next used around 108 million samples from 1 million hours of audio~\citep{ghoshAudioFlamingoNext2026}, and MiDasheng trained on 1.1 million hours~\citep{dinkelMiDashengLMEfficientAudio2025}, both largely drawn from public datasets. We show that by combining public audio datasets with four other groups of sources, such as YouTube and audio derived from Common Crawl, this amount of data can be increased even more. In contrast, recent audio foundation models report training on private datasets of more than 100M hours, which are not publicly available~\citep{teamMiMoAudioAudioLanguage2025,qwenteamQwen35OmniTechnicalReport2026}.

\begin{figure}[t]
\centering\includegraphics[width=\linewidth]{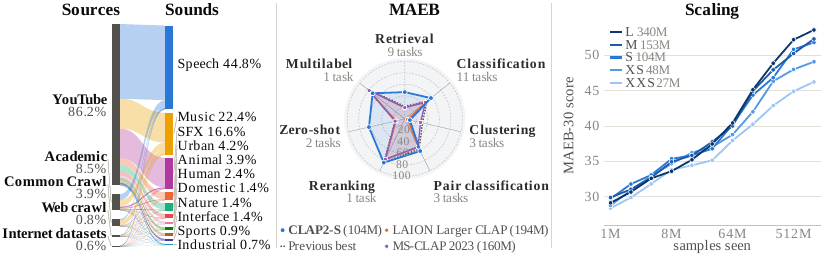}
\vspace{-1.5em}
\caption{Our training dataset consists of many sources, spanning diverse sound categories, and our models achieve strong performance that improves with training scale. \textbf{Left}: the 365M training audio samples from source group to predicted sound category using our trained taxonomy classifier (see Section~\ref{sec:selection}). Ribbon width is proportional to audio samples. \textbf{Middle}: MAEB~\citep{assadiMAEBMassiveAudio2026} of CLAP2-S, LAION Larger CLAP~\citep{wuLargeScaleContrastiveLanguageAudio2023} and MS-CLAP 2023~\citep{elizaldeNaturalLanguageSupervision2024}. Dotted: previous best below 500M parameters. \textbf{Right}: Scaling MAEB-30 against samples seen.}
\label{fig:pipeline}
\vspace{-1em}
\end{figure}

In this work, we collected LAION-Audio-XL, a training corpus of 365 million text-audio pairs from an initial collection of 1.48 billion audio samples drawn from public audio datasets, web data, Common Crawl, and YouTube (Figure~\ref{fig:pipeline}). The dataset is processed through multiple steps, including cleaning, deduplication, annotation, selection and captioning. From a raw pool of 1.48 billion audio samples, the final training corpus is a 365 million dataset of audio-text pairs, with audio of at most 30 seconds. We release the audio of all sources whose license allows redistribution, links with captions and metadata for the audio we crawled ourselves, and our models and code to facilitate future research.

We use this corpus to train CLAP2, a family of contrastive language-audio models. The audio encoder is a NaFlex audio transformer \citep{tschannenSigLIP2Multilingual2025}, which processes variable-length audio. The text encoder is based on ModernBERT \citep{warnerSmarterBetterFaster2025}, but uses a SwiGLU and RMSNorm instead of GeGLU and LayerNorm. We train both encoders from scratch instead of initializing from a pretrained model. The model architecture scales more easily for testing and it allows for better display of scaling laws. Our experiments test 5 different model sizes, from $\approx$27M to $\approx$340M parameters, and plot the results. Training from scratch results in that there is no confounding effects from prior model weights. 

For CLAP2-M (153M parameters), the average score on MAEB-30~\citep{assadiMAEBMassiveAudio2026} increases on average by 2.2 points when the seen samples are doubled between 1M and 1024M samples. Training only on audio of at most 30 seconds improves benchmark performance and requires 3.5 times less compute than training on audio of up to 120 seconds. Our best model, CLAP2-L, reaches 53.6\%, compared to 35.2\% for LAION-CLAP and 44.1\% for GLAP~\citep{dinkelGLAPGeneralContrastive2026}, the best other contrastive audio-text model we evaluated, with 2.5 times more parameters. The model is also the best model trained from scratch on the MAEB-30 leaderboard, where other models rely on backbones whose training data is not released. On HEAR \citep{turianHEARHolisticEvaluation2022}, CLAP2-S to CLAP2-L score higher than all other audio-text models.\medskip

Overall, our contribution can be summarized as follows:
\vspace{0.5em}
\begin{contributions}
\begin{itemize}[leftmargin=1em]
\item We introduce a curated audio-text dataset with 365 million audio-text pairs, selected from 1.48 billion web and dataset audio samples through a fully documented pipeline.
\item We present a scaling study of contrastive audio-text training, showing an average gain of around 1.8 MAEB-30 points per doubling of compute over more than four orders of magnitude.
\item We train CLAP2, five CLAP models from 27M to 340M parameters, whose largest model CLAP2-L reaches 53.6\% on MAEB-30, outperforming models of similar size such as M2D-CLAP (38.0\%) and LAION-CLAP (35.2\%). 
\end{itemize}
\end{contributions}
\vspace{0.5em}

\section{Related work}\label{sec:related}
\textbf{Audio-text datasets.} Earlier datasets such as Clotho \citep{drossosClothoAudioCaptioning2020} and AudioCaps \citep{kimAudioCapsGeneratingCaptions2019} relied on a single source, often YouTube or Freesound. A recent trend shows that models are increasingly using combinations of sources \citep{wijngaardAudioLanguageDatasetsScenes2025}. Earlier examples of encoder models trained on a mixture of data include MS-CLAP trained on four and LAION-CLAP on 11 sources \citep{elizaldeCLAPLearningAudio2023, wuLargeScaleContrastiveLanguageAudio2023}. Later, state-of-the-art models used 20+ sources (see Table \ref{tab:datasets} for an overview). The dataset presented in this work, LAION-Audio-XL, uses a combination of 425 sources, 316 of which are either part of or derived from academic work. Other works enrich audio data with model generated captions, such as AudioSetCaps~\citep{baiAudioSetCapsEnrichedAudioCaption2025}, Sound-VECaps~\citep{yuanSoundvecapsImprovingAudio2025}, and ACAVCaps~\citep{niuACAVCapsEnablingLargeScale2026}. Some models also test varying captioning style, similar to our work, for example AudioVerse and ACAVCaps use a combination of long and short captions, including other variants \citep{xuScalingAudioTextRetrieval2026, niuACAVCapsEnablingLargeScale2026}. CapTalk and LP-MusicCaps do similar splits for speech and music, respectively \citep{suCapTalkUnifiedVoice2026, dohLPMusicCapsLLMBasedPseudo2023}. 

\textbf{Audio scaling laws.} Researchers often release audio-language models in multiple sizes, where based on the scaling law properties, larger models tend to perform better than smaller models. Examples in audio encoder models include AuroLA \citep{xuScalingAudioTextRetrieval2026} testing with audio-language LoRAs on 95M, 322M and 418M trainable parameters, and Dasheng consisting of 3 model sizes, 86M, 0.6B and 1.2B parameters \citep{dinkelScalingMaskedAudio2024}. Similarly, in the audio-language learning field, models are often released in multiple sizes, for example Qwen2.5-Omni with a 3B and 7B model and MOSS-Audio with a 4B and 8B model \citep{xuQwen25OmniTechnicalReport2025, yangMOSSAudioTechnicalReport2026}. While scaling model sizes, it is also important to take into account scaling the amount of data. For audio-language models, SODA \citep{manakulScalingOpenDiscrete2026} shows that the compute-optimal data amount needed grows faster than the amount of training parameters. PE-AV \citep{vyasPushingFrontierAudiovisual2026} separately compares how to scale data and model sizes in audio-visual models. \citet{tsengRevisitingAudioLanguagePretraining2026} train both contrastive and captioning models in pairs in 0.4M, 1M, 4M, and 10M audio-language pairs, and show that on smaller scales having a pretrained model gains the most improvements, but on larger scale training from scratch helps the most. To the best of our knowledge, this work is the first to present how a fixed compute budget should be divided between model sizes and data samples seen for contrastive audio-language models. 

\textbf{Audio taxonomy.} The AudioSet ontology is a taxonomy based on 632 classes, from which the 527 labels of the AudioSet dataset originate. These 632 originated from a list of more than 3000 candidates \citep{gemmekeAudioSetOntology2017}. SALT later proposed an extension to 734 sound labels \citep{stamatiadisSALTStandardizedAudio2024}. Universal Category System is an initiative for standardization of a set of categories of sound, with the January 2024 version spanning 753 subcategories under 82 categories \citep{timnielsenUniversalCategorySystem2024}. This work uses a taxonomy based on 4,416 nodes consisting of 11 root nodes, and a maximum depth of 5. This taxonomy is the basis of a classifier on the set of leaf nodes of 3,718 classes to cluster a raw pool of audio data.

\section{LAION-Audio-XL dataset}
\label{sec:data}
\subsection{Collection}
\begin{figure}[t]
\centering\includegraphics[width=\linewidth]{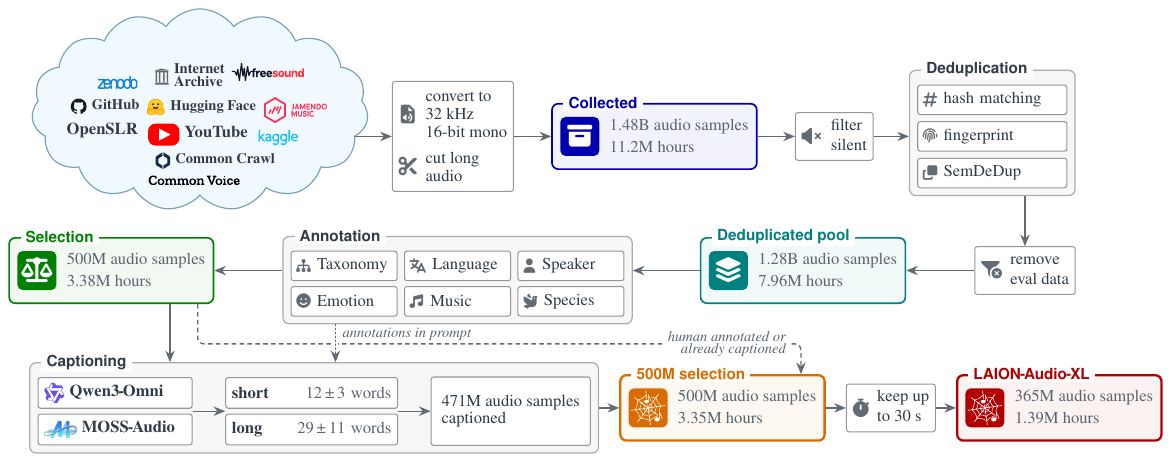}
\vspace{-1.5em}
\caption{\textbf{Overview of the LAION-Audio-XL dataset curation pipeline.} We collected 1.48B audio samples from online platforms, of which 1.47B entered deduplication.
Removing silent audio, duplicates (including SemDeDup \citep{abbasSemDeDupDataefficientLearning2023}), and evaluation data leaves 1.28B audio samples. We then selected 500M audio samples based on categorical taxonomy annotations.
The same annotations were used to prompt Qwen3-Omni \citep{xuQwen3OmniTechnicalReport2025} and MOSS-Audio \citep{yangMOSSAudioTechnicalReport2026} to generate short and long captions for 471M audio samples without textual descriptions, resulting in 500M captioned audio samples in total (the 500M selection). The 365M audio samples that are up to 30 seconds long form LAION-Audio-XL.
}
\label{fig:steps}
\end{figure}

For training contrastive language-audio models, large-scale and diverse datasets are required. For the training of the model presented in this paper, a collection is aggregated over public datasets, YouTube videos, Common Crawl, among others. The final dataset contains 316 academic datasets, each tied to a published paper, preprint, challenge, or other academic release (Appendix~\ref{app:collection}). Most of these datasets include valid descriptions that are also reused in the final dataset, with even a fraction of the full data being human-annotated. Table \ref{tab:datasets} compares the final dataset LAION-Audio-XL with existing audio-text datasets, and Figure~\ref{fig:steps} lists the processing steps outlined in this section.

\begin{table}[t]
\centering\small
\caption{\textbf{LAION-Audio-XL compared to existing audio-text datasets}, including standalone dataset releases and the training data of audio encoders and audio-language models. Our LAION-Audio-XL dataset is sourced from a range of online platforms and websites, and from existing data collections, and datasets (details are provided in Appendix~\ref{app:catalog}).
$^\dagger$Dasheng trains on audio only. $^\ddagger$Not publicly released. Captions counts the text entries each source reports (captions, ASR transcripts, QA pairs); -- under Audio means not reported; $^*$ marks values derived from clip counts and clip lengths.
}
\label{tab:datasets}
\begin{tabular}{@{}lrrrr@{}}
\toprule
\textbf{Dataset} & \textbf{Sources} & \textbf{Audio} & \textbf{Captions} & \textbf{Duration (h)} \\
\midrule
Clotho \citep{drossosClothoAudioCaptioning2020} & 1 & 6K & 30K & 37$^*$ \\
AudioCaps \citep{kimAudioCapsGeneratingCaptions2019} & 1 & 51K & 57K & 143$^*$ \\
WavCaps \citep{meiWavCapsChatGPTAssistedWeaklyLabelled2024} & 4 & 403K & 403K & 7.6K \\
LAION-Audio-630K \citep{wuLargeScaleContrastiveLanguageAudio2023} & 8 & 634K & 634K & 4.3K \\
Auto-ACD \citep{sunAutoACDLargescaleDataset2024} & 2 & 1.9M & 1.9M & 5.3K$^*$ \\
SALMONN-2 \citep{yangSALMONN2AdvancingGeneralPurpose2026} & 21 & -- & 8.9M & 18.2K \\
CaptionStew \citep{tsengRevisitingAudioLanguagePretraining2026} & 8 & 9.3M & 10.7M & 37.3K \\
Audio Flamingo 3 \citep{goelAudioFlamingo32025} & 46 & -- & 26.8M & $\geq$54.4K \\
LAION-BVD (BVD-A-10M) \citep{hochlehnertLAIONBVD10MillionHourOpen2026} & 1 & 10M & 10M & 27K$^*$ \\
AutoReCap-XL \citep{haji-aliTamingDataTransformers2026} & 6 & 47M & 47M & 124K \\
Dasheng mix \citep{dinkelScalingMaskedAudio2024} & 4 & 97M & --$^\dagger$ & 272K \\
SLAP (MovieGen Audio)$^\ddagger$ \citep{meiSLAPScalableLanguageAudio2026} & 1 & 109M & 109M & $\sim$1M~\citep{polyakMovieGenCast2025} \\
Bagpiper \citep{tianBagpiperSolvingOpenEnded2026} & 13 & 392M & 392M & 933K \\
GLAP mix \citep{dinkelGLAPGeneralContrastive2026} & 12 & 445M & 445M & 435K \\
\midrule
500M selection (ours) & \textbf{425} & \textbf{500M} & \textbf{970M} & \textbf{3.35M} \\
LAION-Audio-XL (ours) & \textbf{425} & 365M & 707M & 1.39M \\
\bottomrule
\vspace{-2em}
\end{tabular}
\end{table}

YouTube contains the majority of our dataset. All audio collected on YouTube originates from Creative Commons-licensed videos with explicit reuse permissions. This source produces 1.18B audio samples extracted from 32.66M full videos. YouTube audio is extracted across the full duration of each video. Each video can contribute many audio samples. The audio is segmented at silence boundaries into audio samples between 5 and 120 seconds.

Common Crawl consists of many websites scraped from the open web \citep{commoncrawlfoundationCommonCrawl2024}. A scraper extracts all mentions of audio files with valid audio extensions across all crawls through \textsc{CC-MAIN-2026-08}. The first extraction wave covered 35 crawl directories from 2021 to 2025 and found about 3.3 billion audio links, of which about 423M are unique. The text surrounding the audio on the website is used as a context for subsequent caption generation.

\subsection{Deduplication, signal checks, and benchmark screening}
The pipeline begins with removal of all empty or partial audios, or audios that cannot be read or decoded. Audios are kept if they are between 0.5 and 120 seconds, with an extra check if the audio contains at least 0.5 seconds of audible content. Exact matching groups audio samples by audio duration and the MD5 hash of uncompressed PCM bytes. This step removes 151.1M audio samples. Removing duplicate and silent recordings saves compute \citep{leeDeduplicatingTrainingData2022} and prevents performance degradation and lowers memorization of repeated items \citep{hernandezScalingLawsInterpretability2022}.

Near-duplicate deduplication targets recordings with minor acoustic variations. Two complementary strategies are used: acoustic matching through signal-processing fingerprints directly after exact matching, and semantic matching through self-supervised audio embeddings after the signal gate. Acoustic matching uses Olaf fingerprints \citep{sixOlafLightweightPortable2023} and removes 1.2M audio samples. Semantic near-duplicates are found using SemDeDup \citep{abbasSemDeDupDataefficientLearning2023} based on Dasheng embeddings \citep{dinkelScalingMaskedAudio2024}. This stage removes 18.2M audio samples. Appendix~\ref{app:deduplication} describes both implementations in more detail.

The same deduplication also runs against the evaluation data of MAEB \citep{assadiMAEBMassiveAudio2026}, HEAR \citep{turianHEARHolisticEvaluation2022} and MSEB \citep{heigoldMassiveSoundEmbedding2026}. Source identifiers, acoustic fingerprints and a neural fingerprint find evaluation audio in the corpus, and every match is verified on the waveform (Appendix~\ref{app:quarantine}).

\subsection{Clustering and coverage-aware selection}\label{sec:selection}
After filtering and deduplication, an annotation pipeline gathers information regarding the audio. These serve as filters in manual selection and as extra content for captioning models. Using the addition of this extra information during captioning helps improve the caption quality (see Appendix \ref{app:metadataprompt}). 
A prominent part of our filtering is based on a taxonomy that reduces common content while protecting rare predicted categories. A model is trained with a linear output layer of 3,718 leaf buckets to perform this classification.
In addition to the taxonomy filtering, the metadata collection involves emotion classification, language, speaker age and gender, pitch, music genre and style tags, and species predictions.

The selector combines a binary validity gate, a retention fraction, a cluster-density factor, and a mixture factor. Retention floors protect rare predicted categories, languages, expressive emotions, and species, while cluster-density and mixture factors reduce common content. A deterministic hash determines which candidates are kept. Appendix~\ref{app:selection} gives the full calculation and thresholds.

The deduplicated pool is 75.3\% speech and 11.7\% music by predicted label. After selection and exact-size trimming, 500M base rows are obtained. The final training corpus keeps the 366,006,046 selected audio samples of at most 30 seconds (Appendix~\ref{sec:duration}). This changes the mixture to 44.8\% speech, 22.4\% music, 16.6\% designed sound effects, and 16.2\% other classes.

\subsection{Caption quality and production supervision}
\label{sec:captions}
Caption quality is evaluated with both model judgments and human listening. Seven captioning systems produced short and long descriptions for a shared sample. The two MOSS-Audio-8B configurations, with thinking on and off, ranked highest in both the LLM review and the human study \citep{yangMOSSAudioTechnicalReport2026}, followed by Qwen3-Omni in the LLM review \citep{xuQwen3OmniTechnicalReport2025}. Appendix~\ref{app:captions} reports the model comparison and human listening study.

The captioners \textsc{Qwen3-Omni-30B-A3B-Instruct} and \textsc{MOSS-Audio-8B-Thinking} \citep{xuQwen3OmniTechnicalReport2025,yangMOSSAudioTechnicalReport2026} both caption disjoint halves of the data with a short and a long description for each audio, with 12 and 29 words on average. In total, this stage captions 471,237,555 audio samples with both caption lengths (94.40\% of the full 500M selection).

\begin{table*}[t]
\centering\small
\caption{MAEB benchmark results. Parameter counts (Params) include only the audio and
text parts of each model. Results for our models use the last checkpoint of each size's WSD run: 1024M samples for XXS, XS, S, M, L,
evaluated with un-normalised audio embeddings and with all-caps text inputs lowercased. We report the mean performance over 30 tasks, domain-specific
(18 speech, 6 music, 6 sound tasks), and task-specific results: classification (12,
including one multilabel), zero-shot classification (2), pair classification
(3), clustering (3), and retrieval (10, including one reranking task). All scores are in \%. Models are grouped by size, the best value in each column
within a group is bolded.}
\label{tab:scaling-final}
\setlength{\tabcolsep}{4pt}
\resizebox{\textwidth}{!}{%
\begin{tabular}{l r r rrr rrrrr}
\toprule
& & \multicolumn{9}{c}{\textbf{MAEB}} \\
\cmidrule(lr){3-11}
& & & \multicolumn{3}{c}{\textbf{Domain}} & \multicolumn{5}{c}{\textbf{Task type}} \\
\cmidrule(lr){4-6}\cmidrule(lr){7-11}
\textbf{Model} & \textbf{Params} & \textbf{Mean} & \textbf{Speech} & \textbf{Music} & \textbf{Sound} & \textbf{Clf.} & \textbf{ZS} & \textbf{Pair} & \textbf{Clust.} & \textbf{Retr.} \\
\midrule
\multicolumn{11}{l}{\textit{Large models ($\geq$1B parameters)}} \\
LCO-Embedding-Omni-7B \citep{xiaoScalingLanguageCentricOmnimodal2025} & 7.7B & \textbf{53.5} & 54.8 & \textbf{73.7} & \textbf{29.7} & 57.9 & 64.5 & \textbf{67.3} &  1.7 & 57.6 \\
e5-omni-7B \citep{chenE5OmniExplicitCrossModal2026} & 7.7B & 52.5 & 54.1 & 71.8 & 28.1 & 53.6 & \textbf{68.8} & 65.5 &  1.8 & \textbf{59.1} \\
BidirLM-Omni-2.5B \citep{boizardBidirLMTextOmnimodal2026} & 2.0B & 52.4 & \textbf{58.6} & 66.0 & 20.0 & \textbf{59.9} & 65.8 & 66.7 &  6.2 & 50.2 \\
LCO-Embedding-Omni-3B \citep{xiaoScalingLanguageCentricOmnimodal2025} & 3.7B & 52.3 & 54.0 & 72.0 & 27.3 & 56.1 & 62.2 & 66.7 &  1.3 & 56.6 \\
jina-embeddings-v5-omni-small \citep{honickeJinaEmbeddingsv5OmniGeometryPreservingEmbeddings2026} & 1.3B & 50.4 & 53.7 & 68.0 & 23.1 & 53.4 & 62.2 & 65.4 &  6.1 & 53.2 \\
e5-omni-3B \citep{chenE5OmniExplicitCrossModal2026} & 3.7B & 48.0 & 50.2 & 64.7 & 24.7 & 48.5 & 64.4 & 59.5 &  1.9 & 54.5 \\
OmniEmbed-v0.1 \citep{maTevatron20Unified2025} & 7.8B & 43.9 & 43.5 & 61.7 & 27.2 & 50.7 & 30.2 & 58.9 &  3.2 & 46.1 \\
fusion-embedding-2-2b-preview \cite{tonmoyFusionEmbeddingUnified2026} & 2.4B & 40.2 & 38.8 & 60.6 & 23.9 & 48.4 & 55.2 & 61.5 &  3.7 & 31.9 \\
Qwen2-Audio-7B \citep{chuQwen2AudioTechnicalReport2024} & 7.8B & 34.5 & 32.6 & 61.9 & 12.9 & 56.2 & 12.4 & 56.9 & \textbf{12.7} & 12.8 \\
Qwen2.5-Omni-7B \citep{xuQwen25OmniTechnicalReport2025} & 7.7B & 24.2 & 24.7 & 44.0 &  2.8 & 33.0 & 17.3 & 52.3 &  0.7 & 13.6 \\
\midrule
\addlinespace[1pt]
\midrule
\multicolumn{11}{l}{\textit{Small models ($<$1B parameters)}} \\
jina-embeddings-v5-omni-nano \citep{honickeJinaEmbeddingsv5OmniGeometryPreservingEmbeddings2026} & 855M & 51.8 & 53.5 & 73.8 & 24.7 & 56.8 & 60.2 & 65.5 &  6.9 & 53.4 \\
EmbeddingGemma 2 \cite{EmbeddingGemma2Open2026} & 577M & 49.4 & \textbf{56.6} & 61.4 & 15.7 & 54.1 & 45.3 & \textbf{66.9} &  1.0 & \textbf{53.7} \\
GLAP \citep{dinkelGLAPGeneralContrastive2026} & 855M & 44.1 & 42.4 & 69.6 & 23.4 & 46.0 & 54.5 & 56.4 &  5.9 & 47.3 \\
M2D-CLAP \citep{niizumiM2DCLAPMaskedModeling2024} & 199M & 38.0 & 30.0 & 69.7 & 29.9 & 50.4 & 16.0 & 53.5 & \textbf{31.4} & 24.7 \\
MS-CLAP 2023 \citep{elizaldeNaturalLanguageSupervision2024} & 160M & 37.3 & 29.2 & 68.0 & 30.9 & 48.7 & 17.2 & 53.6 & 28.0 & 25.5 \\
TP-CLAP \citep{liTextPromptedCLAPLearning2026} & 198M & 37.2 & 28.6 & 67.0 & \textbf{33.3} & 47.8 & 18.2 & 54.4 & 23.0 & 27.4 \\
CLAP-IPA small \cite{zhuTasteIPAOpenVocabulary2024} & 174M & 36.5 & 41.4 & 50.5 &  8.1 & 34.7 & 53.5 & 52.9 &  0.6 & 41.2 \\
LAION Larger CLAP \citep{wuLargeScaleContrastiveLanguageAudio2023} & 194M & 36.2 & 27.1 & 70.3 & 29.6 & 52.2 & 15.4 & 54.8 &  5.0 & 25.0 \\
LAION-CLAP HTSAT unfused \citep{wuLargeScaleContrastiveLanguageAudio2023} & 153M & 35.2 & 26.4 & 68.4 & 28.3 & 47.6 & 12.4 & 52.9 & 19.6 & 24.2 \\
MuQ-MuLan-large \cite{zhuMuQSelfSupervisedMusic2025} & 630M & 28.9 & 23.1 & 64.9 & 10.6 & 40.7 & 12.6 & 52.9 &  5.4 & 18.0 \\
wav2clip \citep{wuWav2CLIPLearningRobust2022} &  75M & 28.3 & 24.1 & 58.0 & 11.3 & 39.1 & 10.9 & 57.2 &  9.5 & 15.9 \\
SpeechT5 \citep{aoSpeechT5UnifiedModalEncoderDecoder2022} & 180M & 26.7 & 26.0 & 46.5 &  9.1 & 38.6 & 15.7 & 58.0 &  1.2 & 12.9 \\
\addlinespace[2pt]
\hline
\addlinespace[2pt]
\textit{CLAP2-XXS}              &  27M & 46.2 & 46.0 & 69.2 & 23.9 & 50.9 & 59.9 & 61.4 &  3.1 & 46.2 \\
\textit{CLAP2-XS}               &  48M & 49.1 & 48.8 & 72.1 & 26.9 & 56.2 & 57.8 & 62.0 &  3.6 & 48.6 \\
\textit{CLAP2-S}                & 104M & 51.8 & 52.4 & 74.2 & 27.6 & 59.2 & 63.9 & 62.0 &  9.0 & 50.3 \\
\textit{CLAP2-M}                & 153M & 52.3 & 52.4 & 76.1 & 28.1 & 59.8 & 56.5 & 61.8 &  9.3 & 52.6 \\
\textit{CLAP2-L}                & 340M & \textbf{53.6} & 54.6 & \textbf{76.2} & 27.9 & \textbf{60.6} & \textbf{68.6} & 62.3 & 10.0 & 52.7 \\
\bottomrule
\end{tabular}
}
\end{table*}

\section{Model, training, and evaluation}
\label{sec:model}
\textbf{Model.} Our models are trained from scratch. The \textbf{audio tower} is based on a NaFlex vision transformer \citep{dehghaniPatchNpackNaViT2023,tschannenSigLIP2Multilingual2025} over log-mel patches~\citep{gongASTAudioSpectrogram2021}. The NaViT implementation is based on the timm package~\citep{wightmanPyTorchImageModels2026}. The audio input (32\,kHz) is converted to Mel spectrograms with 128 bins between 50\,Hz and 16\,kHz, and a 320-sample hop. Non-overlapping $16\times16$ patches give 50 tokens per second or eight tokens per 0.16 seconds. We use RoPE~\citep{suRoformerEnhancedTransformer2024, heoRotaryPositionEmbedding2025} to keep the coordinates of each patch. The transformer blocks use RMSNorm \citep{zhangRootMeanSquare2019}, SwiGLU \citep{shazeerGLUVariantsImprove2020} and LayerScale~\citep{touvronGoingDeeperImage2021}. One register token is added, and attention pooling~\citep{zhaiSigmoidLossLanguage2023} followed by two linear layers with a GELU~\citep{hendrycksGaussianErrorLinear2016} activation between them produces the clip embedding. The \textbf{text tower} is a bidirectional transformer with RoPE, RMSNorm, SwiGLU, and query-key normalization~\citep{henryQuerykeyNormalizationTransformers2020}. It has a 50,260-entry byte-pair vocabulary~\citep{radfordLanguageModelsAre2019} and a 256-token context. The CLAP2 models span five sizes, between 27.3M (XXS) and 340.1M trainable parameters (Table~\ref{tab:arch-sizes}). Smaller models enable efficient data filtering.

\textbf{Training} uses a standard CLIP objective on audio and text \citep{oordRepresentationLearningContrastive2019,radfordLearningTransferableVisual2021}. The learnable logit scale is initialized to $1/0.07$ and has an upper bound of $300$. Audio is assigned to a batch of nearest length, with buckets of 6.88, 8.96, 12.96, 22.56, and 30.08 seconds, to reduce padding and speed up training. A GPU batch holds audio samples of one bucket under a budget of 49,152 audio tokens, so a 6.88 second short-clip batch has 136 audio samples and a 30 second batch has 32 per GPU (rounded down to be divisible by eight). Bucket probabilities are set so every bucket supplies about 20\% of the samples, which corresponds to the training distribution.  Audio samples are padded and masked inside their bucket.

All runs use AdamW~\citep{loshchilovDecoupledWeightDecay2019} with $\beta=(0.9,0.98)$, weight decay 0.2, warmup-stable~\citep{zhaiScalingVisionTransformers2022, huMiniCPMUnveilingPotential2024} or cosine~\citep{loshchilovSGDRStochasticGradient2017} learning rate decay, gradient clipping at 1.0, and bfloat16 autocast. Peak learning rate, warmup, and batch size are set per data budget (see Appendix~\ref{app:scaling} for details). The caption is randomly short or long for audio samples that have both. The training compute $C$ is counted as three times the forward cost of both towers~\citep{kaplanScalingLawsNeural2020}, adapted to the distribution mix of the dataset. Table~\ref{tab:arch-sizes} shows that around 80\% of the computation is caused by the audio encoder, due to the quadratic cost of attention on longer inputs.

\textbf{Evaluation: } We evaluated our models on the following benchmarks. Massive Audio Embedding Benchmark (MAEB-30)~\citep{assadiMAEBMassiveAudio2026} is a collection of benchmarks that covers 30 individual tasks across speech, music, and environmental sounds, and spans classification, retrieval, and clustering tasks. The score is the unweighted mean over tasks. A task with several subsets, such as the languages of FLEURS, first averages within task. We group the tasks into speech (18 benchmarks), music (6 benchmarks), and sound (6 benchmarks) (Table~\ref{tab:maeb30-scaling}). Audio samples are truncated to 30 seconds for evaluation to match the length of our training samples.

In addition to MAEB-30, we evaluate our models on the Holistic Evaluation of Audio Representations benchmark (HEAR)~\citep{turianHEARHolisticEvaluation2022}. It  covers speech, music, and environmental sounds. Since this benchmark is designed for the evaluation of audio representations only, not for text-audio evaluations, we use it as a quality measure for our audio embeddings. Additionally, we evaluate on a third, more speech-oriented, Massive Sound Embedding Benchmark (MSEB)~\citep{heigoldMassiveSoundEmbedding2026}. 

\FloatBarrier
\section{Scaling Analysis}
\suppressfloats[t]

\begin{wraptable}{r}{0.5\textwidth}
\centering
\footnotesize
\caption{\textbf{Model tiers for scaling study:} Hidden dimension $d$, parameter counts, and forward-pass compute per encoder. Audio parameters include the MAP pool and projection layers. Text parameters include the pool and output projection but \emph{exclude} the $50{,}260\times d_t$ token-embedding table. Total parameters refer to the full model size, including embeddings. GFLOPs are forward-pass FLOPs per clip at the training length distribution.}

\setlength{\tabcolsep}{2.5pt}
\begin{tabular}{lrrrrrrr}
\toprule
     &     & \multicolumn{3}{c}{Parameters (M)} & \multicolumn{3}{c}{GFLOPs / clip} \\
\cmidrule(lr){3-5} \cmidrule(lr){6-8}
Tier & $d$ & Audio & Text & Total & Audio & Text & Total \\
\midrule
XXS & 224 &   8.3 &   7.8 &  27.3 &  21.6 &  4.6 &  26.1 \\
XS  & 320 &  16.6 &  15.7 &  48.3 &  38.0 &  8.8 &  46.8 \\
S   & 512 &  41.8 &  36.4 & 103.9 &  83.8 & 19.6 & 103.4 \\
M   & 640 &  64.4 &  56.2 & 152.8 & 123.1 & 29.8 & 152.9 \\
L   & 960 & 166.5 & 125.4 & 340.1 & 292.8 & 65.1 & 357.9 \\
\bottomrule
\end{tabular}
\label{tab:arch-sizes}
\end{wraptable}
Before scaling, we fixed the training setup in ablations of our cosine decay model (Appendix~\ref{sec:duration}). Training only on audio of at most 30 seconds scores 4.1 MAEB-30 points higher than training on audio of up to 120 seconds while using 3.5 times less compute, and the caption length has no effect beyond run-to-run variation.

\textbf{Setup: } Our scaling study includes five model sizes ranging from  27.3M to 340.1M parameters (Table~\ref{tab:arch-sizes}). We train these models on our 365M dataset, with samples-seen scales $D = \{1M,\allowbreak\ 2M,\allowbreak\ 4M,\allowbreak\ 8M,\allowbreak\ 16M,\allowbreak\ 32M,\allowbreak\ 64M,\allowbreak\ 128M,\allowbreak\ 256M,\allowbreak\ 512M,\allowbreak\ 1024M\}$, a total of 11 data scales. We do this by sampling the training batches with replacement within each pass over the complete training pool. We train our runs with a warmup-stable-decay learning rate scheduling, which allows separate cooldowns from a shared run~\citep{hageleScalingLawsComputeoptimal2024}. To allow enough optimization steps for smaller data scales, we split the runs into one short (up to 32M samples seen) and one long (up to 1B samples seen) training run. For more details on the hyperparameters used for these runs, refer to Appendix~\ref{app:scaling}.

\begin{figure}[t]
  \centering
  \begin{minipage}[t]{0.49\textwidth}\vspace{0pt}\centering
    \includegraphics[width=\linewidth]{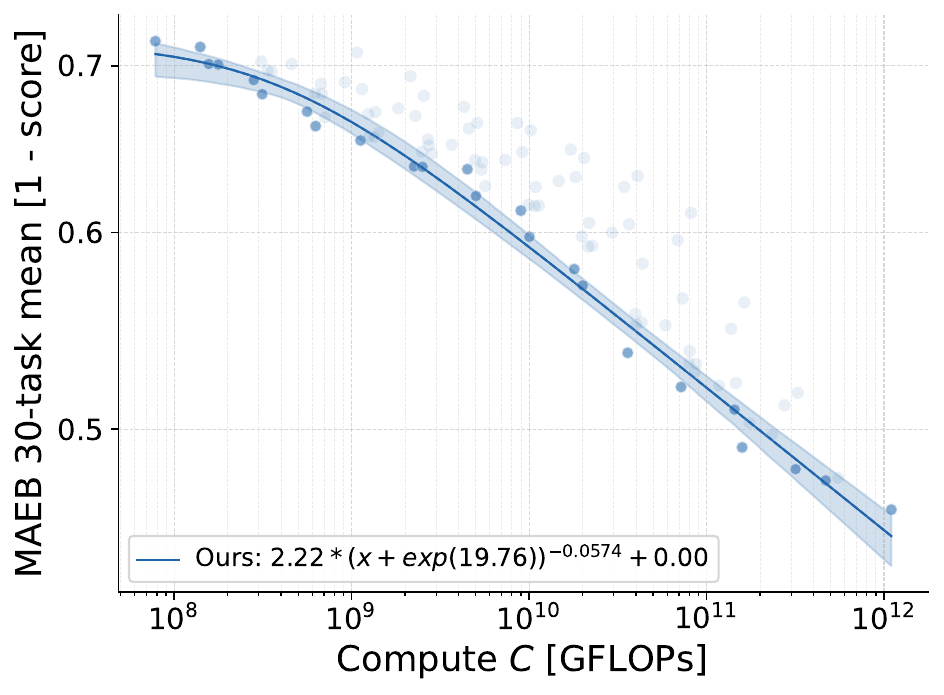}
  \end{minipage}\hfill
  \begin{minipage}[t]{0.49\textwidth}\vspace{0pt}\centering
    \includegraphics[width=\linewidth]{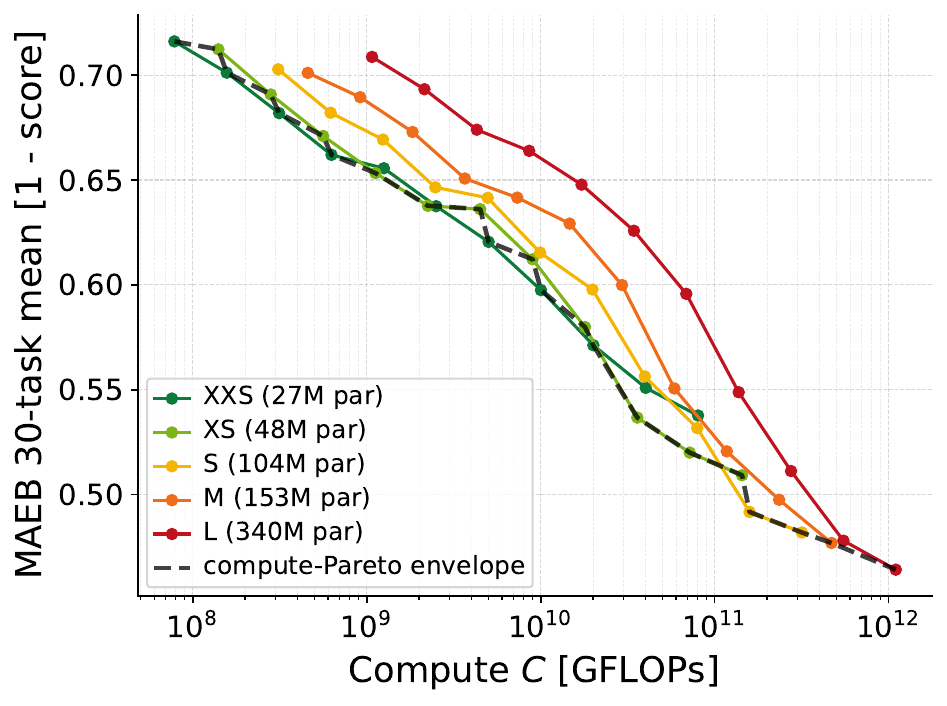}
  \end{minipage}
  \caption{\textbf{Scaling behavior on our LAION-Audio-XL dataset.} MAEB-30 error rate ($1 - \text{mean score}$) per compute $C$ (GFLOPs). \textbf{Left}: the compute-Pareto-optimal runs (solid points) of all warmup-stable-decay and cosine learning-rate-scheduled runs, with our fitted scaling function and its 95\% bootstrap CI (1000 resamples). With $\alpha=0.0574$, the error reduces by 3.9\% for each doubling of compute ($\approx +1.8$ mean MAEB-30 points at the largest budget). \textbf{Right}: each model size, from two WSD runs per size with individual cooldown phases, and the compute-Pareto envelope.}
  \label{fig:scaling_pareto_frontier_maeb}\label{fig:scaling_by_model_size_maeb}
\end{figure}
For the formulation of our scaling study, we follow \citet{nezhurinaScalingLawsRobust2026}, adapting $\mathcal{L} = aC^{b},$ with $C$ being compute and $\mathcal{L}$ referring to the mean MAEB-30 error score or subset of its 30 benchmarks, based on small compute scale saturation ($B_C$). This results in the following requirements for our scaling functions:
\begin{equation}
\begin{aligned}
  \mathcal{L}(C) &> 0 &&\text{(strictly positive)},\\
  \lim_{C \to \infty} \mathcal{L}(C) &= E &&\text{(irreducible error)},\\
  \frac{d\mathcal{L}}{dC} &< 0 &&\text{(monotonic decrease)}.
\end{aligned}
\end{equation}

After taking these three requirements into account, we get the functional form:
\begin{equation}
\mathcal{L}(C) = A_c \cdot (C + B_c)^{-\alpha_c} + E_c, \quad \text{for } \alpha_c > 0,
\label{eq:compute_scaling}
\end{equation}
\vspace{3pt}
with subscript $c$ specifying the compute dependency.

\textbf{Scaling Results: } We fit our experimental data to Eq.~\ref{eq:compute_scaling} as shown in Figure~\ref{fig:scaling_pareto_frontier_maeb}. Our architecture shows a clear scaling trend on our 365M audio-text dataset. The fitted curve through the Pareto frontier shows that with each doubling of compute, MAEB-30 error ($1-\text{mean score}$) is reduced by $\approx4\%$ or $\approx+1.8$ points at the largest budget. We also tested the quality of our scaling prediction by excluding all runs with $C\ge10^{11}$ GFLOPs. We are able to accurately predict the five runs on the compute-Pareto-frontier with $C\ge10^{11}$ GFLOPs (see Figure~\ref{fig:scaling_predictions}).

Figure~\ref{fig:scaling_by_model_size_maeb} shows that the two smallest sizes, XXS and XS, are the compute-optimal choice up to 256M seen samples. Only from 512M samples seen onwards do the larger sizes S, M and L reach the compute-Pareto frontier. Dataset scaling holds up to at least 256M for our smallest model size.

When splitting the MAEB-30 error score into its three domains (speech, music, and sounds) and examining their scaling behavior individually, we observe significant differences among them (Figure~\ref{fig:scaling_maeb_by_domain}). Although sound-related benchmarks saturate after $1.5\times10^{10}$ GFLOPs, speech and music do not saturate within the tested compute range up to $C_{limit} = 1.1\times10^{12}$ GFLOPs. We observe that there is only little improvement on speech benchmarks below $\approx10^{9}$ GFLOPs when scaling compute, while no such warmup is needed for music and sounds (Figure~\ref{fig:scaling_maeb_by_domain}).

\paragraph{Compute-Optimal Allocation}\label{sec:compute_optimal}
\newsavebox{\domainfigbox}
\begin{figure}[t]
  \centering
  \sbox{\domainfigbox}{\includegraphics[width=0.475\textwidth]{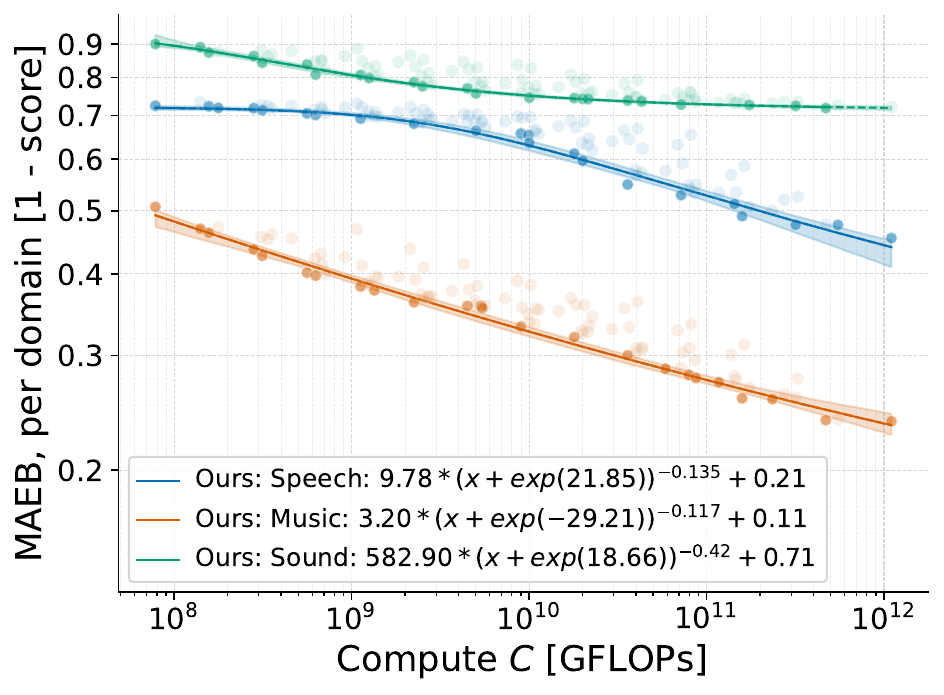}}
  \begin{minipage}[t]{0.475\textwidth}
    \centering
    \usebox{\domainfigbox}
    \caption{\textbf{Domain-specific scaling on our 365M dataset.} MAEB-30 error rate ($1 - \text{mean score}$) per compute $C$ (GFLOPs) per model size and domain. Solid points are the compute-Pareto-optimal WSD and cosine LR runs used for the fit, the band the 95\% bootstrap CI (1000 resamples). Speech and music scale well, sound saturates around $C=1.5\times10^{10}$.}
    \label{fig:scaling_maeb_by_domain}
  \end{minipage}\hfill
  \begin{minipage}[t]{0.505\textwidth}
    \centering
    \parbox[b][\ht\domainfigbox][c]{\linewidth}{\centering\includegraphics[width=\linewidth]{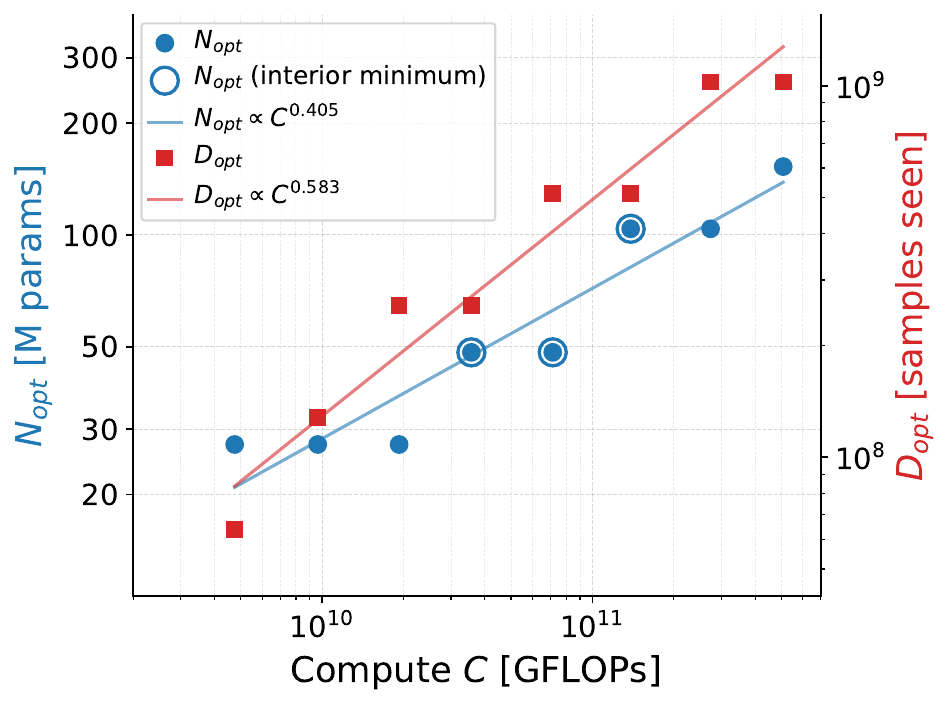}}
    \caption{\textbf{Compute-optimal allocation between model size and samples seen.} For each compute budget $C$ (GFLOPs), the long WSD run with the lowest MAEB-30 error gives the model size $N_{\text{opt}}$ (blue circles, left axis) and samples seen $D_{\text{opt}}$ (red squares, right axis). Circled points mark interior minima. Power-law fits give $N_{\text{opt}} \propto C^{0.405}$ and $D_{\text{opt}} \propto C^{0.583}$: the optimal number of samples grows faster than the optimal model size.}
    \label{fig:compute_optimal}
  \end{minipage}
\end{figure}

To estimate how a compute budget should be split between model size and data, we fit power laws $N_{\text{opt}}\propto C^{a}$ and $D_{\text{opt}}\propto C^{b}$ in log-log space to the compute-Pareto frontier of the long WSD runs for $C\in[5\times10^9, 5.5\times10^{11}]$ GFLOPs (Figure~\ref{fig:compute_optimal}). We exclude the short runs, as their optimum is always our smallest model, an artifact of our model-size range rather than the scaling trend. Even within the long runs, the optimum lies at XXS for the three smallest budgets and at our data cap of $1{,}024$M samples for the two largest, leaving only three interior minima. Since these boundary points are adjacent to the interior minima and a three-point fit is poorly constrained, we include them. Both boundaries bias the model size upwards: the true optimum may be smaller than XXS at low budgets, and a smaller model trained on more samples may win at the data cap. As these biases flatten and steepen the fit respectively, they partly cancel in the exponent of $N_{\text{opt}}$.

Our exponents ($N_{\text{opt}}\propto C^{0.405}$, $D_{\text{opt}}\propto C^{0.583}$) imply that additional compute should go predominantly to more samples rather than larger models. As compute scales with $N \cdot D$, the exponents sum to approximately one ($0.99$), so the optimal number of samples grows about $1.4\times$ faster than the optimal model size. This differs from language models, which scale model size and data roughly equally~\citep{hoffmannTrainingComputeOptimalLarge2022}, and from native multimodal models, which slightly favor model size ($N_{\text{opt}}\propto C^{0.526}$, $D_{\text{opt}}\propto C^{0.473}$)~\citep{shukorScalingLawsNative2025}. It closely matches the discrete audio language model SODA ($N_{\text{opt}}\propto C^{0.367}$, $D_{\text{opt}}\propto C^{0.579}$)~\citep{manakulScalingOpenDiscrete2026}, despite a different objective and fitting procedure. This might be explained by the lower information density of audio compared to text, and possibly by the noise of web-scale audio-text data, as lower data quality shifts the optimal allocation towards data in language models~\citep{deepseek-aiDeepSeekLLMScaling2024}. As our exponents stem from downstream error rather than loss, five discrete model sizes with partly boundary optima, and a compute range comparable to theirs ($\approx5\times10^{18}$--$5.5\times10^{20}$ vs.\ $3\times10^{18}$--$3\times10^{20}$ FLOPs), they are indicative rather than definitive.

Extrapolating the fits, the compute-optimal run for $C=10^{12}$~GFLOPs would use 180M parameters and 1.8B samples, and for $C=2\times10^{12}$~GFLOPs 240M parameters and 2.7B samples. This assumes repeated samples are as useful as new ones: on our 365M dataset, these correspond to about five and seven epochs, whereas repeated data in language models is nearly as useful as new data only for up to about four epochs~\citep{muennighoffScalingDataConstrainedLanguage2023}. Larger budgets would therefore also require more data. Note that this analysis differs from the validation-loss analysis in Figure~\ref{fig:scaling_val_loss_maeb}, which finds the best model size for a fixed number of samples rather than the best split of a fixed compute budget.

\section{Benchmark Results}
The following section discusses the benchmark results of the last checkpoints in our scaling study for all model sizes. All five model sizes were trained on $1{,}024$M data samples. We compare all our models to the leading models on the MAEB-30 leaderboard while distinguishing between smaller models below $1$B parameters and larger models far surpassing our size with more than $1$B parameters. Table~\ref{tab:scaling-final} shows this in terms of mean MAEB benchmark score and split into domain and task type for each benchmark.

\suppressfloats[t]
\begin{figure}[t]
  \centering
  \includegraphics[width=0.99\textwidth]{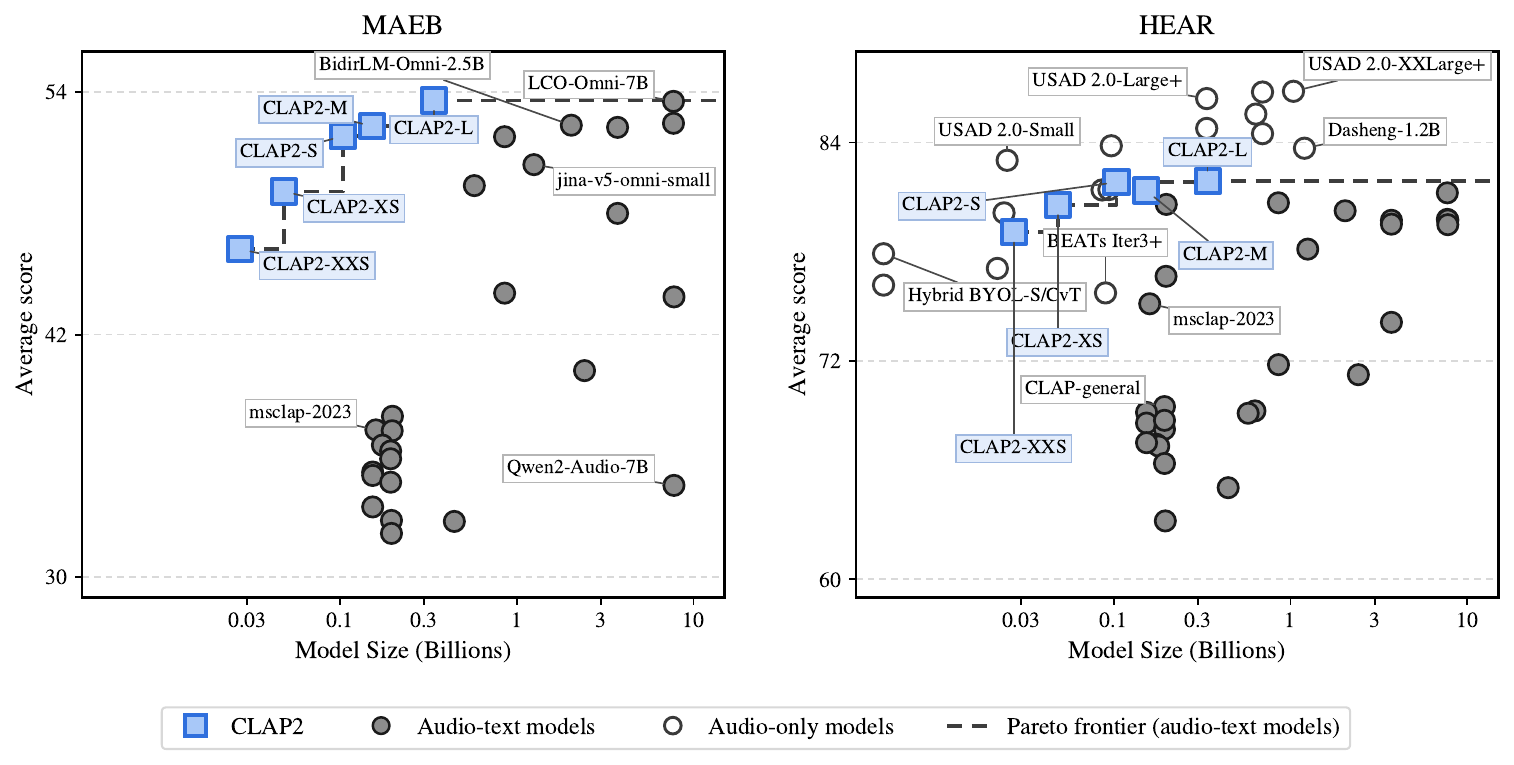}
  \caption{\textbf{Embedding model performance by size.} Average score across the individual tasks of MAEB (all 30 tasks) and HEAR (17 tasks). Blue squares are our own runs. Filled circles are audio-text models, and unfilled circles are audio-only models with no text component. Model sizes count only the audio and text parts of each model.}
  \label{fig:size_vs_score}
\end{figure}

Our models perform strongly across all MAEB domains. Looking just at the mean MAEB performance, we see that our models surpass others several times their size. Figure~\ref{fig:size_vs_score} plots the mean MAEB-30 and HEAR scores against model size. On both benchmarks, our models score higher than other audio-text models of similar size. Especially our smaller models show extraordinary performance for their parameter count. CLAP2-XXS to CLAP2-M are only beaten by models with at least ${8}\times$ more parameters on MAEB-30. With a mean MAEB-30 score of $53.6\%$, CLAP2-L has the highest mean in Table~\ref{tab:scaling-final}, above multi-modal foundation models with billions of parameters such as LCO-Embedding-Omni-7B ($53.5\%$).

We achieve strong performance on speech tasks while beating the best model on music with three of our models (Table~\ref{tab:scaling-final}). Our training data holds much more speech than any other domain. Rebalancing it evenly between speech, music, and other classes lowers the MAEB-30 score at 256M seen samples from 43.0\% to 41.7\% (Appendix~\ref{app:data_domain}). The original data distribution is therefore kept. 
On a per benchmark task breakdown, our models perform well on all of them except clustering which is known to misalign with other benchmarks as seen by the scores from other top-performing models~\citep{assadiMAEBMassiveAudio2026}. For the individual results for all 30 MAEB benchmarks, refer to Table~\ref{tab:maeb30-scaling} in the Appendix.

The appendix reports two additional benchmarks. On HEAR (Table~\ref{tab:hear}), CLAP2-S, CLAP2-M and CLAP2-L score between 81.4 and 81.9, higher than all other audio-text models (at most 81.2). Audio-only encoders such as USAD 2.0 \cite{changUSAD20Scaling2026} score up to 86.8. On MSEB (Table~\ref{tab:mseb}), CLAP2-L scores 30.3, second only to M2D-CLAP (30.7) and higher than all multimodal models with billions of parameters (at most 27.9).

\vspace{0.2em}
\section{Conclusion}
Our LAION-Audio-XL dataset shows that public datasets and web audio can support contrastive audio-language learning at scale with large-scale collection, filtering, deduplication, and captioning pipelines. Our focus on smaller models is backed by the observation that it is computationally more effective to train smaller models on more samples than to increase the parameter count, with the optimal number of samples seen growing $\approx1.4\times$ faster than the optimal model size. Training on audio of at most 30 seconds also produces a better model with less than one-third of the compute. Downstream performance follows a fitted scaling trend, with continued gains in speech and music. These results motivate further data scaling. In particular, training larger models on more than 1B samples would show whether the scaling trend continues. The released audio, links, captions, metadata, model checkpoints, and code allow other researchers to build on both the corpus and the models. CLAP2 is efficient to run, which makes it especially useful for data curation and text-to-audio generation.

\label{sec:mainend}

\clearpage

\section*{Acknowledgements}
This work was in part supported by the BMFTR (FKZ: 16IS24060), and the DFG (SFB 1233, project number: 276693517). Jenia Jitsev and Marianna Nezhurina further acknowledge funding from the European Union Horizon Programme under grant no. 101214398 (ELLIOT), co-funding from the Digital Europe Programme under grant no. 101195233 (openEuroLLM) and grant no. 101198470 (LLMs4EU), funding from the EuroHPC Joint Undertaking under grant no. 101182737 (MINERVA), funding by the German Federal Ministry of Research, Technology and Space (BMFTR) under grant no. 01IS24085C (OPENHAFM), and co-funding under the grant 16HPC117K (MINERVA).

We gratefully acknowledge the  Gauss Centre for Supercomputing e.V. (GCS) for providing computing resources through the John von Neumann Institute for Computing (NIC) on the JUWELS Booster and JUPITER supercomputers at the  Jülich Supercomputing Centre (JSC). We also acknowledge storage resources on JUST operated by JSC and supported by the Helmholtz Data Federation (HDF), computing time granted by JARA and JSC on the JURECA supercomputer, access through the JUREAP (JUPITER Early Access Program), and computing time provided through the Gauss AI Competition (reformo) on JUPITER via GCS and BMFTR.

LAION further acknowledges a public storage grant from  HuggingFace, enabling broad community access to open-source research outputs through HuggingFace repositories. We also thank the teams operating the supercomputing facilities at JSC for their support, especially Björn Hagemeier for assistance with data storage and handling.

\bibliography{bibliography,datasets}

@inproceedings{abbasSemDeDupDataefficientLearning2023,
  title = {{{SemDeDup}}: {{Data}}-Efficient Learning at Web-Scale through Semantic Deduplication},
  shorttitle = {{{SemDeDup}}},
  booktitle = {{{ICLR}} 2023 {{Workshop}} on {{Multimodal Representation Learning}}: {{Perks}} and {{Pitfalls}}},
  author = {Abbas, Amro Kamal Mohamed and Tirumala, Kushal and Simig, Daniel and Ganguli, Surya and Morcos, Ari S.},
  year = 2023,
  langid = {english}
}

@inproceedings{alonso-jimenezEfficientSupervisedTraining2023,
  title = {Efficient {{Supervised Training}} of {{Audio Transformers}} for {{Music Representation Learning}}},
  booktitle = {{{ISMIR}}},
  author = {{Alonso-Jim{\'e}nez}, Pablo and Serra, Xavier and Bogdanov, Dmitry},
  year = 2023,
  eprint = {2309.16418},
  primaryclass = {cs.SD},
  pages = {824--831},
  address = {Milan, Italy},
  doi = {10.5281/ZENODO.10265414},
  urldate = {2026-09-25},
  archiveprefix = {arXiv},
  langid = {english}
}

@inproceedings{aoSpeechT5UnifiedModalEncoderDecoder2022,
  title = {{{SpeechT5}}: {{Unified-Modal Encoder-Decoder Pre-Training}} for {{Spoken Language Processing}}},
  shorttitle = {{{SpeechT5}}},
  booktitle = {{{ACL}}},
  author = {Ao, Junyi and Wang, Rui and Zhou, Long and Wang, Chengyi and Ren, Shuo and Wu, Yu and Liu, Shujie and Ko, Tom and Li, Qing and Zhang, Yu and Wei, Zhihua and Qian, Yao and Li, Jinyu and Wei, Furu},
  editor = {Muresan, Smaranda and Nakov, Preslav and Villavicencio, Aline},
  year = 2022,
  month = may,
  pages = {5723--5738},
  publisher = {Association for Computational Linguistics},
  address = {Dublin, Ireland},
  doi = {10.18653/v1/2022.acl-long.393},
  langid = {english}
}

@misc{assadiMAEBMassiveAudio2026,
  title = {{{MAEB}}: {{Massive Audio Embedding Benchmark}}},
  shorttitle = {{{MAEB}}},
  author = {Assadi, Adnan El and Chung, Isaac and Xiao, Chenghao and Solomatin, Roman and Jha, Animesh and Chand, Rahul and Singh, Silky and Wang, Kaitlyn and Khan, Ali Sartaz and Nasser, Marc Moussa and Fong, Sufen and He, Pengfei and Xiao, Alan and Munot, Ayush Sunil and Shrivastava, Aditya and Gazizov, Artem and Muennighoff, Niklas and Enevoldsen, Kenneth},
  year = 2026,
  month = feb,
  number = {arXiv:2602.16008},
  eprint = {2602.16008},
  primaryclass = {cs},
  publisher = {arXiv},
  doi = {10.48550/arXiv.2602.16008},
  urldate = {2026-02-23},
  archiveprefix = {arXiv},
  langid = {english}
}

@inproceedings{baevskiWav2vec20Framework2020,
  title = {Wav2vec 2.0: {{A Framework}} for {{Self-Supervised Learning}} of {{Speech Representations}}},
  shorttitle = {Wav2vec 2.0},
  booktitle = {{{NeurIPS}}},
  author = {Baevski, Alexei and Zhou, Henry and Mohamed, Abdelrahman and Auli, Michael},
  editor = {Larochelle, H. and Ranzato, M. and Hadsell, R. and Balcan, M.F. and Lin, H.},
  year = 2020,
  volume = {33},
  pages = {12449--12460},
  publisher = {Curran Associates, Inc.},
  langid = {english}
}

@article{baiAudioSetCapsEnrichedAudioCaption2025,
  title = {{{AudioSetCaps}}: {{An Enriched Audio-Caption Dataset Using Automated Generation Pipeline}} with {{Large Audio}} and {{Language Models}}},
  shorttitle = {{{AudioSetCaps}}},
  author = {Bai, Jisheng and Liu, Haohe and Wang, Mou and Shi, Dongyuan and Wang, Wenwu and Plumbley, Mark D and Gan, Woon-Seng and Chen, Jianfeng},
  year = 2025,
  journal = {IEEE Transactions on Audio, Speech and Language Processing},
  volume = {33},
  pages = {2817--2829},
  publisher = {IEEE},
  doi = {10.1109/TASLPRO.2025.3583354},
  langid = {english}
}

@inproceedings{boizardBidirLMTextOmnimodal2026,
  title = {{{BidirLM}}: {{From Text}} to {{Omnimodal Bidirectional Encoders}} by {{Adapting}} and {{Composing Causal LLMs}}},
  shorttitle = {{{BidirLM}}},
  booktitle = {{{COLM}}},
  author = {Boizard, Nicolas and {Deschamps-Berger}, Th{\'e}o and {Gisserot-Boukhlef}, Hippolyte and Hudelot, Celine and Colombo, Pierre},
  year = 2026,
  langid = {english}
}

@article{bradley1952rank,
  title = {Rank {{Analysis}} of {{Incomplete Block Designs}}: {{I}}. the {{Method}} of {{Paired Comparisons}}},
  shorttitle = {Rank {{Analysis}} of {{Incomplete Block Designs}}},
  author = {Bradley, Ralph Allan and Terry, Milton E},
  year = 1952,
  journal = {Biometrika},
  volume = {39},
  number = {3/4},
  pages = {324--345},
  publisher = {JSTOR},
  langid = {english}
}

@inproceedings{burkhardtSpeechBasedAgeGender2023,
  title = {Speech-{{Based Age}} and {{Gender Prediction}} with {{Transformers}}},
  booktitle = {{{ITG Conference}} on {{Speech Communication}}},
  author = {Burkhardt, Felix and Wagner, Johannes and Wierstorf, Hagen and Eyben, Florian and Schuller, Bj{\"o}rn},
  year = 2023,
  pages = {46--50},
  publisher = {VDE},
  langid = {english}
}

@inproceedings{changUSAD20Scaling2026,
  title = {{{USAD}} 2.0: {{Scaling Representation Distillation}} for {{Universal Audio Understanding}}},
  shorttitle = {Usad 2.0},
  booktitle = {Interspeech},
  author = {Chang, Heng-Jui and Liu, Alexander and Bhati, Saurabhchand and Athi, Mrudula and Ratnarajah, Anton and Chhetri, Amit and Glass, James},
  year = 2026,
  pages = {2224--2229},
  issn = {2958-1796},
  doi = {10.21437/Interspeech.2026-1296},
  langid = {english}
}

@inproceedings{changUSADUniversalSpeech2025,
  title = {{{USAD}}: {{Universal Speech}} and {{Audio Representation}} via {{Distillation}}},
  shorttitle = {{{USAD}}},
  booktitle = {{{ASRU}}},
  author = {Chang, Heng-Jui and Bhati, Saurabhchand and Glass, James and Liu, Alexander H.},
  year = 2025,
  month = dec,
  pages = {1--8},
  publisher = {IEEE},
  address = {Honolulu, HI, USA},
  issn = {2997-6995},
  doi = {10.1109/ASRU65441.2025.11434802},
  urldate = {2026-11-03},
  copyright = {https://doi.org/10.15223/policy-029},
  isbn = {979-8-3315-4426-3},
  langid = {english}
}

@inproceedings{chenBEATsAudioPreTraining2023,
  title = {{{BEATs}}: {{Audio Pre-Training}} with {{Acoustic Tokenizers}}},
  shorttitle = {{{BEATs}}},
  booktitle = {{{ICML}}},
  author = {Chen, Sanyuan and Wu, Yu and Wang, Chengyi and Liu, Shujie and Tompkins, Daniel and Chen, Zhuo and Che, Wanxiang and Yu, Xiangzhan and Wei, Furu},
  editor = {Krause, Andreas and Brunskill, Emma and Cho, Kyunghyun and Engelhardt, Barbara and Sabato, Sivan and Scarlett, Jonathan},
  year = 2023,
  month = jul,
  series = {Proceedings of {{Machine Learning Research}}},
  volume = {202},
  pages = {5178--5193},
  publisher = {PMLR},
  langid = {english}
}

@inproceedings{chenE5OmniExplicitCrossModal2026,
  title = {E5-{{Omni}}: {{Explicit Cross-Modal Alignment}} for {{Omni-Modal Embeddings}}},
  shorttitle = {E5-{{Omni}}},
  booktitle = {Findings of {{ACL}}},
  author = {Chen, Haonan and Gao, Sicheng and Timofte, Radu and Sakai, Tetsuya and Dou, Zhicheng},
  editor = {Liakata, Maria and Moreira, Viviane P. and Zhang, Jiajun and Jurgens, David},
  year = 2026,
  month = jul,
  pages = {19430--19443},
  publisher = {Association for Computational Linguistics},
  address = {San Diego, California, United States},
  doi = {10.18653/v1/2026.findings-acl.970},
  urldate = {2026-10-13},
  isbn = {979-8-89176-395-1},
  langid = {english}
}

@inproceedings{chenVGGSoundLargescaleAudiovisual2020,
  title = {{{VGGSound}}: {{A Large-Scale Audio-Visual Dataset}}},
  booktitle = {{{ICASSP}}},
  author = {Chen, Honglie and Xie, Weidi and Vedaldi, Andrea and Zisserman, Andrew},
  year = 2020,
  pages = {721--725},
  publisher = {IEEE}
}

@inproceedings{chertiReproducibleScalingLaws2023,
  title = {Reproducible {{Scaling Laws}} for {{Contrastive Language-Image Learning}}},
  booktitle = {{{CVPR}}},
  author = {Cherti, Mehdi and Beaumont, Romain and Wightman, Ross and Wortsman, Mitchell and Ilharco, Gabriel and Gordon, Cade and Schuhmann, Christoph and Schmidt, Ludwig and Jitsev, Jenia},
  year = 2023,
  pages = {2818--2829},
  publisher = {IEEE},
  doi = {10.1109/CVPR52729.2023.00276},
  langid = {english}
}

@misc{chuQwen2AudioTechnicalReport2024,
  title = {Qwen2-{{Audio Technical Report}}},
  author = {Chu, Yunfei and Xu, Jin and Yang, Qian and Wei, Haojie and Wei, Xipin and Guo, Zhifang and Leng, Yichong and Lv, Yuanjun and He, Jinzheng and Lin, Junyang and Zhou, Chang and Zhou, Jingren},
  year = 2024,
  month = jul,
  number = {arXiv:2407.10759},
  eprint = {2407.10759},
  primaryclass = {eess},
  publisher = {arXiv},
  doi = {10.48550/arXiv.2407.10759},
  urldate = {2026-01-06},
  archiveprefix = {arXiv},
  langid = {english}
}

@misc{commoncrawlfoundationCommonCrawl2024,
  title = {Common {{Crawl}}},
  author = {{Common Crawl Foundation}},
  year = 2024,
  howpublished = {https://commoncrawl.org},
  langid = {english}
}

@inproceedings{cramerLookListenLearn2019,
  title = {Look, {{Listen}}, and {{Learn More}}: {{Design Choices}} for {{Deep Audio Embeddings}}},
  shorttitle = {Look, {{Listen}}, and {{Learn More}}},
  booktitle = {{{ICASSP}}},
  author = {Cramer, Aurora Linh and Wu, Ho-Hsiang and Salamon, Justin and Bello, Juan Pablo},
  year = 2019,
  pages = {3852--3856},
  doi = {10.1109/ICASSP.2019.8682475},
  langid = {english}
}

@misc{deepseek-aiDeepSeekLLMScaling2024,
  title = {{{DeepSeek LLM}}: {{Scaling Open-Source Language Models}} with {{Longtermism}}},
  shorttitle = {{{DeepSeek LLM}}},
  author = {{DeepSeek-AI} and {:} and Bi, Xiao and Chen, Deli and Chen, Guanting and Chen, Shanhuang and Dai, Damai and Deng, Chengqi and Ding, Honghui and Dong, Kai and Du, Qiushi and Fu, Zhe and Gao, Huazuo and Gao, Kaige and Gao, Wenjun and Ge, Ruiqi and Guan, Kang and Guo, Daya and Guo, Jianzhong and Hao, Guangbo and Hao, Zhewen and He, Ying and Hu, Wenjie and Huang, Panpan and Li, Erhang and Li, Guowei and Li, Jiashi and Li, Yao and Li, Y. K. and Liang, Wenfeng and Lin, Fangyun and Liu, A. X. and Liu, Bo and Liu, Wen and Liu, Xiaodong and Liu, Xin and Liu, Yiyuan and Lu, Haoyu and Lu, Shanghao and Luo, Fuli and Ma, Shirong and Nie, Xiaotao and Pei, Tian and Piao, Yishi and Qiu, Junjie and Qu, Hui and Ren, Tongzheng and Ren, Zehui and Ruan, Chong and Sha, Zhangli and Shao, Zhihong and Song, Junxiao and Su, Xuecheng and Sun, Jingxiang and Sun, Yaofeng and Tang, Minghui and Wang, Bingxuan and Wang, Peiyi and Wang, Shiyu and Wang, Yaohui and Wang, Yongji and Wu, Tong and Wu, Y. and Xie, Xin and Xie, Zhenda and Xie, Ziwei and Xiong, Yiliang and Xu, Hanwei and Xu, R. X. and Xu, Yanhong and Yang, Dejian and You, Yuxiang and Yu, Shuiping and Yu, Xingkai and Zhang, B. and Zhang, Haowei and Zhang, Lecong and Zhang, Liyue and Zhang, Mingchuan and Zhang, Minghua and Zhang, Wentao and Zhang, Yichao and Zhao, Chenggang and Zhao, Yao and Zhou, Shangyan and Zhou, Shunfeng and Zhu, Qihao and Zou, Yuheng},
  year = 2024,
  number = {arXiv:2401.02954},
  eprint = {2401.02954},
  primaryclass = {cs.CL},
  publisher = {arXiv},
  doi = {10.48550/ARXIV.2401.02954},
  urldate = {2026-11-03},
  archiveprefix = {arXiv},
  copyright = {arXiv.org perpetual, non-exclusive license},
  langid = {english}
}

@inproceedings{dehghaniPatchNpackNaViT2023,
  title = {Patch {{N}}'pack: {{NaViT}}, a {{Vision Transformer}} for {{Any Aspect Ratio}} and {{Resolution}}},
  shorttitle = {Patch {{N}}'pack},
  booktitle = {{{NeurIPS}}},
  author = {Dehghani, Mostafa and Mustafa, Basil and Djolonga, Josip and Heek, Jonathan and Minderer, Matthias and Caron, Mathilde and Steiner, Andreas and Puigcerver, Joan and Geirhos, Robert and Alabdulmohsin, Ibrahim M and {Others}},
  year = 2023,
  volume = {36},
  pages = {2252--2274},
  doi = {10.52202/075280-0106},
  langid = {english}
}

@inproceedings{dinkelCEDConsistentEnsemble2024,
  title = {{{CED}}: {{Consistent Ensemble Distillation}} for {{Audio Tagging}}},
  shorttitle = {Ced},
  booktitle = {{{ICASSP}}},
  author = {Dinkel, Heinrich and Wang, Yongqing and Yan, Zhiyong and Zhang, Junbo and Wang, Yujun},
  year = 2024,
  month = apr,
  pages = {291--295},
  issn = {2379-190X},
  doi = {10.1109/ICASSP48485.2024.10446348},
  urldate = {2026-10-13},
  langid = {english}
}

@inproceedings{dinkelGLAPGeneralContrastive2026,
  title = {{{GLAP}}: {{General Contrastive Audio-Text Pretraining Across Domains}} and {{Languages}}},
  shorttitle = {{{GLAP}}},
  booktitle = {{{ICASSP}}},
  author = {Dinkel, Heinrich and Yan, Zhiyong and Wang, Tianzi and Wang, Yongqing and Sun, Xingwei and Niu, Yadong and Liu, Jizhong and Li, Gang and Zhang, Junbo and Luan, Jian},
  year = 2026,
  pages = {14737--14741},
  publisher = {IEEE},
  langid = {english}
}

@misc{dinkelMiDashengLMEfficientAudio2025,
  title = {{{MiDashengLM}}: {{Efficient Audio Understanding}} with {{General Audio Captions}}},
  shorttitle = {{{MiDashengLM}}},
  author = {Dinkel, Heinrich and Li, Gang and Liu, Jizhong and Luan, Jian and Niu, Yadong and Sun, Xingwei and Wang, Tianzi and Xiao, Qiyang and Zhang, Junbo and Zhou, Jiahao},
  year = 2025,
  publisher = {ArXiv},
  doi = {10.48550/arXiv.2508.03983},
  langid = {english}
}

@inproceedings{dinkelScalingMaskedAudio2024,
  title = {Scaling up {{Masked Audio Encoder Learning}} for {{General Audio Classification}}},
  booktitle = {Interspeech},
  author = {Dinkel, Heinrich and Yan, Zhiyong and Wang, Yongqing and Zhang, Junbo and Wang, Yujun and Wang, Bin},
  year = 2024,
  pages = {547--551},
  issn = {2958-1796},
  doi = {10.21437/Interspeech.2024-246},
  langid = {english}
}

@inproceedings{dohLPMusicCapsLLMBasedPseudo2023,
  title = {{{LP-MusicCaps}}: {{LLM-Based Pseudo Music Captioning}}},
  shorttitle = {{{LP-MusicCaps}}},
  booktitle = {{{ISMIR}}},
  author = {Doh, Seungheon and Choi, Keunwoo and Lee, Jongpil and Nam, Juhan},
  year = 2023,
  pages = {409--416},
  publisher = {International Society for Music Information Retrieval Conference},
  doi = {10.48550/arxiv.2307.16372},
  langid = {english}
}

@misc{doshiIntroducingGemini362026,
  title = {Introducing {{Gemini}} 3.6 {{Flash}}, 3.5 {{Flash-Lite}}, and 3.5 {{Flash Cyber}}},
  author = {Doshi, Tulsee},
  year = 2026,
  month = jul,
  urldate = {2026-09-25},
  howpublished = {https://blog.google/innovation-and-ai/models-and-research/gemini-models/gemini-3-6-flash-3-5-flash-lite-3-5-flash-cyber/},
  langid = {english}
}

@inproceedings{dosovitskiyImageWorth16x162021,
  title = {An {{Image Is Worth}} 16x16 {{Words}}: {{Transformers}} for {{Image Recognition}} at {{Scale}}},
  shorttitle = {An {{Image Is Worth}} 16x16 {{Words}}},
  booktitle = {{{ICLR}}},
  author = {Dosovitskiy, Alexey and Beyer, Lucas and Kolesnikov, Alexander and Weissenborn, Dirk and Zhai, Xiaohua and Unterthiner, Thomas and Dehghani, Mostafa and Minderer, Matthias and Heigold, Georg and Gelly, Sylvain and Uszkoreit, Jakob and Houlsby, Neil},
  year = 2021,
  langid = {english}
}

@inproceedings{drossosClothoAudioCaptioning2020,
  title = {Clotho: {{An Audio Captioning Dataset}}},
  shorttitle = {Clotho},
  booktitle = {{{ICASSP}}},
  author = {Drossos, Konstantinos and Lipping, Samuel and Virtanen, Tuomas},
  year = 2020,
  month = may,
  pages = {736--740},
  issn = {2379-190X},
  doi = {10.1109/ICASSP40776.2020.9052990},
  langid = {english}
}

@inproceedings{elbannaBYOLSLearningSelfSupervised2022,
  title = {{{BYOL-S}}: {{Learning Self-Supervised Speech Representations}} by {{Bootstrapping}}},
  shorttitle = {Byol-{{S}}},
  booktitle = {{{HEAR}} ({{NeurIPS}} 2021 {{Competition}})},
  author = {Elbanna, Gasser and {Scheidwasser-Clow}, Neil and Kegler, Mikolaj and Beckmann, Pierre and El Hajal, Karl and Cernak, Milos},
  editor = {Turian, Joseph and Schuller, Bj{\"o}rn W. and Herremans, Dorien and Kirchoff, Katrin and Perera, Paola Garcia and Esling, Philippe},
  year = 2021,
  series = {Proceedings of {{Machine Learning Research}}},
  volume = {166},
  pages = {25--47},
  publisher = {PMLR},
  langid = {english}
}

@inproceedings{elbannaHybridHandcraftedLearnable2022,
  title = {Hybrid {{Handcrafted}} and {{Learnable Audio Representation}} for {{Analysis}} of {{Speech Under Cognitive}} and {{Physical Load}}},
  booktitle = {Interspeech},
  author = {Elbanna, Gasser and Biryukov, Alice and {Scheidwasser-Clow}, Neil and Orlandic, Lara and Mainar, Pablo and Kegler, Mikolaj and Beckmann, Pierre and Cernak, Milos},
  year = 2022,
  pages = {386--390},
  issn = {2958-1796},
  doi = {10.21437/Interspeech.2022-10498},
  langid = {english}
}

@inproceedings{elizaldeCLAPLearningAudio2023,
  title = {{{CLAP Learning Audio Concepts}} from {{Natural Language Supervision}}},
  booktitle = {{{ICASSP}}},
  author = {Elizalde, Benjamin and Deshmukh, Soham and Al Ismail, Mahmoud and Wang, Huaming},
  year = 2023,
  month = jun,
  pages = {1--5},
  publisher = {IEEE},
  doi = {10.1109/icassp49357.2023.10095889},
  langid = {english}
}

@inproceedings{elizaldeNaturalLanguageSupervision2024,
  title = {Natural {{Language Supervision}} for {{General-Purpose Audio Representations}}},
  booktitle = {{{ICASSP}}},
  author = {Elizalde, Benjamin and Deshmukh, Soham and Wang, Huaming},
  year = 2024,
  month = apr,
  eprint = {2309.05767},
  primaryclass = {cs, eess},
  pages = {336--340},
  publisher = {IEEE},
  address = {Seoul, Korea, Republic of},
  doi = {10.1109/ICASSP48485.2024.10448504},
  urldate = {2024-11-13},
  archiveprefix = {arXiv},
  copyright = {https://doi.org/10.15223/policy-029},
  isbn = {979-8-3503-4485-1},
  langid = {english}
}

@misc{EmbeddingGemma2Open2026,
  title = {{{EmbeddingGemma}} 2: {{An Open}}, {{Lightweight Multimodal Embedding Model}}},
  shorttitle = {{{EmbeddingGemma}} 2},
  author = {Dua, Sahil and Vera, Henrique Schechter},
  year = 2026,
  month = oct,
  journal = {Google},
  urldate = {2026-10-14},
  howpublished = {https://blog.google/innovation-and-ai/technology/developers-tools/embeddinggemma-2/},
  langid = {english}
}

@inproceedings{gadreDataCompSearchNext2023,
  title = {{{DataComp}}: {{In Search}} of the {{Next Generation}} of {{Multimodal Datasets}}},
  shorttitle = {{{DataComp}}},
  booktitle = {{{NeurIPS}}},
  author = {Gadre, Samir Yitzhak and Ilharco, Gabriel and Fang, Alex and Hayase, Jonathan and Smyrnis, Georgios and Nguyen, Thao and Marten, Ryan and Wortsman, Mitchell and Ghosh, Dhruba and Zhang, Jieyu and {Others}},
  year = 2023,
  volume = {36},
  pages = {27092--27112},
  doi = {10.48550/arXiv.2304.14108},
  langid = {english}
}

@inproceedings{gemmekeAudioSetOntology2017,
  title = {Audio {{Set}}: {{An}} Ontology and Human-Labeled Dataset for Audio Events},
  shorttitle = {Audio {{Set}}},
  booktitle = {{{ICASSP}}},
  author = {Gemmeke, Jort F. and Ellis, Daniel P. W. and Freedman, Dylan and Jansen, Aren and Lawrence, Wade and Moore, R. Channing and Plakal, Manoj and Ritter, Marvin},
  year = 2017,
  month = mar,
  pages = {776--780},
  publisher = {IEEE},
  address = {New Orleans, LA},
  issn = {2379-190X},
  doi = {10.1109/ICASSP.2017.7952261},
  urldate = {2026-11-03},
  isbn = {978-1-5090-4117-6},
  langid = {english}
}

@misc{ghoshAudioFlamingoNext2026,
  title = {Audio {{Flamingo Next}}: {{Next-Generation Open Audio-Language Models}} for {{Speech}}, {{Sound}}, and {{Music}}},
  shorttitle = {Audio {{Flamingo Next}}},
  author = {Ghosh, Sreyan and Goel, Arushi and Jayakumar, Kaousheik and Koroshinadze, Lasha and Anand, Nishit and Kong, Zhifeng and Gururani, Siddharth and Lee, Sang-gil and Kim, Jaehyeon and Aljafari, Aya and {Others}},
  year = 2026,
  publisher = {ArXiv},
  doi = {10.48550/arXiv.2604.10905},
  langid = {english}
}

@inproceedings{ghoshSynthioAugmentingSmallScale2025,
  title = {Synthio: {{Augmenting Small-Scale Audio Classification Datasets}} with {{Synthetic Data}}},
  shorttitle = {Synthio},
  booktitle = {{{ICLR}}},
  author = {Ghosh, Sreyan and Kumar, Sonal and Kong, Zhifeng and Valle, Rafael and Catanzaro, Bryan and Manocha, Dinesh},
  year = 2025,
  volume = {2025},
  pages = {66689--66711},
  langid = {english}
}

@inproceedings{goelAudioFlamingo32025,
  title = {Audio {{Flamingo}} 3: {{Advancing Audio Intelligence With Fully Open Large Audio Language Models}}},
  shorttitle = {Audio {{Flamingo}} 3},
  booktitle = {{{NeurIPS}}},
  author = {Goel, Arushi and Ghosh, Sreyan and Kim, Jaehyeon and Kumar, Sonal and Kong, Zhifeng and Lee, Sang-gil and Yang, Chao-Han Huck and Duraiswami, Ramani and Manocha, Dinesh and Valle, Rafael and Catanzaro, Bryan},
  year = 2025,
  month = jul,
  eprint = {2507.08128},
  primaryclass = {cs},
  doi = {10.48550/arXiv.2507.08128},
  urldate = {2025-09-01},
  archiveprefix = {arXiv},
  langid = {english}
}

@inproceedings{gongASTAudioSpectrogram2021,
  title = {{{AST}}: {{Audio Spectrogram Transformer}}},
  shorttitle = {{{AST}}},
  booktitle = {Interspeech},
  author = {Gong, Yuan and Chung, Yu-An and Glass, James},
  year = 2021,
  pages = {571--575},
  issn = {2958-1796},
  doi = {10.21437/Interspeech.2021-698},
  langid = {english}
}

@inproceedings{hageleScalingLawsComputeoptimal2024,
  title = {Scaling {{Laws}} and {{Compute-Optimal Training Beyond Fixed Training Durations}}},
  booktitle = {{{NeurIPS}}},
  author = {H{\"a}gele, Alexander and Bakouch, Elie and Kosson, Atli and Allal, Loubna B and Von Werra, Leandro and Jaggi, Martin},
  year = 2024,
  volume = {37},
  pages = {76232--76264},
  doi = {10.52202/079017-2427},
  langid = {english}
}

@article{haji-aliTamingDataTransformers2026,
  title = {Taming {{Data}} and {{Transformers}} for {{Audio Generation}}},
  author = {{Haji-Ali}, Moayed and Menapace, Willi and Siarohin, Aliaksandr and Balakrishnan, Guha and Ordonez, Vicente},
  year = 2026,
  journal = {International Journal of Computer Vision},
  volume = {134},
  number = {3},
  pages = {87},
  publisher = {Springer},
  doi = {10.1007/s11263-025-02632-y},
  langid = {english}
}

@inproceedings{heigoldMassiveSoundEmbedding2026,
  title = {Massive {{Sound Embedding Benchmark}} ({{MSEB}})},
  booktitle = {{{NeurIPS}}},
  author = {Heigold, Georg and Variani, Ehsan and Bagby, Tom and Allauzen, Cyril and Ma, Ji and Kumar, Shankar and Riley, Michael D},
  year = 2025,
  volume = {38},
  langid = {english}
}

@misc{hendrycksGaussianErrorLinear2016,
  title = {Gaussian {{Error Linear Units}} ({{GELUs}})},
  author = {Hendrycks, Dan and Gimpel, Kevin},
  year = 2016,
  number = {1606.08415},
  eprint = {1606.08415},
  publisher = {ArXiv},
  doi = {10.48550/arXiv.1606.08415},
  archiveprefix = {arXiv}
}

@inproceedings{henryQuerykeyNormalizationTransformers2020,
  title = {Query-{{Key Normalization}} for {{Transformers}}},
  booktitle = {Findings of {{EMNLP}}},
  author = {Henry, Alex and Dachapally, Prudhvi Raj and Pawar, Shubham Shantaram and Chen, Yuxuan},
  editor = {Cohn, Trevor and He, Yulan and Liu, Yang},
  year = 2020,
  month = nov,
  pages = {4246--4253},
  publisher = {Association for Computational Linguistics},
  address = {Online},
  doi = {10.18653/v1/2020.findings-emnlp.379}
}

@inproceedings{heoRotaryPositionEmbedding2025,
  title = {Rotary {{Position Embedding}} for {{Vision Transformer}}},
  booktitle = {{{ECCV}}},
  author = {Heo, Byeongho and Park, Song and Han, Dongyoon and Yun, Sangdoo},
  editor = {Leonardis, Ale{\v s} and Ricci, Elisa and Roth, Stefan and Russakovsky, Olga and Sattler, Torsten and Varol, G{\"u}l},
  year = 2024,
  pages = {289--305},
  publisher = {Springer Nature Switzerland},
  address = {Cham},
  isbn = {978-3-031-72684-2},
  langid = {english}
}

@misc{hernandezScalingLawsInterpretability2022,
  title = {Scaling {{Laws}} and {{Interpretability}} of {{Learning}} from {{Repeated Data}}},
  author = {Hernandez, Danny and Brown, Tom and Conerly, Tom and DasSarma, Nova and Drain, Dawn and {El-Showk}, Sheer and Elhage, Nelson and {Hatfield-Dodds}, Zac and Henighan, Tom and Hume, Tristan and Johnston, Scott and Mann, Ben and Olah, Chris and Olsson, Catherine and Amodei, Dario and Joseph, Nicholas and Kaplan, Jared and McCandlish, Sam},
  year = 2022,
  month = may,
  number = {arXiv:2205.10487},
  eprint = {2205.10487},
  primaryclass = {cs.LG},
  publisher = {arXiv},
  doi = {10.48550/arXiv.2205.10487},
  urldate = {2026-09-21},
  archiveprefix = {arXiv},
  langid = {english}
}

@misc{hochlehnertLAIONBVD10MillionHourOpen2026,
  title = {{{LAION-BVD}}: {{A}} 10-{{Million-Hour Open Video Dataset}} for {{Multimodal Pre}}-Training},
  shorttitle = {{{LAION-BVD}}},
  author = {Hochlehnert, Andreas and Nezhurina, Marianna and Cherti, Mehdi and Radonjic, Andrej and Wiedemer, Thadd{\"a}us and Schuhmann, Christoph and Beaumont, Romain and Brendel, Wieland and Sch{\"o}lkopf, Bernhard and Koepke, A. Sophia and Jitsev, Jenia and Bethge, Matthias},
  year = 2026,
  month = aug,
  publisher = {ArXiv},
  doi = {10.48550/ARXIV.2608.24845},
  urldate = {2026-11-03},
  copyright = {Creative Commons Attribution 4.0 International},
  langid = {english}
}

@inproceedings{hoffmannTrainingComputeOptimalLarge2022,
  title = {Training {{Compute-Optimal Large Language Models}}},
  booktitle = {{{NeurIPS}}},
  author = {Hoffmann, Jordan and Borgeaud, Sebastian and Mensch, Arthur and Buchatskaya, Elena and Cai, Trevor and Rutherford, Eliza and {de Las Casas}, Diego and Hendricks, Lisa Anne and Welbl, Johannes and Clark, Aidan and Hennigan, Tom and Noland, Eric and Millican, Katie and {van den Driessche}, George and Damoc, Bogdan and Guy, Aurelia and Osindero, Simon and Simonyan, Karen and Elsen, Erich and Vinyals, Oriol and Rae, Jack W. and Sifre, Laurent},
  year = 2022,
  series = {Nips '22},
  publisher = {Curran Associates Inc.},
  address = {Red Hook, NY, USA},
  articleno = {2176},
  isbn = {978-1-7138-7108-8},
  langid = {english}
}

@misc{honickeJinaEmbeddingsv5OmniGeometryPreservingEmbeddings2026,
  title = {Jina-{{Embeddings}}-v5-{{Omni}}: {{Geometry-Preserving Embeddings}} via {{Locked Aligned Towers}}},
  shorttitle = {Jina-{{Embeddings}}-v5-{{Omni}}},
  author = {H{\"o}nicke, Florian and G{\"u}nther, Michael and Koukounas, Andreas and Akram, Mohammad Kalim and Sturua, Saba and Xiao, Han},
  year = 2026,
  month = aug,
  number = {arXiv:2605.08384},
  eprint = {2605.08384},
  primaryclass = {cs.CL},
  publisher = {arXiv},
  doi = {10.48550/arXiv.2605.08384},
  urldate = {2026-10-13},
  archiveprefix = {arXiv},
  langid = {english}
}

@inproceedings{huMiniCPMUnveilingPotential2024,
  title = {{{MiniCPM}}: {{Unveiling}} the {{Potential}} of {{Small Language Models}} with {{Scalable Training Strategies}}},
  shorttitle = {{{MiniCPM}}},
  booktitle = {{{COLM}}},
  author = {Hu, Shengding and Tu, Yuge and Han, Xu and Cui, Ganqu and He, Chaoqun and Zhao, Weilin and Long, Xiang and Zheng, Zhi and Fang, Yewei and Huang, Yuxiang and Zhang, Xinrong and Thai, Zhen Leng and Wang, Chongyi and Yao, Yuan and Zhao, Chenyang and Zhou, Jie and Cai, Jie and Zhai, Zhongwu and Ding, Ning and Jia, Chao and Zeng, Guoyang and Li, Dahai and Liu, Zhiyuan and Sun, Maosong},
  year = 2024,
  langid = {english}
}

@inproceedings{hungTangoFluxSuperFast2026,
  title = {{{TangoFlux}}: {{Super Fast}} and {{Faithful Text}} to {{Audio Generation}} with {{Flow Matching}} and {{CLAP-Ranked Preference Optimization}}},
  shorttitle = {{{TangoFlux}}},
  booktitle = {{{ICLR}}},
  author = {Hung, Chia-Yu and Majumder, Navonil and Kong, Zhifeng and Mehrish, Ambuj and Zadeh, Amir and Li, Chuan and Valle, Rafael and Catanzaro, Bryan and Poria, Soujanya},
  editor = {Vondrick, C. and Hariharan, B. and Raffel, C. and Pinto, L. and Yang, D. and Faust, A.},
  year = 2026,
  volume = {2026},
  pages = {150793--150816},
  langid = {english}
}

@misc{kaplanScalingLawsNeural2020,
  title = {Scaling {{Laws}} for {{Neural Language Models}}},
  author = {Kaplan, Jared and McCandlish, Sam and Henighan, Tom and Brown, Tom B. and Chess, Benjamin and Child, Rewon and Gray, Scott and Radford, Alec and Wu, Jeffrey and Amodei, Dario},
  year = 2020,
  number = {arXiv:2001.08361},
  eprint = {2001.08361},
  primaryclass = {cs},
  publisher = {arXiv},
  doi = {10.48550/ARXIV.2001.08361},
  urldate = {2026-11-03},
  archiveprefix = {arXiv},
  copyright = {arXiv.org perpetual, non-exclusive license},
  langid = {english}
}

@inproceedings{kimAudioCapsGeneratingCaptions2019,
  title = {{{AudioCaps}}: {{Generating Captions}} for {{Audios}} in {{The Wild}}},
  shorttitle = {{{AudioCaps}}},
  booktitle = {{{NAACL}}},
  author = {Kim, Chris Dongjoo and Kim, Byeongchang and Lee, Hyunmin and Kim, Gunhee},
  year = 2019,
  month = jun,
  pages = {119--132},
  publisher = {Association for Computational Linguistics},
  address = {Minneapolis, Minnesota},
  doi = {10.18653/v1/N19-1011},
  urldate = {2022-08-24},
  langid = {english}
}

@inproceedings{kimCREPEConvolutionalRepresentation2018,
  title = {{{CREPE}}: {{A Convolutional Representation}} for {{Pitch Estimation}}},
  shorttitle = {{{CREPE}}},
  booktitle = {{{ICASSP}}},
  author = {Kim, Jong Wook and Salamon, Justin and Li, Peter and Bello, Juan Pablo},
  year = 2018,
  pages = {161--165},
  publisher = {IEEE},
  doi = {10.1109/ICASSP.2018.8461329},
  langid = {english}
}

@article{koepkeAudioRetrievalNatural2023,
  title = {Audio {{Retrieval}} with {{Natural Language Queries}}: {{A Benchmark Study}}},
  shorttitle = {Audio {{Retrieval}} with {{Natural Language Queries}}},
  author = {Koepke, A Sophia and Oncescu, Andreea-Maria and Henriques, Jo{\~a}o F and Akata, Zeynep and Albanie, Samuel},
  year = 2023,
  journal = {IEEE Transactions on Multimedia},
  volume = {25},
  pages = {2675--2685},
  publisher = {IEEE},
  doi = {10.1109/tmm.2022.3149712},
  langid = {english}
}

@misc{kongUnifiedAudioIntelligence2026,
  title = {Unified {{Audio Intelligence Without Regressing}} on {{Text Intelligence}}},
  author = {Kong, Zhifeng and Lee, Sang-gil and Kim, Jaehyeon and Wang, Boxin and Liu, Zihan and Kim, Sungwon and Chen, Yang and Goel, Arushi and Roy, Rajarshi and Dai, Wenliang and Yang, Zhuolin and Chen, Yangyi and Jiang, Dongfu and Ghosh, Sreyan and Rintamaki, Tuomas and Tao, Andrew and Raiman, Jonathan and Shoeybi, Mohammad and Catanzaro, Bryan and Ping, Wei},
  year = 2026,
  month = jul,
  number = {arXiv:2607.05196},
  eprint = {2607.05196},
  primaryclass = {cs.CL},
  publisher = {arXiv},
  doi = {10.48550/arXiv.2607.05196},
  urldate = {2026-09-25},
  archiveprefix = {arXiv},
  langid = {english}
}

@inproceedings{koutiniEfficientTrainingAudio2022,
  title = {Efficient {{Training}} of {{Audio Transformers}} with {{Patchout}}},
  booktitle = {Interspeech},
  author = {Koutini, Khaled and Schl{\"u}ter, Jan and {Eghbal-zadeh}, Hamid and Widmer, Gerhard},
  year = 2022,
  pages = {2753--2757},
  issn = {2958-1796},
  doi = {10.21437/Interspeech.2022-227},
  langid = {english}
}

@inproceedings{leeDeduplicatingTrainingData2022,
  title = {Deduplicating {{Training Data Makes Language Models Better}}},
  booktitle = {{{ACL}}},
  author = {Lee, Katherine and Ippolito, Daphne and Nystrom, Andrew and Zhang, Chiyuan and Eck, Douglas and {Callison-Burch}, Chris and Carlini, Nicholas},
  editor = {Muresan, Smaranda and Nakov, Preslav and Villavicencio, Aline},
  year = 2022,
  month = may,
  pages = {8424--8445},
  publisher = {Association for Computational Linguistics},
  address = {Dublin, Ireland},
  doi = {10.18653/v1/2022.acl-long.577},
  urldate = {2024-05-06},
  langid = {english}
}

@misc{liTextPromptedCLAPLearning2026,
  title = {Text-{{Prompted CLAP}}: {{Learning Text-Conditioned Audio Representations}} via {{Contrastive Learning}}},
  shorttitle = {Text-{{Prompted CLAP}}},
  author = {Li, Mohan and Doddipatla, Rama and Woodland, Philip C.},
  year = 2026,
  month = sep,
  number = {arXiv:2607.25085},
  eprint = {2607.25085},
  primaryclass = {eess.AS},
  publisher = {arXiv},
  doi = {10.48550/ARXIV.2607.25085},
  urldate = {2026-11-03},
  archiveprefix = {arXiv},
  copyright = {Creative Commons Attribution 4.0 International},
  langid = {english}
}

@inproceedings{liuAudioLDMTexttoAudioGeneration2023,
  title = {{{AudioLDM}}: {{Text}}-to-{{Audio Generation}} with {{Latent Diffusion Models}}},
  shorttitle = {{{AudioLDM}}},
  booktitle = {{{ICML}}},
  author = {Liu, Haohe and Chen, Zehua and Yuan, Yi and Mei, Xinhao and Liu, Xubo and Mandic, Danilo and Wang, Wenwu and Plumbley, Mark D.},
  year = 2023,
  month = jul,
  series = {Proceedings of {{Machine Learning Research}}},
  volume = {202},
  eprint = {2301.12503},
  primaryclass = {cs, eess},
  pages = {21450--21474},
  publisher = {PMLR},
  issn = {2640-3498},
  urldate = {2024-11-13},
  archiveprefix = {arXiv},
  langid = {english}
}

@inproceedings{loshchilovDecoupledWeightDecay2019,
  title = {Decoupled {{Weight Decay Regularization}}},
  booktitle = {{{ICLR}}},
  author = {Loshchilov, Ilya and Hutter, Frank},
  year = 2019,
  langid = {english}
}

@inproceedings{loshchilovSGDRStochasticGradient2017,
  title = {{{SGDR}}: {{Stochastic Gradient Descent}} with {{Warm Restarts}}},
  shorttitle = {Sgdr},
  booktitle = {{{ICLR}}},
  author = {Loshchilov, Ilya and Hutter, Frank},
  year = 2017,
  langid = {english}
}

@article{maCLAPSepLeveragingContrastive2024,
  title = {{{CLAPSep}}: {{Leveraging Contrastive Pre-Trained Model}} for {{Multi-Modal Query-Conditioned Target Sound Extraction}}},
  shorttitle = {{{CLAPSep}}},
  author = {Ma, Hao and Peng, Zhiyuan and Li, Xu and Shao, Mingjie and Wu, Xixin and Liu, Ju},
  year = 2024,
  journal = {IEEE/ACM Transactions on Audio, Speech, and Language Processing},
  volume = {32},
  pages = {4945--4960},
  issn = {2329-9290, 2329-9304},
  doi = {10.1109/TASLP.2024.3497586},
  urldate = {2025-10-01},
  copyright = {https://ieeexplore.ieee.org/Xplorehelp/downloads/license-information/IEEE.html},
  langid = {english}
}

@inproceedings{maEmotion2vecSelfSupervisedPreTraining2024,
  title = {Emotion2vec: {{Self-Supervised Pre-Training}} for {{Speech Emotion Representation}}},
  shorttitle = {Emotion2vec},
  booktitle = {Findings of {{ACL}}},
  author = {Ma, Ziyang and Zheng, Zhisheng and Ye, Jiaxin and Li, Jinchao and Gao, Zhifu and Zhang, Shiliang and Chen, Xie},
  year = 2024,
  pages = {15747--15760},
  doi = {10.18653/v1/2024.findings-acl.931},
  langid = {english}
}

@misc{manakulScalingOpenDiscrete2026,
  title = {Scaling {{Open Discrete Audio Foundation Models}} with {{Interleaved Semantic}}, {{Acoustic}}, and {{Text Tokens}}},
  author = {Manakul, Potsawee and Gan, Woody Haosheng and Bartelds, Martijn and Sun, Guangzhi and Held, William and Yang, Diyi},
  year = 2026,
  publisher = {ArXiv},
  langid = {english}
}

@inproceedings{maTevatron20Unified2025,
  title = {Tevatron 2.0: {{Unified Document Retrieval Toolkit Across Scale}}, {{Language}}, and {{Modality}}},
  shorttitle = {Tevatron 2.0},
  booktitle = {{{SIGIR}}},
  author = {Ma, Xueguang and Gao, Luyu and Zhuang, Shengyao and Zhan, Jiaqi Samantha and Callan, Jamie and Lin, Jimmy},
  year = 2025,
  series = {Sigir '25},
  pages = {4061--4065},
  publisher = {Association for Computing Machinery},
  address = {New York, NY, USA},
  doi = {10.1145/3726302.3730135},
  isbn = {979-8-4007-1592-1},
  langid = {english}
}

@inproceedings{meiSLAPScalableLanguageAudio2026,
  title = {{{SLAP}}: {{Scalable Language-Audio Pretraining}} with {{Variable-Duration Audio}} and {{Multi-Objective Training}}},
  shorttitle = {{{SLAP}}},
  booktitle = {{{ICASSP}}},
  author = {Mei, Xinhao and Le Lan, Gael and Liu, Haohe and Ni, Zhaoheng and Nagaraja, Varun and Liu, Yang and Shi, Yangyang and Chandra, Vikas},
  year = 2026,
  pages = {14517--14521},
  publisher = {IEEE},
  langid = {english}
}

@article{meiWavCapsChatGPTAssistedWeaklyLabelled2024,
  title = {{{WavCaps}}: {{A ChatGPT-Assisted Weakly-Labelled Audio Captioning Dataset}} for {{Audio-Language Multimodal Research}}},
  shorttitle = {{{WavCaps}}},
  author = {Mei, Xinhao and Meng, Chutong and Liu, Haohe and Kong, Qiuqiang and Ko, Tom and Zhao, Chengqi and Plumbley, Mark D. and Zou, Yuexian and Wang, Wenwu},
  year = 2024,
  journal = {IEEE/ACM Transactions on Audio, Speech, and Language Processing},
  volume = {32},
  pages = {3339--3354},
  issn = {2329-9290, 2329-9304},
  doi = {10.1109/TASLP.2024.3419446},
  urldate = {2026-11-03},
  copyright = {https://ieeexplore.ieee.org/Xplorehelp/downloads/license-information/IEEE.html},
  langid = {english}
}

@inproceedings{muennighoffMtebMassiveText2023,
  title = {{{MTEB}}: {{Massive Text Embedding Benchmark}}},
  booktitle = {{{EACL}}},
  author = {Muennighoff, Niklas and Tazi, Nouamane and Magne, Lo{\"\i}c and Reimers, Nils},
  year = 2023,
  pages = {2014--2037}
}

@inproceedings{muennighoffScalingDataConstrainedLanguage2023,
  title = {Scaling {{Data-Constrained Language Models}}},
  booktitle = {{{NeurIPS}}},
  author = {Muennighoff, Niklas and Rush, Alexander and Barak, Boaz and Le Scao, Teven and Tazi, Nouamane and Piktus, Aleksandra and Pyysalo, Sampo and Wolf, Thomas and Raffel, Colin},
  year = 2023,
  volume = {36},
  pages = {50358--50376},
  doi = {10.52202/075280-2191},
  langid = {english}
}

@inproceedings{nezhurinaScalingLawsRobust2026,
  title = {Scaling {{Laws}} for {{Robust Comparison}} of {{Open Foundation Language-Vision Models}} and {{Datasets}}},
  booktitle = {{{NeurIPS}}},
  author = {Nezhurina, Marianna and Porian, Tomer and Puccetti, Giovanni and Kerssies, Tommie and Beaumont, Romain and Cherti, Mehdi and Jitsev, Jenia},
  year = 2025,
  volume = {38},
  pages = {153979--154018},
  langid = {english}
}

@inproceedings{niizumiM2DCLAPMaskedModeling2024,
  title = {{{M2D-CLAP}}: {{Masked Modeling Duo Meets CLAP}} for {{Learning General-Purpose Audio-Language Representation}}},
  shorttitle = {{{M2D-CLAP}}},
  booktitle = {Interspeech},
  author = {Niizumi, Daisuke and Takeuchi, Daiki and Ohishi, Yasunori and Harada, Noboru and Yasuda, Masahiro and Tsubaki, Shunsuke and Imoto, Keisuke},
  year = 2024,
  month = jun,
  eprint = {2406.02032},
  pages = {57--61},
  doi = {10.48550/arXiv.2406.02032},
  urldate = {2024-10-25},
  archiveprefix = {arXiv},
  langid = {english}
}

@inproceedings{niuACAVCapsEnablingLargeScale2026,
  title = {{{ACAVCaps}}: {{Enabling Large-Scale Training}} for {{Fine-Grained}} and {{Diverse Audio Understanding}}},
  shorttitle = {{{ACAVCaps}}},
  booktitle = {{{ICASSP}}},
  author = {Niu, Yadong and Wang, Tianzi and Dinkel, Heinrich and Sun, Xingwei and Zhou, Jiahao and Li, Gang and Liu, Jizhong and Zhang, Junbo and Luan, Jian},
  year = 2026,
  pages = {15347--15351},
  publisher = {IEEE},
  langid = {english}
}

@misc{nvidiaNemotron3Nano2026,
  title = {Nemotron 3 {{Nano Omni}}: {{Efficient}} and {{Open Multimodal Intelligence}}},
  shorttitle = {Nemotron 3 {{Nano Omni}}},
  author = {{Nvidia} and Deshmukh, Amala Sanjay and Chumachenko, Kateryna and Rintamaki, Tuomas and Le, Matthieu and Poon, Tyler and Taheri, Danial Mohseni and Karmanov, Ilia and Liu, Guilin and Seppanen, Jarno and Goel, Arushi and Ranzinger, Mike and Heinrich, Greg and Chen, Guo and Voegtle, Lukas and Fischer, Philipp and Roman, Timo and Sapra, Karan and McCarthy, Collin and Zhang, Shaokun and Liu, Fuxiao and Ye, Hanrong and Dong, Yi and Liu, Mingjie and Peng, Yifan and Zelasko, Piotr and Chen, Zhehuai and Koluguri, Nithin Rao and Tadevosyan, Nune and Grigoryan, Lilit and Asl, Ehsan Hosseini and Biswas, Pritam and Tavabi, Leili and Su, Yuanhang and Yu, Zhiding and Jin, Peter and Milesi, Alexandre and Haber, Netanel and Xu, Yao and Amiraslani, Sarah and Mulepati, Nabin and Tramel, Eric and Jung, Jaehun and Lu, Ximing and Cui, Brandon and Xu, Jin and Li, Zhiqi and Wang, Shihao and Kuang, Yuanguo and Zhang, Shaokun and Yang, Huck and Li, Boyi and Yin, Hongxu and Han, Song and Kartal, Bilal and Molchanov, Pavlo and Renduchintala, Adi and Wang, Charles and Mosallanezhad, David and Singhal, Soumye and Vega, Luis and Cheung, Katherine and Ghosh, Sreyan and Zhang, Yian and Bukharin, Alexander and Srinivasan, Venkat and Greco, Johnny and Manoel, Andre and Segbroeck, Maarten Van and Panguliri, Suseella and Watve, Rohit and Kakwani, Divyanshu and Pachori, Shubham and Glick, Jeffrey and {Sri-Tharan}, Radha and Zaman, Aileen and Nguyen, Khanh and Chen, Shi and Fang, Jiaheng and Miao, Qing and Zhou, Wenfei and Wang, Yu and Bhat, Zaid Pervaiz and Praveen, Varun and Jain, Arihant and Arunachalam, Ramanathan and Kornuta, Tomasz and Sharabiani, Ashton and Shen, Amy and Huang, Wei and Wu, Yi-Fu and Ghias, Ali Roshan and Li, Huiying and Yu, Brian and Tajbakhsh, Nima and Cui, Chen and Gao, Wenwen and Ding, Li and Kong, Terry and Kilaru, Manoj and Bhiwandiwalla, Anahita and Wawrzos, Marek and Korzekwa, Daniel and Ribalta, Pablo and Chlebus, Grzegorz and Nushi, Besmira and Dobrowolska, Ewa and Mikulski, Maciej Jakub and Dhawan, Kunal and Huang, Steve and Balam, Jagadeesh and Wang, Yongqiang and Karpov, Nikolay and Mendelev, Valentin and Zelenfroynd, George and Mkrtchyan, Meline and Miao, Qing and Almog, Omri and Pawar, Bhavesh and Shivbhakta, Rameshwar and Sabnis, Sudeep and Sharabiani, Ashrton and Habibi, Negar and Venkataramani, Geethapriya and Peng, Pamela and Rodney, Prerit and Panev, Serge and Mazzarese, Richard and Liu, Nicky and Fukuyama, Michael and Skliar, Andrii and Waleffe, Roger and Riach, Duncan and Zou, Yunheng and Hu, Jian and Zhang, Hao and Xu, Binfeng and Yang, Yuhao and Ahmed, Zuhair and Milesi, Alexandre and del Mundo, Carlo and Voegele, Chad and Cheng, Zhiyu and Assaf, Nave and Skliar, Andrii and Afrimi, Daniel and Bagrov, Natan and Zilberstein, Ran and Masad, Ofri and Khvedchenia, Eugene and Bagrov, Natan and Tymchenko, Borys and Asida, Tomer and Afrimi, Daniel and Mannan, Parth and Cui, Victor and Evans, Michael and Luna, Katherine and Lou, Jie and Xu, Pinky and Huang, Guyue and Habibi, Negar and Boone, Michael and Thalasta, Pradeep and Adesoba, Adeola and Yared, Dina and Parisien, Christopher and Derczynski, Leon and Ghosh, Shaona and Feely, Wes and Schaffer, Micah and {Sri-Tharan}, Radha and Glick, Jeffrey and Simkin, Barnaby and Zelenfroynd, George and Grzegorzek, Tomasz and Garg, Rishabh and Jhunjhunwala, Aastha and Kolchenko, Sergei and Memarian, Farzan and Kumar, Haran and Kumar, Shiv and Hulseman, Isabel and Shah, Anjali and Briski, Kari and Subramanian, Padmavathy and Conway, Joey and Karpas, Udi and Scowcroft, Jane Polak and Surla, Annie and Ammireddy, Shilpa and Evans, Ellie and Oliver, Jesse and Balough, Tom and Chen, Chia-Chih and Bhaskar, Sandip and Rico, Alejandra and Sadeghi, Bardiya and Mard, Seph and Cheung, Katherine and Price, Meredith and Sleiman, Laya and Kaji, Saori and Helmholz, Wesley and Quan, Wendy and Lightstone, Michael and Cohen, Jonathan and Zhang, Jian and Kuchaiev, Oleksii and Ginsburg, Boris and Kautz, Jan and Long, Eileen and Shoeybi, Mohammad and Patwary, Mostofa and Olabiyi, Oluwatobi and Tao, Andrew and Catanzaro, Bryan and Karpas, Udi},
  year = 2026,
  month = may,
  number = {arXiv:2604.24954},
  eprint = {2604.24954},
  primaryclass = {cs.LG},
  publisher = {arXiv},
  doi = {10.48550/arXiv.2604.24954},
  urldate = {2026-09-26},
  archiveprefix = {arXiv},
  langid = {english}
}

@inproceedings{oncescuAudioRetrievalNatural2021,
  title = {Audio {{Retrieval}} with {{Natural Language Queries}}},
  booktitle = {Interspeech},
  author = {Oncescu, Andreea-Maria and Koepke, A. Sophia and Henriques, Jo{\~a}o F. and Akata, Zeynep and Albanie, Samuel},
  year = 2021,
  pages = {2411--2415},
  issn = {2958-1796},
  doi = {10.21437/Interspeech.2021-2227},
  langid = {english}
}

@misc{oordRepresentationLearningContrastive2019,
  title = {Representation {{Learning}} with {{Contrastive Predictive Coding}}},
  author = {van den Oord, Aaron and Li, Yazhe and Vinyals, Oriol},
  year = 2019,
  month = jan,
  number = {arXiv:1807.03748},
  eprint = {1807.03748},
  primaryclass = {cs.LG},
  publisher = {arXiv},
  doi = {10.48550/arXiv.1807.03748},
  urldate = {2026-09-26},
  archiveprefix = {arXiv},
  langid = {english}
}

@misc{polyakMovieGenCast2025,
  title = {Movie {{Gen}}: {{A Cast}} of {{Media Foundation Models}}},
  shorttitle = {Movie {{Gen}}},
  author = {Polyak, Adam and Zohar, Amit and Brown, Andrew and Tjandra, Andros and Sinha, Animesh and Lee, Ann and Vyas, Apoorv and Shi, Bowen and Ma, Chih-Yao and Chuang, Ching-Yao and Yan, David and Choudhary, Dhruv and Wang, Dingkang and Sethi, Geet and Pang, Guan and Ma, Haoyu and Misra, Ishan and Hou, Ji and Wang, Jialiang and Jagadeesh, Kiran and Li, Kunpeng and Zhang, Luxin and Singh, Mannat and Williamson, Mary and Le, Matt and Yu, Matthew and Singh, Mitesh Kumar and Zhang, Peizhao and Vajda, Peter and Duval, Quentin and Girdhar, Rohit and Sumbaly, Roshan and Rambhatla, Sai Saketh and Tsai, Sam and Azadi, Samaneh and Datta, Samyak and Chen, Sanyuan and Bell, Sean and Ramaswamy, Sharadh and Sheynin, Shelly and Bhattacharya, Siddharth and Motwani, Simran and Xu, Tao and Li, Tianhe and Hou, Tingbo and Hsu, Wei-Ning and Yin, Xi and Dai, Xiaoliang and Taigman, Yaniv and Luo, Yaqiao and Liu, Yen-Cheng and Wu, Yi-Chiao and Zhao, Yue and Kirstain, Yuval and He, Zecheng and He, Zijian and Pumarola, Albert and Thabet, Ali and Sanakoyeu, Artsiom and Mallya, Arun and Guo, Baishan and Araya, Boris and Kerr, Breena and Wood, Carleigh and Liu, Ce and Peng, Cen and Vengertsev, Dimitry and Schonfeld, Edgar and Blanchard, Elliot and {Juefei-Xu}, Felix and Nord, Fraylie and Liang, Jeff and Hoffman, John and Kohler, Jonas and Fire, Kaolin and Sivakumar, Karthik and Chen, Lawrence and Yu, Licheng and Gao, Luya and Georgopoulos, Markos and Moritz, Rashel and Sampson, Sara K. and Li, Shikai and Parmeggiani, Simone and Fine, Steve and Fowler, Tara and Petrovic, Vladan and Du, Yuming},
  year = 2024,
  number = {arXiv:2410.13720},
  eprint = {2410.13720},
  primaryclass = {cs.CV},
  publisher = {arXiv},
  doi = {10.48550/ARXIV.2410.13720},
  urldate = {2026-11-03},
  archiveprefix = {arXiv},
  copyright = {arXiv.org perpetual, non-exclusive license},
  langid = {english}
}

@article{pratapScalingSpeechTechnology2024,
  title = {Scaling {{Speech Technology}} to 1,000+ {{Languages}}},
  author = {Pratap, Vineel and Tjandra, Andros and Shi, Bowen and Tomasello, Paden and Babu, Arun and Kundu, Sayani and Elkahky, Ali and Ni, Zhaoheng and Vyas, Apoorv and {Fazel-Zarandi}, Maryam and {Others}},
  year = 2024,
  journal = {Journal of Machine Learning Research},
  volume = {25},
  number = {97},
  pages = {1--52},
  langid = {english}
}

@misc{qwenteamQwen35OmniTechnicalReport2026,
  title = {Qwen3.5-{{Omni Technical Report}}},
  author = {{Qwen Team}},
  year = 2026,
  eprint = {2604.15804},
  primaryclass = {cs.CL},
  archiveprefix = {arXiv},
  langid = {english}
}

@article{radfordLanguageModelsAre2019,
  title = {Language {{Models Are Unsupervised Multitask Learners}}},
  author = {Radford, Alec and Wu, Jeffrey and Child, Rewon and Luan, David and Amodei, Dario and Sutskever, Ilya},
  year = 2019,
  journal = {OpenAI Blog},
  volume = {1},
  number = {8},
  pages = {9},
  langid = {english}
}

@inproceedings{radfordLearningTransferableVisual2021,
  title = {Learning {{Transferable Visual Models From Natural Language Supervision}}},
  booktitle = {{{ICML}}},
  author = {Radford, Alec and Kim, Jong Wook and Hallacy, Chris and Ramesh, Aditya and Goh, Gabriel and Agarwal, Sandhini and Sastry, Girish and Askell, Amanda and Mishkin, Pamela and Clark, Jack and Krueger, Gretchen and Sutskever, Ilya},
  year = 2021,
  pages = {8748--8763},
  publisher = {PmLR},
  doi = {10.48550/arxiv.2103.00020},
  langid = {english}
}

@inproceedings{schuhmannLaion5bOpenLargeScale2022,
  title = {{{LAION-5B}}: {{An Open Large-Scale Dataset}} for {{Training Next Generation Image-Text Models}}},
  shorttitle = {Laion-5b},
  booktitle = {{{NeurIPS}}},
  author = {Schuhmann, Christoph and Beaumont, Romain and Vencu, Richard and Gordon, Cade and Wightman, Ross and Cherti, Mehdi and Coombes, Theo and Katta, Aarush and Mullis, Clayton and Wortsman, Mitchell and {Others}},
  year = 2022,
  volume = {35},
  pages = {25278--25294},
  doi = {10.52202/068431-1833},
  langid = {english}
}

@misc{shazeerGLUVariantsImprove2020,
  title = {{{GLU Variants Improve Transformer}}},
  author = {Shazeer, Noam},
  year = 2020,
  month = feb,
  number = {arXiv:2002.05202},
  eprint = {2002.05202},
  primaryclass = {cs.LG},
  publisher = {arXiv},
  doi = {10.48550/arXiv.2002.05202},
  urldate = {2026-09-25},
  archiveprefix = {arXiv},
  langid = {english}
}

@inproceedings{shukorScalingLawsNative2025,
  title = {Scaling {{Laws}} for {{Native Multimodal Models}}},
  booktitle = {{{ICCV}}},
  author = {Shukor, Mustafa and Fini, Enrico and {Guilherme Turrisi da Costa}, Victor and Cord, Matthieu and Susskind, Joshua and {El-Nouby}, Alaaeldin},
  year = 2025,
  pages = {12--23},
  doi = {10.1109/ICCV51701.2025.00009},
  langid = {english}
}

@article{sixOlafLightweightPortable2023,
  title = {Olaf: {{A Lightweight}}, {{Portable Audio Search System}}},
  shorttitle = {Olaf},
  author = {Six, Joren},
  year = 2023,
  journal = {Journal of Open Source Software},
  volume = {8},
  number = {87},
  eprint = {1854/LU-01H54YQ2WNG5BS3CY3GFBWXT68},
  eprinttype = {hdl},
  issn = {2475-9066},
  doi = {10.21105/joss.05459},
  urldate = {2024-04-18},
  copyright = {Creative Commons Attribution 4.0 International Public License (CC-BY 4.0)},
  langid = {english}
}

@inproceedings{stamatiadisSALTStandardizedAudio2024,
  title = {{{SALT}}: {{Standardized Audio Event Label Taxonomy}}},
  shorttitle = {{{SALT}}},
  booktitle = {{{DCASE}}},
  author = {Stamatiadis, Paraskevas and Olvera, Michel and Essid, Slim},
  year = 2024,
  month = oct,
  address = {Tokyo},
  langid = {english}
}

@misc{suCapTalkUnifiedVoice2026,
  title = {{{CapTalk}}: {{Unified Voice Design}} for {{Single-Utterance}} and {{Dialogue Speech Generation}}},
  shorttitle = {{{CapTalk}}},
  author = {Su, Xiaosu and Sun, Zihan and Jia, Peilei and Gao, Jun},
  year = 2026,
  month = apr,
  number = {arXiv:2604.08363},
  eprint = {2604.08363},
  primaryclass = {cs.SD},
  publisher = {arXiv},
  doi = {10.48550/arXiv.2604.08363},
  urldate = {2026-09-24},
  archiveprefix = {arXiv},
  langid = {english}
}

@inproceedings{sunAutoACDLargescaleDataset2024,
  title = {Auto-{{ACD}}: {{A Large}}-Scale {{Dataset}} for {{Audio-Language Representation Learning}}},
  shorttitle = {Auto-{{ACD}}},
  booktitle = {{{ACM MM}}},
  author = {Sun, Luoyi and Xu, Xuenan and Wu, Mengyue and Xie, Weidi},
  editor = {Cai, Jianfei and Kankanhalli, Mohan S. and Prabhakaran, Balakrishnan and Boll, Susanne and Subramanian, Ramanathan and Zheng, Liang and Singh, Vivek K. and C{\'e}sar, Pablo and Xie, Lexing and Xu, Dong},
  year = 2024,
  eprint = {2309.11500},
  primaryclass = {cs, eess},
  pages = {5025--5034},
  publisher = {ACM},
  doi = {10.1145/3664647.3681472},
  archiveprefix = {arXiv},
  langid = {english}
}

@article{suRoformerEnhancedTransformer2024,
  title = {{{RoFormer}}: {{Enhanced Transformer}} with {{Rotary Position Embedding}}},
  shorttitle = {Roformer},
  author = {Su, Jianlin and Ahmed, Murtadha and Lu, Yu and Pan, Shengfeng and Bo, Wen and Liu, Yunfeng},
  year = 2024,
  journal = {Neurocomputing},
  volume = {568},
  pages = {127063},
  publisher = {Elsevier},
  doi = {10.1016/j.neucom.2023.127063},
  langid = {english}
}

@misc{teamMiMoAudioAudioLanguage2025,
  title = {{{MiMo-Audio}}: {{Audio Language Models}} Are {{Few-Shot Learners}}},
  shorttitle = {{{MiMo-Audio}}},
  author = {Team, Xiaomi LLM-Core and Zhang, Dong and Wang, Gang and Xue, Jinlong and Fang, Kai and Zhao, Liang and Ma, Rui and Ren, Shuhuai and Liu, Shuo and Guo, Tao and Zhuang, Weiji and Zhang, Xin and Song, Xingchen and Yan, Yihan and He, Yongzhe and Cici and Shen, Bowen and Zhu, Chengxuan and Ma, Chong and Chen, Chun and Chen, Heyu and Li, Jiawei and Li, Lei and Zhu, Menghang and Li, Peidian and Wang, Qiying and Deng, Sirui and Xiong, Weimin and Huang, Wenshan and Yang, Wenyu and Jiang, Yilin and Yang, Yixin and Tian, Yuanyuan and Ma, Yue and Yu, Yue and Zhang, Zihan and Yue, Zihao and Xiao, Bangjun and Xia, Bingquan and Gao, Bofei and Ye, Bowen and Cai, Can and Liu, Chang and He, Chenhong and Li, Chunan and Zhu, Dawei and Zhang, Duo and Shi, Fengyuan and Wang, Guoan and Zhang, Hailin and Lv, Hanglong and Li, Hanyu and Tian, Hao and Qu, Heng and Xu, Hongshen and Zhang, Houbin and Liu, Huaqiu and Duo, Jiangshan and Zuo, Jianguang and Wei, Jianyu and Xiao, Jiebao and Dong, Jinhao and Shi, Jun and Hu, Junhao and Bao, Kainan and Zhou, Kang and Zhang, Linghao and Chen, Meng and Chen, Nuo and Zhang, Peng and Chen, Qianli and Wang, Qiantong and Li, Rang and Liu, Shaohui and Wang, Shengfan and Li, Shicheng and Yu, Shihua and Cao, Shijie and Chen, Shimao and Gu, Shuhao and Wang, Weikun and Ma, Wenhan and Deng, Xiangwei and Yong, Xing and Zhang, Xing and Wang, Xu and Song, Yifan and Zhao, Yihao and Zhao, Yingbo and Gao, Yizhao and Cheng, Yu and Tu, Yu and Wang, Yudong and Huang, Zhaojun and Tang, Zhengju and Lin, Zhenru and Song, Zhichao and Xu, Zhipeng and Zheng, Zhixian and Jiang, Zihan},
  year = 2025,
  month = dec,
  number = {arXiv:2512.23808},
  eprint = {2512.23808},
  primaryclass = {cs},
  publisher = {arXiv},
  doi = {10.48550/arXiv.2512.23808},
  urldate = {2026-02-10},
  archiveprefix = {arXiv},
  langid = {english}
}

@inproceedings{tianBagpiperSolvingOpenEnded2026,
  title = {Bagpiper: {{Solving Open-Ended Audio Tasks Via Rich Captions}}},
  shorttitle = {Bagpiper},
  booktitle = {{{COLM}}},
  author = {Tian, Jinchuan and Wang, Haoran and Su, Bo-Hao and Huang, Chien-yu and Wang, Qingzheng and Shi, Jiatong and Chen, William and Gong, Xun and Arora, Siddhant and Li, Chin-Jou and Someki, Masao and Maekaku, Takashi and Shinohara, Yusuke and Sakuma, Jin and Goto, Keita and Yang, Chao-Han Huck and Watanabe, Shinji},
  year = 2026,
  langid = {english}
}

@misc{timnielsenUniversalCategorySystem2024,
  title = {Universal {{Category System}}},
  author = {Nielsen, Tim and Drury, Justin and Paquin, Kai},
  year = 2024,
  month = jan,
  urldate = {2026-09-23},
  howpublished = {https://universalcategorysystem.com/},
  langid = {english}
}

@misc{tonmoyFusionEmbeddingUnified2026,
  title = {Fusion {{Embedding}}: {{A Unified Embedding Space}} for {{Text}}, {{Image}}, {{Video}}, and {{Audio}}},
  shorttitle = {Fusion {{Embedding}}},
  author = {Tonmoy, Abdul Basit and Hoque, Kazi Fardinul and Arham, Md Shahrier Islam and Luthra, Arman},
  year = 2026,
  month = jul,
  number = {arXiv:2607.18666},
  eprint = {2607.18666},
  primaryclass = {cs.CL},
  publisher = {arXiv},
  doi = {10.48550/arXiv.2607.18666},
  urldate = {2026-10-13},
  archiveprefix = {arXiv},
  langid = {english}
}

@inproceedings{touvronGoingDeeperImage2021,
  title = {Going {{Deeper}} with {{Image Transformers}}},
  booktitle = {{{ICCV}}},
  author = {Touvron, Hugo and Cord, Matthieu and Sablayrolles, Alexandre and Synnaeve, Gabriel and J{\'e}gou, Herv{\'e}},
  year = 2021,
  pages = {32--42},
  doi = {10.1109/ICCV48922.2021.00010},
  langid = {english}
}

@misc{tschannenSigLIP2Multilingual2025,
  title = {{{SigLIP}} 2: {{Multilingual Vision-Language Encoders}} with {{Improved Semantic Understanding}}, {{Localization}}, and {{Dense Features}}},
  shorttitle = {{{SigLIP}} 2},
  author = {Tschannen, Michael and Gritsenko, Alexey and Wang, Xiao and Naeem, Muhammad Ferjad and Alabdulmohsin, Ibrahim and Parthasarathy, Nikhil and Evans, Talfan and Beyer, Lucas and Xia, Ye and Mustafa, Basil and H{\'e}naff, Olivier and Harmsen, Jeremiah and Steiner, Andreas and Zhai, Xiaohua},
  year = 2025,
  month = feb,
  number = {arXiv:2502.14786},
  eprint = {2502.14786},
  primaryclass = {cs.CV},
  publisher = {arXiv},
  doi = {10.48550/arXiv.2502.14786},
  urldate = {2026-09-21},
  archiveprefix = {arXiv},
  langid = {english}
}

@inproceedings{tsengRevisitingAudioLanguagePretraining2026,
  title = {Revisiting {{Audio-Language Pretraining}} for {{Learning General-Purpose Audio Representation}}},
  booktitle = {{{ACL}}},
  author = {Tseng, Wei-Cheng and Zhou, Xuanru and Huo, Mingyue and Shao, Yiwen and Zhang, Hao and Yu, Dong},
  year = 2026,
  pages = {34244--34263},
  doi = {10.18653/v1/2026.acl-long.1581},
  langid = {english}
}

@inproceedings{turianHEARHolisticEvaluation2022,
  title = {{{HEAR}}: {{Holistic Evaluation}} of {{Audio Representations}}},
  shorttitle = {{{HEAR}}},
  booktitle = {{{NeurIPS Competitions}} and {{Demonstrations Track}}},
  author = {Turian, Joseph and Shier, Jordie and Khan, Humair Raj and Raj, Bhiksha and Schuller, Bj{\"o}rn W. and Steinmetz, Christian J. and Malloy, Colin and Tzanetakis, George and Velarde, Gissel and McNally, Kirk and Henry, Max and Pinto, Nicolas and Noufi, Camille and Clough, Christian and Herremans, Dorien and Fonseca, Eduardo and Engel, Jesse and Salamon, Justin and Esling, Philippe and Manocha, Pranay and Watanabe, Shinji and Jin, Zeyu and Bisk, Yonatan},
  year = 2022,
  month = jul,
  pages = {125--145},
  publisher = {PMLR},
  issn = {2640-3498},
  urldate = {2026-09-22},
  langid = {english}
}

@inproceedings{valleFugatto1Foundational2025,
  title = {Fugatto 1: {{Foundational Generative Audio Transformer Opus}} 1},
  shorttitle = {Fugatto 1},
  booktitle = {{{ICLR}}},
  author = {Valle, Rafael and Badlani, Rohan and Kong, Zhifeng and Lee, Sang-gil and Goel, Arushi and Kim, Sungwon and Santos, Joao and Dai, Shuqi and Gururani, Siddharth and Aljafari, Aya and {Others}},
  year = 2025,
  volume = {2025},
  pages = {4325--4353},
  langid = {english}
}

@misc{vanmerrienboerPerch20Bittern2025,
  title = {Perch 2.0: {{The Bittern Lesson}} for {{Bioacoustics}}},
  shorttitle = {Perch 2.0},
  author = {{van Merri{\"e}nboer}, Bart and Dumoulin, Vincent and Hamer, Jenny and Harrell, Lauren and Burns, Andrea and Denton, Tom},
  year = 2025,
  number = {2508.04665},
  eprint = {2508.04665},
  publisher = {ArXiv},
  archiveprefix = {arXiv},
  langid = {english}
}

@inproceedings{vyasPushingFrontierAudiovisual2026,
  title = {Pushing the {{Frontier}} of {{Audiovisual Perception}} with {{Large-Scale Multimodal Correspondence Learning}}},
  booktitle = {{{CVPR}}},
  author = {Vyas, Apoorv and Chang, Heng-Jui and Yang, Cheng-Fu and Huang, Po-Yao and Gao, Luya and Richter, Julius and Chen, Sanyuan and Le, Matthew and Doll{\'a}r, Piotr and Feichtenhofer, Christoph and {Others}},
  year = 2026,
  pages = {30172--30182},
  langid = {english}
}

@inproceedings{warnerSmarterBetterFaster2025,
  title = {Smarter, {{Better}}, {{Faster}}, {{Longer}}: {{A Modern Bidirectional Encoder}} for {{Fast}}, {{Memory Efficient}}, and {{Long Context Finetuning}} and {{Inference}}},
  shorttitle = {Smarter, {{Better}}, {{Faster}}, {{Longer}}},
  booktitle = {{{ACL}}},
  author = {Warner, Benjamin and Chaffin, Antoine and Clavi{\'e}, Benjamin and Weller, Orion and Hallstr{\"o}m, Oskar and Taghadouini, Said and Gallagher, Alexis and Biswas, Raja and Ladhak, Faisal and Aarsen, Tom and {Others}},
  year = 2025,
  pages = {2526--2547},
  doi = {10.18653/v1/2025.acl-long.127},
  langid = {english}
}

@misc{wightmanPyTorchImageModels2026,
  title = {{{PyTorch}} Image Models},
  author = {Wightman, Ross},
  year = 2019,
  doi = {10.5281/zenodo.4414861},
  copyright = {Apache 2.0}
}

@article{wijngaardAudioLanguageDatasetsScenes2025,
  title = {Audio-{{Language Datasets}} of {{Scenes}} and {{Events}}: {{A Survey}}},
  shorttitle = {Audio-{{Language Datasets}} of {{Scenes}} and {{Events}}},
  author = {Wijngaard, Gijs and Formisano, Elia and Esposito, Michele and Dumontier, Michel},
  year = 2025,
  month = jan,
  journal = {IEEE Access},
  volume = {13},
  pages = {20328--20360},
  doi = {10.1109/ACCESS.2025.3534621},
  langid = {english}
}

@inproceedings{wuEfficacySelfSupervisedSpeech2022,
  title = {The {{Efficacy}} of {{Self-Supervised Speech Models}} for {{Audio Representations}}},
  booktitle = {{{HEAR}} ({{NeurIPS}} 2021 {{Competition}})},
  author = {Wu, Tung-Yu and Hsu, Tsu-Yuan and Li, Chen-An and Lin, Tzu-Han and Lee, Hung-yi},
  editor = {Turian, Joseph and Schuller, Bj{\"o}rn W. and Herremans, Dorien and Kirchoff, Katrin and Perera, Paola Garcia and Esling, Philippe},
  year = 2021,
  series = {Proceedings of {{Machine Learning Research}}},
  volume = {166},
  pages = {90--110},
  publisher = {PMLR},
  langid = {english}
}

@inproceedings{wuLargeScaleContrastiveLanguageAudio2023,
  title = {Large-{{Scale Contrastive Language-Audio Pretraining With Feature Fusion}} and {{Keyword}}-to-{{Caption Augmentation}}},
  booktitle = {{{ICASSP}}},
  author = {Wu, Yusong and Chen, Ke and Zhang, Tianyu and Hui, Yuchen and {Berg-Kirkpatrick}, Taylor and Dubnov, Shlomo},
  year = 2023,
  month = jun,
  eprint = {2211.06687},
  primaryclass = {cs, eess},
  pages = {1--5},
  publisher = {IEEE},
  address = {Rhodes Island, Greece},
  doi = {10.1109/ICASSP49357.2023.10095969},
  urldate = {2024-11-13},
  archiveprefix = {arXiv},
  copyright = {https://doi.org/10.15223/policy-029},
  isbn = {978-1-7281-6327-7},
  langid = {english}
}

@inproceedings{wuWav2CLIPLearningRobust2022,
  title = {{{Wav2CLIP}}: {{Learning Robust Audio Representations}} from {{Clip}}},
  shorttitle = {{{Wav2CLIP}}},
  booktitle = {{{ICASSP}}},
  author = {Wu, Ho-Hsiang and Seetharaman, Prem and Kumar, Kundan and Bello, Juan Pablo},
  year = 2022,
  pages = {4563--4567},
  doi = {10.1109/ICASSP43922.2022.9747669},
  langid = {english}
}

@inproceedings{xiaoScalingLanguageCentricOmnimodal2025,
  title = {Scaling {{Language-Centric Omnimodal Representation Learning}}},
  booktitle = {{{NeurIPS}}},
  author = {Xiao, Chenghao and Chan, Hou Pong (Ken) and Zhang, Hao and Xu, Weiwen and Aljunied, Mahani and Rong, Yu},
  editor = {Belgrave, D. and Zhang, C. and Lin, H. and Pascanu, R. and Koniusz, P. and Ghassemi, M. and Chen, N.},
  year = 2025,
  volume = {38, Main Conference},
  pages = {158370--158401},
  publisher = {Curran Associates, Inc.},
  doi = {10.52202/085713-5291},
  langid = {english}
}

@misc{xuQwen25OmniTechnicalReport2025,
  title = {Qwen2.5-{{Omni Technical Report}}},
  author = {Xu, Jin and Guo, Zhifang and He, Jinzheng and Hu, Hangrui and He, Ting and Bai, Shuai and Chen, Keqin and Wang, Jialin and Fan, Yang and Dang, Kai and Zhang, Bin and Wang, Xiong and Chu, Yunfei and Lin, Junyang},
  year = 2025,
  month = mar,
  number = {arXiv:2503.20215},
  eprint = {2503.20215},
  primaryclass = {cs},
  publisher = {arXiv},
  doi = {10.48550/arXiv.2503.20215},
  urldate = {2026-01-06},
  archiveprefix = {arXiv},
  langid = {english}
}

@misc{xuQwen3OmniTechnicalReport2025,
  title = {Qwen3-{{Omni Technical Report}}},
  author = {Xu, Jin and Guo, Zhifang and Hu, Hangrui and Chu, Yunfei and Wang, Xiong and He, Jinzheng and Wang, Yuxuan and Shi, Xian and He, Ting and Zhu, Xinfa and {Others}},
  year = 2025,
  publisher = {ArXiv},
  doi = {10.48550/arXiv.2509.17765},
  langid = {english}
}

@inproceedings{xuScalingAudioTextRetrieval2026,
  title = {Re-Purposing {{Multimodal Large Language Models}} for {{Audio-Text Retrieval}}},
  booktitle = {{{CBMI}}},
  author = {Xu, Jilan and Thom{\'e}, Carl and Horak, Danijela and Xie, Weidi and Zisserman, Andrew},
  year = 2026,
  eprint = {2602.18010},
  primaryclass = {cs},
  address = {Toulouse, France},
  doi = {10.48550/arXiv.2602.18010},
  urldate = {2026-02-23},
  archiveprefix = {arXiv},
  langid = {english}
}

@misc{yangMOSSAudioTechnicalReport2026,
  title = {{{MOSS-Audio Technical Report}}},
  author = {Yang, Chen and Yu, Chufan and Chen, Hanfu and Zhu, Jie and Chen, Jingqi and Chen, Ke and Wang, Wenxuan and Wang, Yang and Jiang, Yaozhou and Jiang, Yi and {Others}},
  year = 2026,
  publisher = {ArXiv},
  doi = {10.48550/arXiv.2606.01802},
  langid = {english}
}

@misc{yangSALMONN2AdvancingGeneralPurpose2026,
  title = {{{SALMONN-2}}: {{Advancing General-Purpose Hearing Abilities}} with {{Self-Supervised Representations}}},
  shorttitle = {{{SALMONN-2}}},
  author = {Yang, Xiaoyu and Xu, Xuenan and Yu, Wenyi and Wang, Siyin and Tang, Changli and Chiba, Terumi and Hou, Siyuan and Zhang, Ziyang and Wu, Wen and Li, Baoxiang and Sun, Guangzhi and Zhang, Chao and Woodland, Philip},
  year = 2026,
  month = jul,
  number = {arXiv:2607.17079},
  eprint = {2607.17079},
  primaryclass = {eess.AS},
  publisher = {arXiv},
  doi = {10.48550/arXiv.2607.17079},
  urldate = {2026-09-24},
  archiveprefix = {arXiv},
  langid = {english}
}

@inproceedings{yuanSoundvecapsImprovingAudio2025,
  title = {Sound-{{VECaps}}: {{Improving Audio Generation}} with {{Visually Enhanced Captions}}},
  shorttitle = {Sound-{{VECaps}}},
  booktitle = {{{ICASSP}}},
  author = {Yuan, Yi and Jia, Dongya and Zhuang, Xiaobin and Chen, Yuanzhe and Chen, Zhuo and Wang, Yuping and Wang, Yuxuan and Liu, Xubo and Kang, Xiyuan and Plumbley, Mark D. and Wang, Wenwu},
  year = 2025,
  month = apr,
  pages = {1--5},
  publisher = {IEEE},
  address = {Hyderabad, India},
  doi = {10.1109/ICASSP49660.2025.10889473},
  urldate = {2026-11-03},
  copyright = {https://doi.org/10.15223/policy-029},
  isbn = {979-8-3503-6874-1},
  langid = {english}
}

@inproceedings{zhaiScalingVisionTransformers2022,
  title = {Scaling {{Vision Transformers}}},
  booktitle = {{{CVPR}}},
  author = {Zhai, Xiaohua and Kolesnikov, Alexander and Houlsby, Neil and Beyer, Lucas},
  year = 2022,
  pages = {1204--1213},
  doi = {10.1109/CVPR52688.2022.01179},
  langid = {english}
}

@inproceedings{zhaiSigmoidLossLanguage2023,
  title = {Sigmoid {{Loss}} for {{Language Image Pre-Training}}},
  booktitle = {{{ICCV}}},
  author = {Zhai, Xiaohua and Mustafa, Basil and Kolesnikov, Alexander and Beyer, Lucas},
  year = 2023,
  pages = {11941--11952},
  publisher = {IEEE},
  doi = {10.1109/ICCV51070.2023.01100},
  langid = {english}
}

@inproceedings{zhangRootMeanSquare2019,
  title = {Root {{Mean Square Layer Normalization}}},
  booktitle = {{{NeurIPS}}},
  author = {Zhang, Biao and Sennrich, Rico},
  year = 2019,
  volume = {32},
  langid = {english}
}

@article{zhuMuQSelfSupervisedMusic2025,
  title = {{{MuQ}}: {{Self-Supervised Music Representation Learning}} with {{Mel Residual Vector Quantization}}},
  shorttitle = {{{MuQ}}},
  author = {Zhu, Haina and Zhou, Yizhi and Chen, Hangting and Yu, Jianwei and Ma, Ziyang and Gu, Rongzhi and Luo, Yi and Tan, Wei and Chen, Xie},
  year = 2025,
  journal = {IEEE Transactions on Audio, Speech and Language Processing},
  volume = {33},
  pages = {3653--3664},
  issn = {2998-4173},
  doi = {10.1109/TASLPRO.2025.3602320},
  urldate = {2026-10-14},
  langid = {english}
}

@inproceedings{zhuTasteIPAOpenVocabulary2024,
  title = {The {{Taste}} of {{IPA}}: {{Towards Open-Vocabulary Keyword Spotting}} and {{Forced Alignment}} in {{Any Language}}},
  shorttitle = {The {{Taste}} of {{IPA}}},
  booktitle = {{{NAACL}}},
  author = {Zhu, Jian and Yang, Changbing and Samir, Farhan and Islam, Jahurul},
  editor = {Duh, Kevin and Gomez, Helena and Bethard, Steven},
  year = 2024,
  month = jun,
  pages = {750--772},
  publisher = {Association for Computational Linguistics},
  address = {Mexico City, Mexico},
  doi = {10.18653/v1/2024.naacl-long.43},
  langid = {english}
}

@inproceedings{gomezgutierrezCanteFlamencoTech2018,
  title = {Cante {{Flamenco Tech}}: {{Aprendizaje Del Cante Con La Ayuda De La Tecnolog{\'\i}a}}},
  shorttitle = {Cante {{Flamenco Tech}}},
  booktitle = {V {{Congreso Interdisciplinar Investigaci\'on}} y {{Flamenco}} ({{INFLA}})},
  author = {G{\'o}mez Guti{\'e}rrez, Emilia and Guerrero, Alba and Rodr{\'\i}guez, Sonia},
  year = 2018,
  month = jun,
  address = {Sevilla, Espa\~na},
  langid = {english}
}

@article{ostermannAAMDatasetArtificial2023,
  title = {{{AAM}}: A Dataset of {{Artificial Audio Multitracks}} for Diverse Music Information Retrieval Tasks},
  shorttitle = {{{AAM}}},
  author = {Ostermann, Fabian and Vatolkin, Igor and Ebeling, Martin},
  year = 2023,
  month = mar,
  journal = {EURASIP Journal on Audio, Speech, and Music Processing},
  volume = {2023},
  number = {1},
  pages = {13},
  issn = {1687-4722},
  doi = {10.1186/s13636-023-00278-7},
  urldate = {2026-09-21},
  langid = {english}
}

@article{pandaNovelAudioFeatures2020,
  title = {Novel {{Audio Features}} for {{Music Emotion Recognition}}},
  author = {Panda, Renato and Malheiro, Ricardo and Paiva, Rui Pedro},
  year = 2020,
  month = oct,
  journal = {IEEE Transactions on Affective Computing},
  volume = {11},
  number = {4},
  pages = {614--626},
  publisher = {IEEE Computer Society Press},
  doi = {10.1109/TAFFC.2018.2820691},
  urldate = {2026-09-21},
  langid = {english}
}

@inproceedings{peetersPerceptualTempoEstimation2012,
  title = {Perceptual {{Tempo Estimation Using GMM-Regression}}},
  booktitle = {{{ACM Workshop}} on {{Music Information Retrieval}} with {{User-Centered}} and {{Multimodal Strategies}}},
  author = {Peeters, Geoffroy and {Flocon-Cholet}, Joachim},
  year = 2012,
  month = nov,
  series = {{{MIRUM}} '12},
  pages = {45--50},
  publisher = {Association for Computing Machinery},
  address = {New York, NY, USA},
  doi = {10.1145/2390848.2390861},
  urldate = {2026-09-25},
  isbn = {978-1-4503-1591-3},
  langid = {english}
}

@inproceedings{takahashiDeepConvolutionalNeural2016,
  title = {Deep {{Convolutional Neural Networks}} and {{Data Augmentation}} for {{Acoustic Event Recognition}}},
  booktitle = {Interspeech},
  author = {Takahashi, Naoya and Gygli, Michael and Pfister, Beat and Gool, Luc Van},
  year = 2016,
  pages = {2982--2986},
  issn = {2958-1796},
  doi = {10.21437/Interspeech.2016-805},
  langid = {english}
}

@article{vryzasSpeechEmotionRecognition2018,
  title = {Speech {{Emotion Recognition}} for {{Performance Interaction}}},
  author = {Vryzas, Nikolaos and Kotsakis, Rigas and Liatsou, Aikaterini and Dimoulas, Charalampos A and Kalliris, George},
  year = 2018,
  journal = {Journal of the Audio Engineering Society},
  volume = {66},
  number = {6},
  pages = {457--467},
  publisher = {Audio Engineering Society},
  doi = {10.17743/jaes.2018.0036},
  langid = {english}
}

\bibliographystyle{iclr2027_conference_idmax20}
\clearpage
\appendix

\etocdepthtag.toc{appendix}
\etocsettagdepth{mainbody}{none}
\etocsettagdepth{appendix}{subsection}
\renewcommand{\contentsname}{Appendix}
\tableofcontents

\section{Collection and source accounting}
\label{app:collection}
\subsection{Public datasets and web audio}
For public datasets, a search across platforms such as Hugging Face, Zenodo, and the literature focused on gathering a large number of permissively licensed collections. This work invests in reusing existing academic datasets. These datasets often bring cleaner audio and have been tested in practice.

Academic datasets form the largest group by number of sources, accounting for 316 of the 425 sources listed in Table~\ref{tab:catalog}. YouTube and Common Crawl-derived data contribute most of the data by both clip count and audio duration. Table~\ref{tab:catalog} displays these datasets. Sources with fewer than ten LAION-Audio-XL audio samples were removed from both corpora.

YouTube audio is extracted across the full duration of each video. Each video can contribute many audio samples.

Common Crawl, a non-profit that maintains a free, open repository of web crawl data, releases multiple crawls per year, often containing roughly 2--3 billion web pages each. These crawls consist of WARC response records, which preserve the captured HTML, including the text surrounding audio links. Audio elements and direct audio-file links are extracted from this HTML, together with the page URL and up to five sentences before and five sentences after each link. When an audio element offers multiple sources, one is chosen by a fixed format preference that prioritizes WAV, FLAC, and AIFF. The audio URLs are deduplicated and then downloaded, after which the audio is decoded and its duration checked. The first extraction wave covered 35 crawl directories from 2021 to 2025 and found about 3.3 billion audio links, of which about 423M are unique. A second wave added crawls from 2013 to 2020 and from mid-2025 to early 2026. Candidate URLs are not successful downloads: about 54\% downloaded and 44\% of the downloads converted to audio. Files larger than 50\,MB were discarded before download to prioritize shorter recordings and accelerate retrieval. Downloads that took longer than 30 seconds were ignored. Websites that did not connect or send any packet back to the scraper within 5 seconds were skipped, after a retry 2 seconds later. Files with a reported duration above 120 seconds were rejected before conversion. Downloaded recordings that still decoded to more than 120 seconds were subsequently segmented as described in Appendix~\ref{app:segment}. The deduplication input includes 141,523,038 Common Crawl audio samples, of which 51,453,212 remain after exact and fingerprint deduplication.

\subsection{Audio segmentation}
\label{app:segment}
\begin{wrapfigure}{R}{0.55\textwidth}
\centering
\input{figures/segmentation_timing_note.tex}
\includegraphics[width=\linewidth]{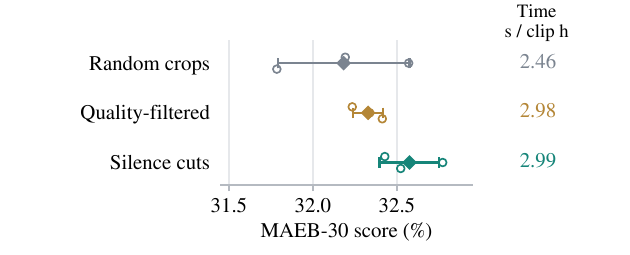}
\caption{Segmentation comparison after 8M samples per run. Diamonds and whiskers show the MAEB-30 mean and standard deviation over three seeds. Open circles show individual runs.\segmentationtimingnote}
\label{fig:segmentation-comparison}
\end{wrapfigure}
An experiment compared three ways of cutting long recordings into clips: random crops, quality-filtered crops, and silence cuts (Figure~\ref{fig:segmentation-comparison}). Each method used approximately 1M clips from the same YouTube source pool, captioned with Qwen3-Omni, to train our XS model with three seeds per method and approximately 8M samples seen per run. Among these three methods, silence cuts scored highest on MAEB-30 with $32.57\pm0.18$, followed by quality-filtered crops with $32.33\pm0.09$ and random crops with $32.18\pm0.39$ (mean $\pm$ standard deviation). All long recordings are therefore segmented with silence cuts, which only need CPU frame-energy scans of 8\,kHz audio.

Silence cuts occur below $-40$\,dB lasting at least 0.5 seconds, at the midpoints of these regions. Audio samples shorter than 5 seconds merge with adjacent segments. Segments longer than 120 seconds split at quiet frames within ten seconds of 30-second target positions. Beyond YouTube, another pass over the other data converted 3.04M long recordings into 7.15M segments.

\subsection{Decoding and signal checks}
\label{app:signal}
The pipeline removes audios that libraries such as \texttt{FFmpeg} and \texttt{soundfile} are unable to read or decode. It also records 40 statistics per clip: decoded duration, original sample rate and channels, RMS and peak level, dynamic range, DC offset, the proportion and position of silent frames at $-60$\,dB, clipping proportion, bandwidth, spectral centroid and flatness, band energies, and the three 10-second windows with most energy.

\subsection{Deduplication}
\label{app:deduplication}
\paragraph{Exact identity.}
All data are converted to 16\,kHz mono signed 16-bit PCM signal for exact-match representation. The key combines the MD5 hash of its bytes with the audio length. This checksum is used only for full exact matching. Audios that have the same semantic content, but are in different encodings or formats, will not match exactly. Exact matching of MD5 hash removes 151,056,336 audio samples in total. Common Crawl keeps 51.5M of its 141.5M audio samples after exact and fingerprint deduplication. A small number of repeated files dominate these duplicates: ten files account for 73.6M removals. The raw collection also contained 16.3M audio samples of pure silence.

\paragraph{Acoustic fingerprint matching.}
The acoustic stage links copies of the same recording after re-encoding, level changes, cropping, or padding. It uses Olaf \citep{sixOlafLightweightPortable2023}, a fingerprint technique similar to that used in Shazam. A fingerprint encodes the geometry of three spectral peaks, and two clips match when many fingerprints agree on one time offset. Fingerprints are computed on the full clip at 16\,kHz mono, using a 64\,ms Hamming window and an 8\,ms hop. Hashes exclude absolute time but include absolute frequency, so pitch-shifted or speed-changed copies are not matched.

Each shared hash votes for the offset between its occurrences, quantized to 32\,ms. The match count $m$ counts distinct shared hashes at the offset with most votes. Chance collisions between unrelated audio spread over many offsets. For audio with $n_a$ and $n_b$ fingerprints, the score is
\begin{equation}
 q(a,b)=\frac{m}{\sqrt{n_a n_b}},
\end{equation}
and pairs with $q\ge0.50$ are linked. Audio with equal fingerprint counts must share at least half their fingerprints at one offset. A fully matching shorter clip must contain at least a quarter of the longer clip's fingerprints. This limits matches caused by short shared fragments.

Matching uses a distributed hash join over 100B+ fingerprint rows from 1.47 billion audio samples. Rows are partitioned by hash into 512 buckets and joined on shared hashes, aggregated by clip pair and offset. Counts are summed across buckets before choosing each pair's best offset. Hashes occurring more than 500 times within one source track are removed from that track before the join. This limits the quadratic join cost of frequent hashes. 

Duplicate groups are connected components of edges with $q\ge0.50$. One audio per group is kept, preferring public datasets first. When audio comes from the same dataset group, it prefers audio in order of model-generated captions first, then longer text, more channels, a higher sample rate, and the lowest identifier. This stage removes 1,196,096 audio samples beyond exact matching.

\paragraph{Semantic near-duplicate screening.}
The semantic deduplication step is there to remove duplicates that the fingerprint stage misses. It follows SemDeDup \citep{abbasSemDeDupDataefficientLearning2023}: embed every clip, cluster the embeddings with $k$-means, and compare pairs only inside a cluster. The encoder is Dasheng-base \citep{dinkelScalingMaskedAudio2024}, a masked audio encoder trained without text. Because this encoder is separate from the CLAP model, the deduplication does not depend on the representations of the CLAP model being trained. In a small sample of 735K audio samples, taken from tar files that contain many similar recordings, 30,569 pairs reached the threshold. None had been removed by the earlier stages, and only 4.6\% had any fingerprint edge (14.5\% among pairs of audio samples of at least 2 seconds). Most new pairs were very short audio samples such as wake words and single utterances. Other cases were one file crawled twice and a clip next to its own re-segmented copy.

Each clip is decoded to 16\,kHz mono, cut to its first 10 seconds where audio after 10 seconds is not compared in this step. Audio is zero-padded if shorter than 10 seconds. Dasheng-base returns a sequence of 768-dimensional frame features, which is mean pooled over time and stored in 16-bit floating point. Embeddings exist for 1,412,277,595 stored audio samples. The screen runs on the 1,263,840,901 pool audio samples of at least 2 seconds. Shorter audio samples are not compared.

The embeddings are normalized to unit length and clustered with $k$-means in faiss on one GPU with $k=65{,}536$, 20 iterations, a training sample of 20M random audio samples, seed 0, and assignment of every clip to its nearest centroid. The update is the standard Euclidean one. Centroids are not re-normalized. A target of one cluster per 2,000 audio samples would need about 632,000 clusters, and assignment cost grows with $Nk$, so $k$ was capped at 65,536. The median cluster has 18,226 audio samples and the largest 95,813. The cluster sizes of this run are the density signal $n_{k(i)}$ in Equation~\ref{eq:selection}. At the selector cap of $2^{14.5}$, 17,870 clusters exceed the cap and hold 42.6\% of the audio samples.

Inside each cluster, the pairwise cosine matrix of the members is computed on the GPU. This cosine similarity is set relatively high, as earlier results show false positives. Mostly only pairs at 0.995 and above in the listening sets were repeated recordings. Clusters with more than 20,000 members are subsampled to 20,000 random members before the comparison. Within each k-means cluster, near-duplicate samples with >0.995 cosine similarity form groups, where only 1 sample per group is kept. The keep order is by group: public datasets with captions are more important than noisy YouTube data. SemDeDup removes 18.2M audio samples.

\subsection{Evaluation quarantine}
\label{app:quarantine}
All audio in the 500M selection and LAION-Audio-XL is screened against the evaluation data of MAEB, HEAR and MSEB. The screen uses the acoustic and semantic matching of the corpus deduplication together with identifier matching. Identifier matching compares source identifiers, file names, Freesound and YouTube identifiers with their time windows, and test-split metadata with the benchmark records. Acoustic matching uses two signal fingerprints, spectral-peak landmarks as in Olaf and binary band-energy sub-fingerprints, which find copies that were re-encoded, trimmed or embedded in longer recordings. Semantic matching uses a neural fingerprint of 1-second windows, which also covers codec and mixing changes. An exhaustive search of every evaluation item over the corpus waveforms confirms each match with a test of statistical significance, and every candidate pair is verified on the waveform by cross-correlation. An audio sample is removed when it holds the same recording as an evaluation item or an excerpt of it. Short generic sounds that match by chance are kept.

\subsection{Storage format}
The 500M selection and LAION-Audio-XL use WebDataset shards pairing FLAC audio with JSON metadata under a shared key, with numeric archive names and at most 4,096 samples per shard. Audio above 32\,kHz was re-encoded to 32\,kHz, reducing storage by approximately 16\%. This choice was supported by measurements showing that 99.68\% of samples contained at least 95\% of their spectral energy below 13\,kHz. 

\section{Taxonomy, enrichment, and deterministic selection}
\label{app:taxonomy}
\subsection{Taxonomy construction and classifier}
\begin{figure}[!htb]
\centering
\includegraphics[width=0.94\textwidth]{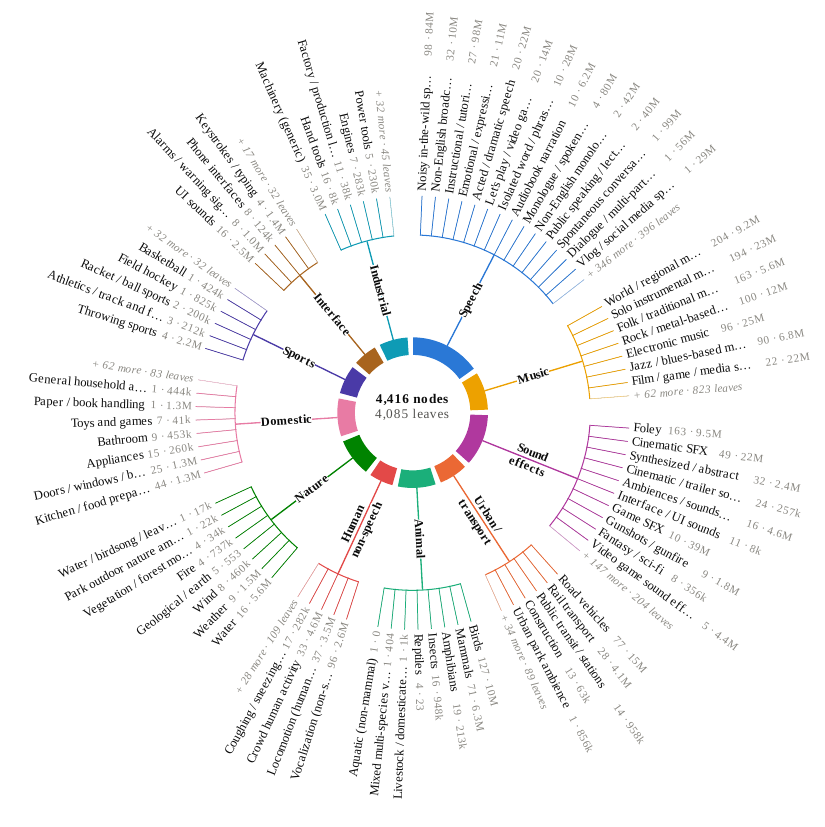}
\caption{Visualization of the sound taxonomy. The five levels contain 11, 837, 3,150, 393, and 25 nodes, respectively, for a total of 4,416 nodes, including 4,085 leaves. Coloured inner arcs denote the 11 top-level groups, ordered clockwise from speech at the top. Branches show 77 of the 837 second-level categories, selected by descendant leaf count and audio sample count in the deduplicated pool of 1.28B audio samples (grey numbers, respectively). Italic rows summarize omitted categories and their leaves.}
\label{fig:taxonomy_radial}
\end{figure}

Figure~\ref{fig:taxonomy_radial} shows the top-level groups and a selection of second-level categories of the taxonomy.
The taxonomy was expanded through caption-classification batches. An LLM assigned each caption to a leaf and proposed additional categories. Merging removed duplicates and also rejected categories consisting of multiple events. The final hierarchy contains 4,416 nodes, including 4,085 leaves and 11 top-level groups, with depth up to five. The top-level groups are speech, music, designed sound effects, urban/transport, animal/bioacoustics, human non-speech, nature/environmental, domestic/household, interface/electronic, sports/recreation, and industrial/mechanical. For the classifier, only depths up to 3 were kept (instead of the full hierarchy of 5), resulting in a smaller prediction vocabulary of 3,718.

The taxonomy classifier is a multilayer perceptron (MLP) trained on 1,280-dimensional mean-pooled features from a frozen Dasheng-0.6B encoder. A shared 1,024-unit hidden layer applies a linear transformation, batch normalization, GELU activation, and dropout. Separate linear output layers map the hidden representation to either leaf or top-level or second-level class logits. The labelled set has 232,728 caption-distilled audio samples, split by archive into 185,006 training, 22,404 validation, and 25,318 test audio samples. The MLP classifier is optimized for 40 epochs with AdamW, batch size 8,192, learning rate $10^{-3}$, and weight decay $10^{-4}$. Labels were refined with audio-language models; the released head reaches 85.6\% top-level accuracy on the held-out test set.

\subsection{Specialist enrichment}
\label{sub:specialist_enrichment}
Before our selection process, we gather extra data to make a better informed decision. Specialist enrichment is routed by the predicted top-level taxonomy class. Speech samples receive language, emotion, speaker age and gender, and pitch predictions. Music samples receive genre and style tags, tempo, and musical key. Animal samples receive bioacoustic class predictions. These outputs accompany the taxonomy labels and signal statistics, including duration, loudness, silence fraction, clipping fraction, and bandwidth.

For speech, MMS-LID-4017 \citep{pratapScalingSpeechTechnology2024} predicts one of 4,017 language labels, using the highest-scoring label and its probability. Emotion2vec+ large \citep{maEmotion2vecSelfSupervisedPreTraining2024} predicts one of nine labels: angry, disgusted, fearful, happy, neutral, other, sad, surprised, and unknown. The wav2vec2~\citep{baevskiWav2vec20Framework2020} age and gender model \citep{burkhardtSpeechBasedAgeGender2023} provides a continuous age estimate in years and a label of either female, male, or child. CREPE \citep{kimCREPEConvolutionalRepresentation2018} estimates the pitch of the voice (its fundamental frequency) between 50 and 550\,Hz every 200\,ms. The stored pitch is the median of these estimates, counting only the moments where the voice is clearly periodic (periodicity above 0.5).

For music, MAEST \citep{alonso-jimenezEfficientSupervisedTraining2023} predicts out of 400 Discogs genre and style labels. Signal processing with librosa estimates continuous tempo in beats per minute and one of 24 musical keys, of 12 pitch classes in major or minor mode. For bioacoustics, Perch 2.0 \citep{vanmerrienboerPerch20Bittern2025} has 14,795 output classes, consisting of 14,597 species labels and 198 general sound-event classes. The five highest-scoring classes and their logits are used.

\subsection{Selection parameters and exact-size pass}\label{app:selection}
For candidate clip $i$, the selection probability is $p_i$. This depends on whether the clip passes the quality checks ($g_i$), its minimum selection probability ($r_i$), the size of its acoustic cluster ($c_i$), and the adjustment for the target class proportions ($w_i$):
\begin{equation}
 p_i=g_i\left[r_i+(1-r_i)c_iw_i\right],\qquad
 c_i=\min\left(1,\frac{2^{14.5}}{n_{k(i)}}\right).
 \label{eq:selection}
\end{equation}
Here, $n_{k(i)}$ is the number of clips in the acoustic cluster. The factor $c_i$ reduces the selection probability for clips in large clusters. For the 500M selection, each clip must pass all of the following checks: the audio must decode successfully, the clip must contain at least 1\,s of content and its duration must be between 0.5 and 120\,s. Clips with both silence fraction $\geq0.98$ and RMS $\leq-80$\,dB are excluded. We set $g_i=1$ if the clip passes all checks, and $g_i=0$ otherwise. For these correct clips, the rules below set a minimum selection probability. The value $r_i$ is the largest probability assigned by these rules, or zero if none applies. A value of 1 means the clip is always kept.

For each category at the most specific level of the taxonomy, valid candidates are sorted from highest to lowest prediction confidence. Each receives a weight of 1 at confidence $\geq0.5$, or 0.3 otherwise. Starting at the top of the list, we add these weights until their sum reaches or exceeds 2,000. Each clip included in this sum gets its weight as a minimum selection probability. If the sum for the whole list is below 2,000, this applies to all clips in the list.

We set $r_i=1$ for up to 50,000 valid candidates for each predicted language, chosen from highest to lowest language confidence. The same applies to up to 1M candidates per emotion label and 500 per species label, chosen from lowest to highest hash value. The emotion labels covered by this rule are angry, disgusted, fearful, happy, sad, and surprised. We also set $r_i=1$ for valid clips whose second top-level prediction has probability $\geq0.25$ for music and designed sound effects, or $\geq0.10$ for the remaining categories other than speech. This rule also covers valid clips from CommonVoice\_22\_0 and EarSAVAS, to promote speaker and language diversity.

The target class proportions are applied to expected clip counts, where a candidate selected with probability 0.3 contributes 0.3 to the count. The initial budget is $T=660$M before the final size adjustment, with shares $(0.37,0.29,0.34)$ for speech, music, and all remaining classes ($(244.2,191.4,224.4)$M). Unused budget in one group is not moved to another. For second-level category $h$, the protected expected count $P_h$ comes from the minimum probabilities $r_i$. The adjustable expected count $D_h$ is the additional contribution after applying the cluster-size factor $c_i$, before adjusting the class proportions:
\begin{equation}
 P_h=\sum_{i\in h}g_ir_i,\qquad D_h=\sum_{i\in h}g_i(1-r_i)c_i.
\end{equation}
For group $b$ with share $a_b$, the required reduction in expected count is $E_b=\max(0,\sum_{h\in b}(P_h+D_h)-a_bT)$. Let $\mathcal H_b$ initially contain its categories other than language/accent and condition-specific speech. We reduce the largest adjustable category counts first, bringing them toward a common limit until the required reduction is reached or no adjustable count remains. This limit $\lambda_b$ satisfies
\begin{equation}
 \sum_{h\in\mathcal H_b}(D_h-\lambda_b)_+
 =\min\left(E_b,\sum_{h\in\mathcal H_b}D_h\right),
 \qquad X_h=(D_h-\lambda_b)_+,
\end{equation}
where $(x)_+=\max(x,0)$ and $X_h$ is the reduction in expected count for category $h$. If a further reduction is needed, the same calculation is applied to the categories left out of the first step. The protected expected count $P_h$ is never reduced. The implementation skips reductions of at most 0.5 in expected clip count.

Within each category requiring a reduction, candidates with a positive adjustable contribution $d_i=g_i(1-r_i)c_i$ are sorted by decreasing removal score
\begin{equation}
 s_i=4G_i+2q_i+\frac{8}{170}\log_2 n_{k(i)}+2A_i-2E_i.
\end{equation}
Here $G_i$ marks a clip from YouTube or another large web collection. $q_i$ is top-level prediction confidence. $A_i$ marks an original caption, a model caption, or text of at least 25 characters. $E_i$ marks one of the emotion labels listed above. Missing confidence and missing cluster density contribute zero to the score. Candidates with the same score are ordered by $\operatorname{SplitMix64}(\mathrm{audio\_id}_i)$, from lowest to highest. Adding $d_i$ in this order, let $j$ be the first candidate for which the sum reaches $X_h$. The factor for adjusting the class proportions is
\begin{equation}
 w_i=\begin{cases}
 0,&i\text{ precedes }j,\\
 1-\bigl(X_h-\sum_{\ell\prec j}d_\ell\bigr)/d_j,&i=j,\\
 1,&i\text{ follows }j.
 \end{cases}
\end{equation}
Categories that do not require reduction have $w_i=1$. To decide which clips are kept, we turn each clip's audio ID into a random number between 0 and 1 with a hash, based on a fixed seed. A clip is kept if this number is below its selection probability $p_i$. The number depends only on the audio ID, so every run selects the same clips. To reach 500M clips, we randomly remove speech from the selection excluding the sets we want to keep. 

\section{Dataset Captioning}
\label{app:captions}

\subsection{Captioning Model}
For captioning the dataset, we put effort into gathering high quality and diverse captions through a pipeline with fast throughput. Moreover, if the captions often contain incorrect details or are very similar to one another, we end up with a weak training signal. To pick the best models for our use case, we tested seven configurations covering six model candidates: \textsc{MOSS-Audio-8B} (thinking on and off)~\citep{yangMOSSAudioTechnicalReport2026}, \textsc{Qwen3-Omni-30B-A3B-Instruct}~\citep{xuQwen3OmniTechnicalReport2025}, \textsc{Qwen2.5-Omni-7B}~\citep{xuQwen25OmniTechnicalReport2025}, \textsc{AudioFlamingoNext}~\citep{ghoshAudioFlamingoNext2026}, \textsc{Nemotron-3-Nano-Omni-30B-A3B-Reasoning-FP8}~\citep{nvidiaNemotron3Nano2026}, and \textsc{Nemotron-Labs-Audex-30B-A3B}~\citep{kongUnifiedAudioIntelligence2026}. Model names in the following might be shortened.

\begin{table}[t]
\centering
\newcommand{\ci}[1]{\,{\scriptsize$\pm$#1}}
\newcommand{\hd}[1]{\begin{tabular}[b]{@{}c@{}}#1\end{tabular}}
\newcommand{\up}{\,$\uparrow$}
\newcommand{\down}{\,$\downarrow$}
\setlength{\tabcolsep}{3pt}
\begin{tabular*}{\textwidth}{@{\extracolsep{\fill}}lcccccc@{}}
\toprule
& speed\up & \multicolumn{2}{c}{diversity\up} & \multicolumn{3}{c}{quality} \\
\cmidrule(lr){2-2} \cmidrule(lr){3-4} \cmidrule(lr){5-7}
Captioner & \hd{samples/s\\per node} & \hd{vocab\\@80k} & \hd{vocab\\@400k} & \hd{human\\rank\down} & \hd{LLM\\rank\down} & \hd{LLM\\BT\up} \\
\midrule
MOSS-Audio-8B-Thinking (think-ON) & 22.9 & \textbf{5{,}488} & \textbf{16{,}613} & \textbf{2.91}\ci{0.39} & 2.76\ci{0.50} & 1.694 \\
MOSS-Audio-8B-Thinking (think-OFF) & 19.2 & 5{,}054 & 15{,}768 & 3.24\ci{0.36} & \textbf{2.61}\ci{0.40} & \textbf{1.755} \\
Qwen2.5-Omni-7B & 23.0 & 4{,}652 & 8{,}660 & 4.10\ci{0.45} & 4.43\ci{0.50} & 0.873 \\
AudioFlamingoNext & 55.5 & 4{,}743 & 9{,}858 & 4.26\ci{0.41} & 4.51\ci{0.53} & 0.714 \\
Qwen3-Omni-30B-A3B & 75.9 & 2{,}693 & 6{,}619 & 4.32\ci{0.45} & 3.43\ci{0.51} & 1.093 \\
Nemotron-3-Nano-Omni & \textbf{96.5} & 2{,}284 & 7{,}078 & 4.57\ci{0.47} & 5.16\ci{0.40} & 0.459 \\
Audex-30B-A3B & 36.5 & 5{,}395 & 9{,}697 & 4.60\ci{0.45} & 5.10\ci{0.57} & 0.411 \\
\bottomrule
\end{tabular*}
\caption{\textbf{Caption Model Comparison}. Arrows indicate whether higher ($\uparrow$) or lower ($\downarrow$) is better. \textbf{Speed:} Throughput is given in samples/s per node (4x GH200 GPUs). \textbf{Diversity:} \emph{Vocab@80k} and \emph{vocab@400k} are the number of unique words across 80k short and 400k long caption tokens, respectively. \textbf{Quality:} \emph{Human rank} and \emph{LLM rank (Gemini 3.6 Flash)} report the mean rank among the 7 captioners for the 50 shared audio samples (1 = best, 7 = worst) $\pm$ 95\% CI. LLM BT is the Bradley--Terry strength for the Gemini 3.6 Flash judge on the full 8997 test audio samples.}
\label{tab:caption_model_performance}
\end{table}

\textbf{Quality:} In the following study (Table~\ref{tab:caption_model_performance}), we compare the caption quality of the candidate models. We use an automated LLM review in which all model captions are ranked against one another for each clip. Gemini 3.6 Flash \citep{doshiIntroducingGemini362026} ranked anonymized and shuffled candidates according to faithfulness, coverage, and specificity. Additional permutations, pairwise comparisons, and a claim audit measured judge consistency. We picked a subset of 8997 audio samples matching our data distribution. After generating one caption per candidate, we calculate the Bradley-Terry strength~\citep{bradley1952rank} from the automated rankings. Table~\ref{tab:caption_model_performance} shows that MOSS Audio 8B produces the best results by showing the highest Bradley-Terry strength in the LLM review regardless of the thinking setting. In contrast, Audex and Nemotron-3 are the two weakest models across the evaluation.

To ground these results in human perception, we rank a subset of 50 audio samples (25 short, 25 long captions) with two human annotators  and compare the results with the LLM judge. The labeling interface anonymizes models and shuffles the captions. The annotators only agreed 12 out of 50 times on their top choice, with a mean per-clip Kendall $\tau_b$ of $+0.295$ $[+0.202, +0.387]$. Compared to the inter-annotator agreement, the LLM judge falls within the range of a human rater: $\tau_b = +0.345$ $[+0.250, +0.436]$ with one annotator and $+0.265$ $[+0.166, +0.362]$ with the other. Brackets are $95\%$ percentile bootstrap confidence intervals from $20{,}000$ resamples over clips. Pooling the two annotators, humans and judge produce nearly the same ordering of the seven captioners (Table~\ref{tab:caption_model_performance}). Both put the two MOSS Audio configurations on top and Nemotron-3-Nano-Omni and Audex-30B-A3B at the bottom. The largest single disagreement is Qwen3-Omni-30B-A3B, which the judge places 3rd while the ranking done by humans places it 5th.

Note that the set of $n = 50$ human-annotated audio samples is meant to validate the LLM review, not to serve as a leaderboard. It can only give a rough impression of the top and bottom models while checking that the LLM judge behaves similarly to a human rater.

\textbf{Speed:} Another important aspect when selecting captioning models is inference speed. When captioning several hundred million samples, even slightly slower models can lead to a significant increase in compute cost. To optimize the captioning speed of each model, we considered serving layout, hyperparameter settings (e.g., maximum number of sequences), NUMA bindings, caching, output format, and more. We ended up with large differences of up to \textasciitilde$5\times$ between our fastest (Nemotron, 96.5 samples/s/node) and slowest (MOSS think-OFF, 19.2 samples/s/node) model (Table~\ref{tab:caption_model_performance}). This puts the theoretical compute estimate for 471 million audio samples between 1,350 and 6,800 node-hours.

\textbf{Diversity:} Table~\ref{tab:caption_model_performance} shows the number of unique words within a limited number of text tokens. We evaluate this metric on short and long captions with 80k and 400k token limits, respectively. This way, the metric does not depend on the verbosity of each model. There is a \textasciitilde$2.0$--$2.5\times$ difference between the most repetitive (Nemotron-3-Nano-Omni and Qwen3-Omni-30B-A3B) and the most diverse (MOSS Audio think-ON) model.

Based on the results of this study and our three main criteria (quality, diversity, and computational cost), two models stood out to us. MOSS Audio think-ON stood out for its high diversity and quality despite being slow, and Qwen3-Omni for its speed while still performing strongly on the Bradley-Terry strength metric. We decided to use a combination of Qwen3-Omni and MOSS Audio think-ON. This combination achieves a caption diversity of \textasciitilde4,395 and \textasciitilde11,976 unique words per 80k/400k text tokens, respectively, while balancing the quality and compute cost of MOSS Audio think-ON with the speed of Qwen3-Omni-30B-A3B. Both models then captioned disjoint halves of the \textasciitilde471 million audio samples, balanced within each source dataset. This took 3,884 node-hours ($4\times$ GH200 per node).

\subsection{Metadata Prompt Expansion}
\label{app:metadataprompt}
\begin{table}[t]
  \centering
\begin{tabular}{cllcc}
    \toprule
    \# & Model & Caption type & Mean rank $\downarrow$ & Win rate (\%) $\uparrow$ \\
    \midrule
    1 & Qwen3-Omni & Raw JSON         & $\mathbf{3.20} \pm 0.10$ & $\mathbf{19.9} \pm 2.5$ \\
    2 & MOSS Audio & Raw JSON         & $3.35 \pm 0.10$ & $16.4 \pm 2.4$ \\
    3 & Qwen3-Omni & Natural sounding & $3.46 \pm 0.10$ & $14.7 \pm 2.2$ \\
    4 & MOSS Audio & Natural sounding & $3.56 \pm 0.10$ & $12.9 \pm 2.1$ \\
    5 & MOSS Audio & Audio only       & $3.58 \pm 0.12$ & $19.6 \pm 2.5$ \\
    6 & Qwen3-Omni & Audio only       & $3.84 \pm 0.12$ & $16.6 \pm 2.4$ \\
    \bottomrule
  \end{tabular}
        \caption{\textbf{Attaching raw metadata strings leads to the highest accuracy.} Overall ranking of two captioning models with three prompt types across the same 976 audio samples, evaluated with Gemini 3.6 Flash (thinking: low). Mean rank ranges from 1 (best) to 6 (worst). Win rate refers to being the best caption for a clip (higher-is-better). The $\pm$ values give 95\% percentile confidence intervals over $B=20{,}000$ resamples of the 976 audio samples.}
          \label{tab:metadata_prompt_expansion_results}

\end{table}

When captioning audio samples, the prompt given to the captioning model directly affects the quality of the produced caption. In addition to the general instructions, we identified three primary approaches to include each sample's metadata generated by specialist models (Appendix~\ref{sub:specialist_enrichment}). They are (1)~passing only general instructions with the audio, (2)~attaching the whole raw JSON metadata string, or (3)~adding only selected JSON fields converted to natural language to the prompt. We tested these three approaches in combination with our two captioning models (Qwen3-Omni and MOSS Audio) on 976 audio samples, selected to cover the dataset's duration and domains, resulting in six long captions per clip. Gemini 3.6 Flash then ranked all 6 captions for the same audio sample by accuracy, completeness, temporal details, specificity, and writing. The judge received only the audio and the six randomly ordered captions, never the metadata, so it could not simply confirm our specialists' predictions. 

The results in Table~\ref{tab:metadata_prompt_expansion_results} show that adding the raw metadata JSON string results in the best average performance across models when ranking them. This observation also holds for almost all audio lengths and domains. We attribute this to the quality of the metadata and the ability of modern large audio-language models to process JSON strings well. Considering these results, we include the raw metadata JSON string in the prompts for our captioning process.

\section{Architecture}

\subsection{Architecture Settings}
\label{app:architecture}
All models used in our experiments share a single dual-encoder layout. The layout is a contrastive audio-text model in the CLAP family: a spectrogram Vision Transformer with native variable-length input (NaFlex)~\citep{tschannenSigLIP2Multilingual2025} encodes audio, a bidirectional Transformer encodes text, and both are mapped into a shared embedding space trained with a standard CLIP objective on audio and text.

\begin{table}[t]
\centering
\small
\setlength{\tabcolsep}{4pt}
\begin{tabular}{lcccrrrrrr}
\toprule
Tier & Audio & Text & $d$ & A-Size & T-Size & A:T & $N$ & Total & GFLOPs \\
     & $d_a{\times}L_a$, $h$, $d_{\text{ff}}$ & $d_t{\times}L_t$, $h$, $d_{\text{ff}}$ & & (M) & (M) & & (M) & (M) & /clip \\
\midrule
XXS & $224{\times}12$, 4, 640   & $224{\times}12$, 4, 640   & 224 &   8.3 &   7.8 & 1.06 &  16.0 &  27.3 &  26.1 \\
XS  & $320{\times}12$, 5, 896   & $320{\times}12$, 5, 896   & 320 &  16.6 &  15.7 & 1.06 &  32.2 &  48.3 &  46.8 \\
S   & $512{\times}12$, 8, 1408  & $512{\times}11$, 8, 1408  & 512 &  41.8 &  36.4 & 1.15 &  78.2 & 103.9 & 103.4 \\
M   & $640{\times}12$, 10, 1728 & $640{\times}11$, 10, 1728 & 640 &  64.4 &  56.2 & 1.15 & 120.6 & 152.8 & 152.9 \\
L   & $1024{\times}12$, 16, 2816 & $960{\times}11$, 15, 2560 & 960 & 166.5 & 125.4 & 1.33 & 291.8 & 340.1 & 357.9 \\
\bottomrule
\end{tabular}
\caption{\textbf{Model tiers for WSD scaling runs:} Audio and text columns give
width $\times$ depth, number of attention heads, and SwiGLU hidden size per encoder. A-Size and T-Size are the respective encoder parameter counts in millions. Audio counts include the MAP pool and projection layers. Text counts include the pool and output projection but \emph{exclude} the $50{,}260\times d_t$ token-embedding table. $N$ (non-embedding) is the sum of A-Size and T-Size. GFLOPs are total (audio + text tower) forward-pass FLOPs per clip at the training length distribution. }
\label{tab:arch-dimensions}
\end{table}

For the architectural design, we aimed to align with established encoder widths and depths while offering a broad range of sizes. We ended up with five model scales from XXS (27.3M parameters) up to L (340M parameters) that are aligned with known small-scale ViT dimensions~\citep{dosovitskiyImageWorth16x162021} and optimized for hardware alignment as seen in Table~\ref{tab:arch-dimensions}. For more details on the audio-to-text encoder ratio, refer to Appendix~\ref{app:at_ratio}. With this setup, we cover the computational range per forward pass from an average of 26.1 GFLOPs up to 357.9 GFLOPs per clip, given our training distribution.

\begin{table}[t]
\centering
\caption{Architecture settings shared by all tiers in the scaling study. For details on architecture shape refer to Table~\ref{tab:arch-dimensions}~\&~\ref{tab:arch-dimensions-cosine}.}
\small
\begin{tabular}{ll}
\toprule
\multicolumn{2}{l}{\emph{Audio front-end}} \\
Sample rate / mel bins          & 32\,kHz / 128 (50\,Hz--16\,kHz) \\
Window / hop                    & 1024 / 320 samples (100 frames/s) \\
Patch (freq $\times$ time)      & $16\times16$ (160\,ms), 50 tokens/s \\
Max.\ clip length               & 30\,s (1504 tokens), train and eval \\
Length buckets (tokens)         & 344 / 448 / 648 / 1128 / 1504 \\
Audio tokens per GPU step       & 49{,}152 \\
\midrule
\multicolumn{2}{l}{\emph{Audio encoder}} \\
Positional encoding             & 2-D axial RoPE (freq, time), no learned pos.\ emb. \\
Norm / MLP                      & pre-norm RMSNorm / SwiGLU \\
Extra tokens                    & 1 register, no class token \\
LayerScale init                 & $10^{-5}$ \\
Pooling / projection                  & MAP / Linear--GELU--Linear \\
\midrule
\multicolumn{2}{l}{\emph{Text encoder}} \\
Tokenizer / vocab / context     & GPT-2 BPE (\texttt{r50k\_base}) / 50{,}260 / 256 \\
Attention                       & bidirectional, QK-norm, no biases \\
Positional encoding             & 1-D RoPE \\
Norm / MLP                      & pre-norm RMSNorm / SwiGLU \\
Pooling / projection                  & MAP / bias-free linear \\
\midrule
\multicolumn{2}{l}{\emph{Joint space}} \\
Temperature                     & learned, init $1/0.07$, clamp 300 \\
Loss                            & symmetric InfoNCE over global batch \\
Precision                       & bf16 mixed precision \\
\bottomrule
\end{tabular}
\label{tab:arch-shared}
\end{table}

For the architectural hyperparameters and mechanisms, we adopt proven approaches such as RoPE and SwiGLU, while motivating others through our own experiments in Appendix~\ref{sec:duration}. An overview of key design choices can be found in Table~\ref{tab:arch-shared}.

\subsection{Audio to Text Encoder Parameter Ratio}
\label{app:at_ratio}
A main component of the architecture design is the parameter ratio between the audio and text encoders. This ratio depends on many factors, including downstream benchmarks, model size, and the training corpus. In the following, all reported ratios and parameter shares exclude parameters related to text token embedding, which is a fixed lookup cost. When comparing other contrastive pre-trained architectures, one finds substantial discrepancies among them. While OpenCLIP~\citep{chertiReproducibleScalingLaws2023} and SigLIP2~\citep{tschannenSigLIP2Multilingual2025} allocate up to \textasciitilde78\% of the model parameters to the vision encoder, SigLIP1~\citep{zhaiSigmoidLossLanguage2023} splits evenly, and LAION-CLAP~\citep{wuLargeScaleContrastiveLanguageAudio2023} and MS-CLAP~\citep{elizaldeCLAPLearningAudio2023} do the opposite, by assigning between \textasciitilde51\% and \textasciitilde73\% of parameters to the text encoder. The difference is that vision-based contrastive models are trained on large vision-language datasets that provide substantial visual variety while offering rather compressed text captions. Therefore, scaling the vision encoder results in higher benchmarks. Older CLAP models, on the other hand, struggle with overfitting due to data scarcity while depending on pre-trained component sizes.

\begin{figure}[t]
  \centering
  \begin{minipage}[c]{0.46\textwidth}
    \includegraphics[width=\linewidth]{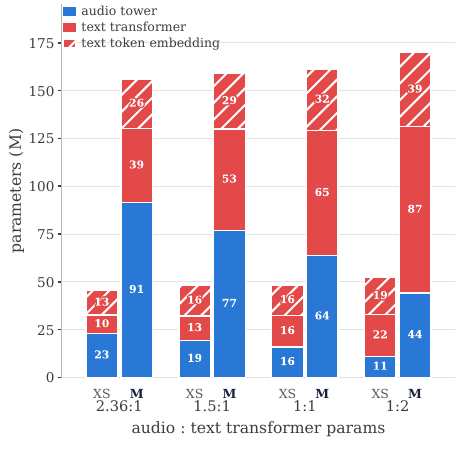}
  \end{minipage}\hfill
  \begin{minipage}[c]{0.50\textwidth}
    \centering
    \small
    \setlength{\tabcolsep}{3pt}
    \begin{tabular}{@{\quad}lrrrr@{}}
\toprule
& \multicolumn{4}{c}{audio\,:\,text} \\
\cmidrule(l){2-5}
$D$ & 2.36\,:\,1 & 1.5\,:\,1 & 1\,:\,1 & 1\,:\,2 \\
\midrule
\multicolumn{5}{@{}l}{\itshape XS tier} \\
16M & \textbf{34.6} & \textbf{34.6} & 34.4 & 33.5 \\
32M & 36.4 & 36.2 & \textbf{36.5} & 35.9 \\
64M & 38.5 & \textbf{38.8} & 37.9 & 37.7 \\
512M & \textbf{47.3} & -- & 45.1 & -- \\
\addlinespace[2pt]
\multicolumn{5}{@{}l}{\itshape M tier} \\
16M & 35.4 & 35.7 & \textbf{36.1} & 34.6 \\
32M & 36.8 & 36.7 & \textbf{37.3} & 36.3 \\
64M & 39.5 & 38.7 & \textbf{39.6} & 39.3 \\
\bottomrule
\end{tabular}
  \end{minipage}
  \caption{\textbf{Audio:text capacity choice does not significantly affect performance up to 64M samples.}
  Left: parameter composition of the four capacity splits for model sizes XS and M. Within a tier, the non-embedding total is held constant. The 50k-vocab token embedding is adapted to the individual layer dimensions and is a fixed lookup cost, not encoder capacity. Right: Mean MAEB percentage scores over 30 tasks per training sample seen $D$ show that the audio-to-text tower ratio does not have a significant effect up to 64M samples.}
  \label{fig:ta_ratio}
\end{figure}

We tested four different parameter allocations on different data scales and model sizes, as shown in Figure~\ref{fig:ta_ratio}. Changing the audio-to-text ratio from 2.36:1 to 1:2 does not show significant differences in the MAEB mean score with $D \le 64M$. This observation does not hold for the two XS parameter ratios we tested on $D = 512M$, with a difference of 2.2 points. Performance differences might appear as $D$ increases further. For computational reasons, we stopped after the runs shown in Figure~\ref{fig:ta_ratio}. In general, shifting parameters to the text encoder results in faster inference speeds given that the encoder widths remain hardware-aligned. Across our model range, this results in up to \textasciitilde22\% faster inference speed. Considering these aspects, we chose conservative CLIP-inspired ratios by allocating \textasciitilde51\% to \textasciitilde57\% of parameters to the audio encoder. The split also accounts for our new data scale and the design freedom we gain by avoiding pre-trained components. For more details on the architecture, refer to Appendix~\ref{app:architecture}.

\section{Model Scaling}
\label{app:scaling}

\textbf{Setup and hyperparameters:} All our scaling plots are based on two batches of runs. One is trained with warmup-stable-decay (WSD), and the other with cosine learning-rate scheduling. Both of these batches are consistent with each other but differ in the scheduler, warmup, learning rate, batch size in addition to altering the architecture proportions. While the older cosine-scheduled runs were done a bit earlier in development, they have a more audio-encoder-heavy parameter allocation and slightly change the total number of parameters (Table~\ref{tab:arch-dimensions-cosine}). We mark them with a $C$ after the model size, like XS-C. Since we compare the error rate to compute while varying hyperparameters like model size, we can include both in the scaling plots without issue. For all other plots and tables, unless explicitly mentioned otherwise, we always refer to the warmup-stable-decay scheduled runs, as they almost always outperform the cosine runs. You can find the respective training settings for the WSD, and the cosine runs in Table~\ref{tab:wsd_hypers} and~\ref{tab:cosine-hparams} respectively. We did some initial learning rate and batch size experiments on 32M, 64M, and 128M data scales and extrapolated them to other scales following ~\cite{nezhurinaScalingLawsRobust2026}. We then adapted the learning rate for WSD with a factor of $\times0.75$ and $\times0.80$ while choosing a data scale appropriate batch size.

\begin{table}[t]
  \centering
   \caption{\textbf{Model tiers for cosine scheduled scaling runs.} Audio and text columns give
width~$\times$~depth, number of attention heads, and SwiGLU hidden size per encoder. A-Size and T-Size are the respective encoder parameter counts in millions. Audio counts include the MAP pool and projection layers. Text counts include the pool and output projection but \emph{exclude} the $50{,}260\times d_t$ token-embedding table. $N$ (non-embedding) is the model size used in all scaling fits while total refers to the actual model size. GFLOPs are total (audio + text tower) forward-pass FLOPs per clip at the training length distribution.}
  \small
  \setlength{\tabcolsep}{4pt}
  \resizebox{\linewidth}{!}{%
  \begin{tabular}{lllrrrrrrr}
  \toprule
  Tier & Audio & Text & $d$ & A-Size & T-Size & A:T & $N$ & Total & GFLOPs \\
   & $d_a{\times}L_a$, $h$, $d_{ff}$ & $d_t{\times}L_t$, $h$, $d_{ff}$ & & (M) & (M) & & (M) & (M) & /clip \\
  \midrule
  XXS-C & 256$\times$12, 4, 704   & 192$\times$12, 3, 512  & 192 & 10.5  & 5.5  & 1.92 & 15.9  & 25.6  & 29.2  \\
  XS-C  & 384$\times$12, 6, 1024  & 256$\times$12, 4, 682  & 256 & 22.9  & 9.7  & 2.36 & 32.6  & 45.5  & 58.7  \\
  S-C   & 576$\times$12, 9, 1536  & 384$\times$12, 6, 1024 & 384 & 51.5  & 21.8 & 2.36 & 73.3  & 92.6  & 117.1 \\
  M-C   & 768$\times$12, 12, 2048 & 512$\times$12, 8, 1365 & 512 & 91.4  & 38.8 & 2.36 & 130.2 & 156.0 & 194.9 \\
  L-C   & 896$\times$21, 14, 2389 & 640$\times$12, 10, 1706 & 640 & 211.3 & 60.6 & 3.49 & 271.9 & 304.1 & 431.1 \\
  \bottomrule
  \end{tabular}
  }
  \label{tab:arch-dimensions-cosine}
  \end{table}

\begin{table}[t]
\centering
\small
\setlength{\tabcolsep}{4pt}
\caption{\textbf{Training configuration of the cooled-down WSD endpoints.} ``XXS--L'' denotes
all five of our model sizes (XXS, XS, S, M, L); Samples is the number of training
samples seen. Every model size shares the schedule at a given budget. Warmup and
cooldown are in optimizer steps, and the cooldown decays the learning rate linearly
to zero over the final 20\,\% of the budget. Batch size varies based on the length
buckets. L at 1,024M ran on twice the GPUs with half the per-GPU token budget (24,576), which gives the same global batches.}
\label{tab:wsd_hypers}
\begin{tabular}{lrrrrrrc}
\toprule
 & & & & \multicolumn{3}{c}{Global Batch} & \\
\cmidrule(lr){5-7}
Models & Samples & Warmup & Cooldown & Min & Max & Avg & LR \\
\midrule
XXS--L & 1M     & 350   & 224    & 512   & 2,176 & 916   & $4.0 \times 10^{-4}$ \\
XXS--L & 2M     & 350   & 448    & 512   & 2,176 & 916   & $4.0 \times 10^{-4}$ \\
XXS--L & 4M     & 350   & 896    & 512   & 2,176 & 916   & $4.0 \times 10^{-4}$ \\
XXS--L & 8M     & 350   & 1,792  & 512   & 2,176 & 916   & $4.0 \times 10^{-4}$ \\
XXS--L & 16M    & 350   & 3,584  & 512   & 2,176 & 916   & $4.0 \times 10^{-4}$ \\
XXS--L & 32M    & 350   & 7,168  & 512   & 2,176 & 916   & $4.0 \times 10^{-4}$ \\
\midrule
XXS--L & 32M    & 2,800 & 1,744  & 2,048 & 8,704 & 3,683 & $7.5 \times 10^{-4}$ \\
XXS--L & 64M    & 2,800 & 3,488  & 2,048 & 8,704 & 3,683 & $7.5 \times 10^{-4}$ \\
XXS--L & 128M   & 2,800 & 6,976  & 2,048 & 8,704 & 3,683 & $7.5 \times 10^{-4}$ \\
XXS--L & 256M   & 2,800 & 13,952 & 2,048 & 8,704 & 3,683 & $7.5 \times 10^{-4}$ \\
XXS--L & 512M   & 2,800 & 27,904 & 2,048 & 8,704 & 3,683 & $7.5 \times 10^{-4}$ \\
XXS--L & 1,024M & 2,800 & 55,808 & 2,048 & 8,704 & 3,683 & $7.5 \times 10^{-4}$ \\
\bottomrule
\end{tabular}
\end{table}

\begin{table}[t]
  \centering
  \caption{\textbf{Training hyperparameters for the cosine learning rate scheduled scaling runs.} ``(XXS-L)-C'' denotes
  all five of our model sizes (XXS-C, XS-C, S-C, M-C, L-C); Samples is the number of training
  samples seen. Batch size varies based on the length buckets.}
  \small
  \setlength{\tabcolsep}{4pt}
  \begin{tabular}{lrrrrrrc}
  \toprule
   & & & & \multicolumn{3}{c}{Global Batch} & \\
  \cmidrule(lr){5-7}
  Models & Samples & Warmup & Nodes & Min & Max & Avg & LR \\
  \midrule
  (XXS-L)-C  & 2M   & 472   & 2  & 256   & 1,088 & 461   & $2.5 \times 10^{-4}$ \\
  (XXS-L)-C  & 4M   & 472   & 4  & 512   & 2,176 & 922   & $2.5 \times 10^{-4}$ \\
  (XXS-L)-C  & 8M   & 943   & 4  & 512   & 2,176 & 922   & $5.0 \times 10^{-4}$ \\
  (XXS-L)-C  & 16M  & 943   & 8  & 1,024 & 4,352 & 1,843 & $5.0 \times 10^{-4}$ \\
  (XXS-L)-C  & 32M  & 1,183 & 12 & 1,536 & 6,528 & 2,765 & $1.0 \times 10^{-3}$ \\
  (XXS-L)-C  & 64M  & 2,367 & 12 & 1,536 & 6,528 & 2,765 & $1.0 \times 10^{-3}$ \\
  (XXS-L)-C  & 128M & 3,774 & 16 & 2,048 & 8,704 & 3,686 & $1.0 \times 10^{-3}$ \\
  (XXS-L)-C  & 256M & 5,000 & 16 & 2,048 & 8,704 & 3,686 & $1.0 \times 10^{-3}$ \\
  XS-C, S-C   & 512M & 5,000 & 16 & 2,048 & 8,704 & 3,686 & $1.0 \times 10^{-3}$ \\
  \bottomrule
  \end{tabular}
  \label{tab:cosine-hparams}
\end{table}

\begin{figure}[htbp]
  \centering
  \begin{subfigure}[b]{0.32\textwidth}
    \centering
    \includegraphics[width=\textwidth]{figures/scaling_law/maeb_full_curve_fit_saturation_both_clap_1.00e_13_maeb_main_score_clap.pdf}
    \caption{MAEB (full, 30 tasks)}
    \label{fig:scaling_maeb_full}
  \end{subfigure}%
  \hfill
  \begin{subfigure}[b]{0.32\textwidth}
    \centering
    \includegraphics[width=\textwidth]{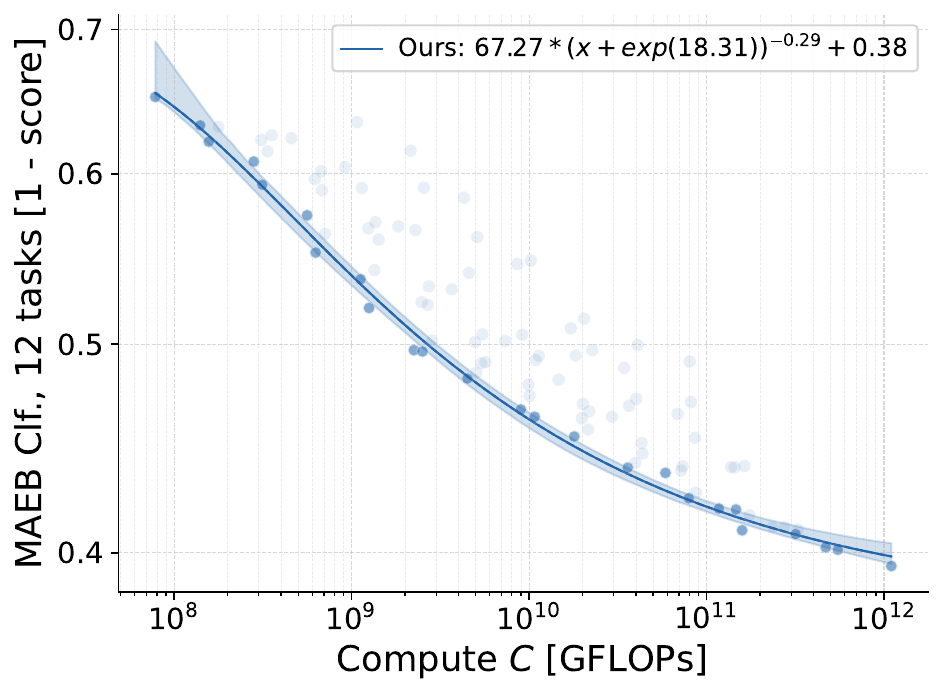}
    \caption{Classification (12 tasks)}
    \label{fig:scaling_maeb_clf}
  \end{subfigure}%
  \hfill
  \begin{subfigure}[b]{0.32\textwidth}
    \centering
    \includegraphics[width=\textwidth]{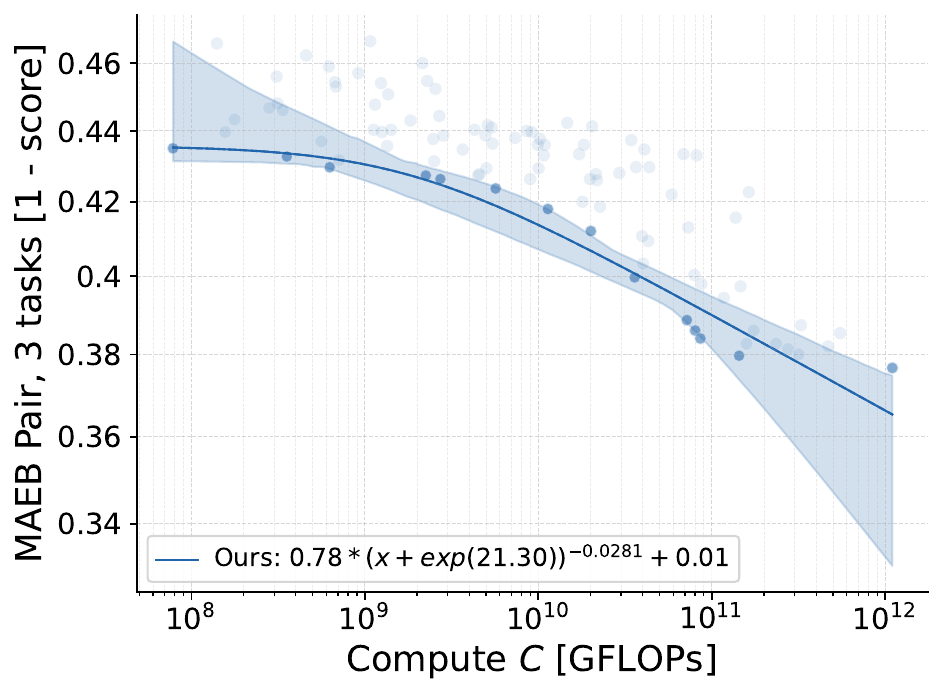}
    \caption{Pair Classification (3 tasks)}
    \label{fig:scaling_maeb_pair}
  \end{subfigure}

  \vspace{1em}

  \centering
  \begin{subfigure}[b]{0.32\textwidth}
    \centering
    \includegraphics[width=\textwidth]{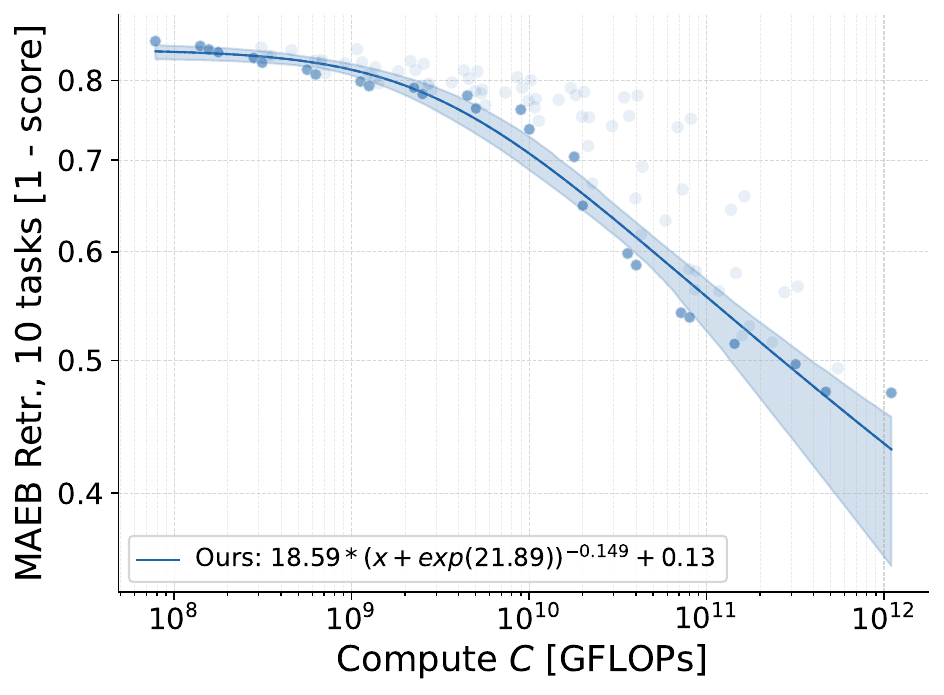}
    \caption{Retrieval (10 tasks)}
    \label{fig:scaling_maeb_retr}
  \end{subfigure}%
  \hspace{0.04\textwidth}%
  \begin{subfigure}[b]{0.32\textwidth}
    \centering
    \includegraphics[width=\textwidth]{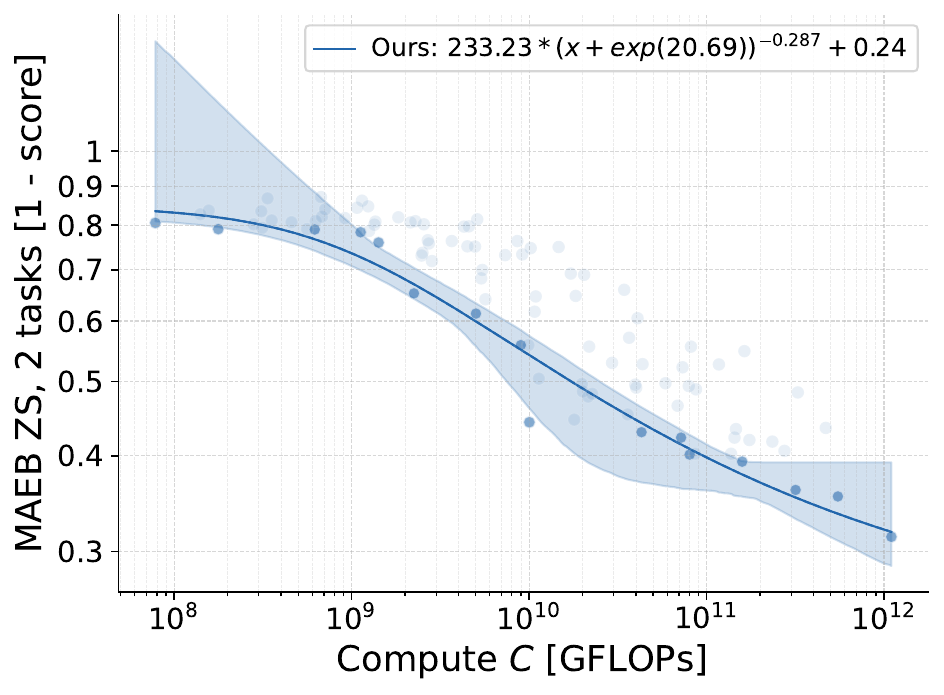}
    \caption{ZS Classification (2 tasks)}
    \label{fig:scaling_maeb_zs}
  \end{subfigure}
  \caption{\textbf{Scaling behavior on the full MAEB benchmark and per task type.} Error rate ($1 - \text{mean score}$) of the full MAEB-30 and of each task-type subset per compute C (GFLOPs). The band marks the 95\% bootstrap CI with 1000 resamples. It shows clear trends in compute scaling for zero-shot (ZS) classification, pair classification, and retrieval. In contrast, it isolates the saturating classification task.}
  \label{fig:scaling_maeb_tasks}
\end{figure}

\textbf{Additional scaling results:} Figure~\ref{fig:scaling_maeb_tasks} shows the scaling behavior of our model in contrast to different tasks of MAEB-30. It shows that, in addition to the mean MAEB-30 error rate, pair classification, retrieval, and zero-shot classification also scale well with compute. But what is not shown on the mean score scale is that classification tasks saturate at the end of our compute budget. Even more extreme are the clustering benchmarks, which scale inversely to the compute, so we do not fit them and leave them out of Figure~\ref{fig:scaling_maeb_tasks}. This is a known challenge of the MAEB benchmark as noted in their conclusion~\citep{assadiMAEBMassiveAudio2026}.

\begin{table}[t]
  \centering\small
  \caption{MAEB-30 task scores (\%) of the final scaling-grid models, each the
    cooled-down checkpoint of its tier at 1024M samples seen. Parameter counts are
    totals over both towers, including embeddings. All models are evaluated with native
    MTEB~2.21.0~\citep{muennighoffMtebMassiveText2023} on the same embedding wrapper, with audio capped at 30\,s (1504 tokens), un-normalised audio embeddings, and all-caps text inputs lowercased.
    Multi-subset tasks average all subsets.
    Bold marks the best value per row.}
  \label{tab:maeb30-scaling}
  \setlength{\tabcolsep}{4pt}
  \begin{tabular}{llrrrrr}
  \toprule
  Task                  & Type      &      XXS (27M) &       XS (48M) &       S (104M) &       M (153M) &       L (340M) \\
  \textit{Samples seen} &           &          1024M &          1024M &          1024M &          1024M &          1024M \\
  \midrule
  \multicolumn{7}{l}{\textit{Speech (18 tasks)}} \\
  SpokenSQuAD T2A       & Retr.     &           34.3 &           39.0 &           38.0 &           41.7 &  \textbf{45.0} \\
  CommonVoiceMini21 T2A & Retr.     &           60.8 &           65.6 &           68.0 &  \textbf{71.9} &           70.7 \\
  FLEURS T2A            & Retr.     &           47.3 &           50.7 &           55.2 &  \textbf{61.8} &           59.4 \\
  GigaSpeech T2A        & Retr.     &           59.1 &           64.3 &           71.6 &           70.4 &  \textbf{72.3} \\
  MInDS-14              & Clf.      &           47.1 &           61.9 &           70.6 &           71.7 &  \textbf{73.8} \\
  VoxCeleb SA           & Clf.      &           29.4 &           30.7 &           33.4 &           33.3 &  \textbf{34.2} \\
  IEMOCAP gender        & Clf.      &           80.1 &           85.9 &           90.4 &           90.6 &  \textbf{93.9} \\
  VoxPopuli language ID & Clf.      &           92.5 &           96.4 &           98.0 &           98.1 &  \textbf{98.2} \\
  CREMA-D               & Clf.      &           30.8 &           34.2 &           37.4 &           38.4 &  \textbf{39.6} \\
  CommonLanguage age    & Clf.      &           15.4 &           15.6 &           16.0 &           15.8 &  \textbf{16.7} \\
  VoxPopuli gender      & Clust.    &            5.6 &            8.1 &           23.8 &           23.6 &  \textbf{26.5} \\
  CREMA-D (clustering)  & Clust.    &            0.7 &            0.6 &            0.6 &            0.6 &   \textbf{0.8} \\
  SIB-FLEURS            & Clf.      &           21.3 &           23.6 &           26.8 &           27.2 &  \textbf{27.6} \\
  NMSQA                 & Pair      &           75.6 &           76.8 &           76.9 &           76.4 &  \textbf{78.0} \\
  VoxPopuli accent      & Pair      &           53.6 &  \textbf{53.9} &           53.5 &           53.1 &           53.2 \\
  CREMA-D (pair)        & Pair      &           55.0 &           55.4 &           55.6 &  \textbf{55.9} &           55.8 \\
  RAVDESS               & Zero-shot &           28.4 &           21.4 &           32.2 &           21.0 &  \textbf{39.7} \\
  SpeechCommands v0.02  & Zero-shot &           91.3 &           94.1 &           95.6 &           92.0 &  \textbf{97.6} \\
  \addlinespace[1pt]
  \textit{Mean}         &           &           46.0 &           48.8 &           52.4 &           52.4 &  \textbf{54.6} \\
  \midrule
  \multicolumn{7}{l}{\textit{Music (6 tasks)}} \\
  JamAlt artist A2A     & Retr.     &           88.3 &           88.7 &           88.9 &  \textbf{89.5} &           88.9 \\
  JamAlt lyrics A2T     & Retr.     &           41.3 &           44.2 &           47.4 &  \textbf{54.8} &           54.2 \\
  Beijing Opera         & Clf.      &           85.2 &           91.3 &           93.3 &           94.5 &  \textbf{94.6} \\
  GTZAN genre           & Clf.      &           83.6 &           87.2 &           87.1 &           86.9 &  \textbf{87.6} \\
  Mridangam tonic       & Clf.      &           32.8 &           37.7 &           44.0 &           46.6 &  \textbf{47.8} \\
  GTZAN reranking       & Rerank    &           84.0 &           83.4 &           84.2 &  \textbf{84.5} &           84.3 \\
  \addlinespace[1pt]
  \textit{Mean}         &           &           69.2 &           72.1 &           74.2 &           76.1 &  \textbf{76.2} \\
  \midrule
  \multicolumn{7}{l}{\textit{Sound (6 tasks)}} \\
  Clotho T2A            & Retr.     &           28.3 &           31.1 &           31.2 &  \textbf{31.6} &           30.9 \\
  MACS T2A              & Retr.     &           18.3 &           18.1 &           17.8 &           18.6 &  \textbf{20.1} \\
  UrbanSound8K T2A      & Retr.     &            0.8 &            0.8 &   \textbf{0.9} &   \textbf{0.9} &            0.8 \\
  BirdCLEF              & Clf.      &           31.4 &           39.4 &           41.8 &  \textbf{42.9} &           42.6 \\
  VehicleSound          & Clust.    &            3.1 &            2.1 &            2.6 &   \textbf{3.8} &            2.6 \\
  FSDKaggle2019         & Multi.    &           61.5 &           70.0 &  \textbf{71.4} &           71.1 &           70.2 \\
  \addlinespace[1pt]
  \textit{Mean}         &           &           23.9 &           26.9 &           27.6 &  \textbf{28.1} &           27.9 \\
  \midrule
  \textbf{MAEB-30 mean} &           &          46.23 &          49.08 &          51.81 &          52.31 & \textbf{53.58} \\
  \bottomrule
  \end{tabular}
\end{table}

\begin{figure}[htbp]
  \centering
  \begin{minipage}[t]{0.48\textwidth}
    \centering
    \includegraphics[width=\linewidth]{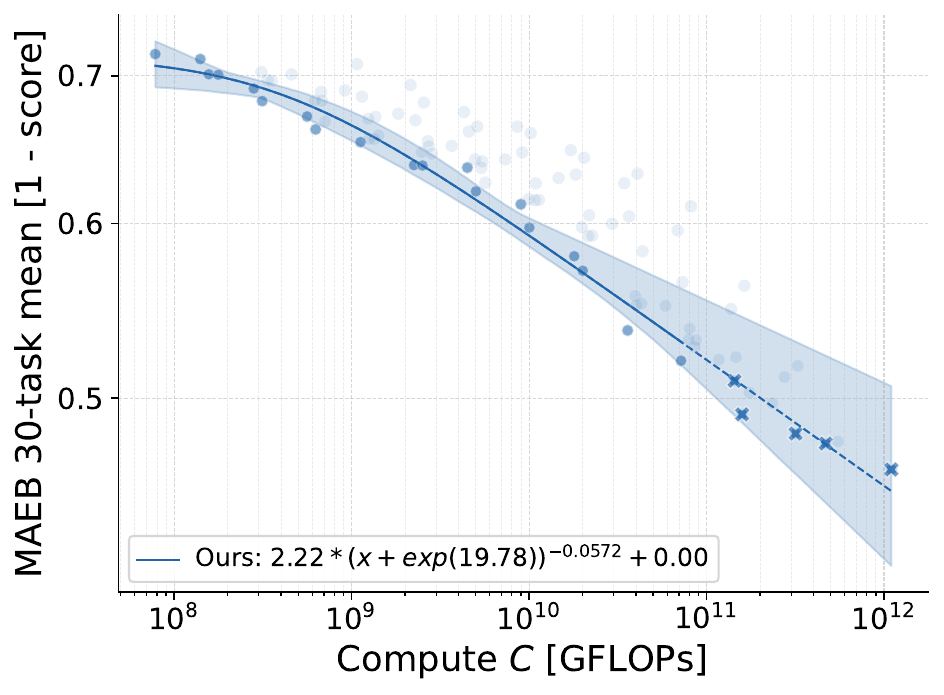}
    \caption{\textbf{Scaling predictions} We are able to accurately predict our best performing frontier models with more than $10^{11}$ GFLOPs of compute. We can do this with a mean squared error of $7.35\times10^{-5}$ on the 0--1 error scale or a mean absolute error of $6.4\times10^{-3}$.}
    \label{fig:scaling_predictions}
  \end{minipage}
  \hfill
  \begin{minipage}[t]{0.48\textwidth}
    \centering
    \includegraphics[width=\linewidth]{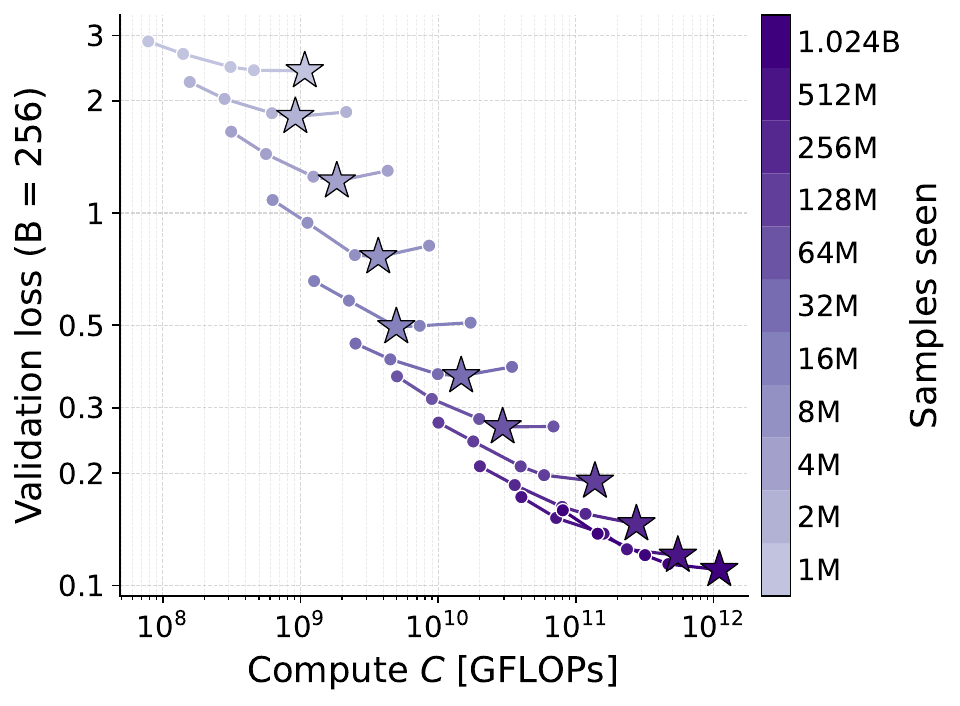}
    \caption{\textbf{Validation loss per compute $C$.} We plot our models based on compute and data scale against their validation loss with batch size $B=256$. We can see clear valleys in the loss at smaller compute scales, indicating the optimal compute for a given data scale (warmup-stable-decay only).}
    \label{fig:scaling_val_loss_maeb}
  \end{minipage}
\end{figure}

We also tested the quality of our scaling prediction by excluding all runs with $C\ge10^{11}$ GFLOPs. As shown in Figure~\ref{fig:scaling_predictions}, we are able to accurately predict the five runs on the compute-Pareto-frontier with $C\ge10^{11}$ GFLOPs. The predicted runs are marked with an $\times$ in the figure. Our predictions have a mean absolute error of 0.64 MAEB-30 points and a mean squared error of $7.35\times10^{-5}$ on the 0--1 error scale.

Last but not least, we plot the compute against our validation loss. Our validation set consists of $\approx1.5$k samples that were not used in training. Figure~\ref{fig:scaling_val_loss_maeb} shows isolines with the same data scale, while marking the lowest loss with a star. From these valleys, we can determine the most sample efficient model size that is neither over- nor under-trained. An example for this would be $D=32M$, where size M offers the best compromise between model size and data scale for minimizing the validation loss. To predict this for a larger $D$, we would need to train models larger than L, with up to $D=1{,}024M$ samples seen, since there is no valley for these data scales yet. One important note is that validation loss does not directly transfer to downstream benchmark performance as shown in Figure~\ref{fig:scaling_by_model_size_maeb}.

\FloatBarrier
\section{Design studies}
\label{sec:duration}
\label{app:data_domain}

\begin{table}[!htb]
\centering\small
\caption{Impact of training on different audio lengths, and caption types on mean MAEB-30 performance for our model of size M-C. Domain columns average the 18 speech, 6 music, and 6 sound tasks. All models were trained on the 500M selection with a 120 second audio length limit unless stated otherwise.}
\label{tab:ablations}
\setlength{\tabcolsep}{4pt}
\resizebox{\textwidth}{!}{%
\begin{tabular}{lrlrrrrr}
\toprule
 & Samples & Variant & FLOPs ($10^{20}$) & MAEB-30 & Speech & Music & Sound \\
\midrule
\textbf{Baseline} & 256M & up to 120 s & 4.97 & 43.0 & 40.2 & 68.4 & 26.2 \\
\midrule
\textbf{Audio length} & 256M & drop audio above 30 s & \textbf{1.44} & \textbf{47.1} & \textbf{45.8} & \textbf{72.3} & 26.1 \\
 & 256M & crop to 30 s & 1.76 & 46.5 & 45.1 & 70.6 & \textbf{26.4} \\
 & 256M & short-clip oversampling & 3.18 & 41.9 & 39.0 & 67.3 & 24.8 \\
\midrule\midrule
\textbf{Captions} & 128M & 50\% short, 50\% long captions & 2.48 & \textbf{39.1} & \textbf{34.9} & 66.3 & \textbf{24.6} \\
 & 128M & 100\% short captions & 2.48 & 38.6 & 34.1 & \textbf{66.7} & 23.9 \\
 & 128M & 100\% long captions & 2.48 & 38.5 & 34.5 & 65.7 & 23.4 \\
\bottomrule
\end{tabular}
}
\end{table}
\textbf{Audio duration:} Training on audio of at most 30 seconds produces a better model with less than one-third of the compute. At 256 million training samples seen, dropping longer audio samples scores 47.1\%, cropping them to 30 seconds scores 46.5\%, and training on audio up to 120 seconds scores only 43.0\% on the mean MAEB benchmark (Table~\ref{tab:ablations}). For this evaluation clips were cut off at 120 seconds. Allowing batches to hold up to 120-second audio samples leads to far more audio tokens per clip. At 50 audio tokens per second, the maximum number of audio tokens increases from 1504 (30 seconds) to 6008 (120 seconds), leading to smaller batches per GPU and slower training. The 120-second bucket holds only 8 audio samples per GPU batch but accounts for about three-quarters of the training FLOPs because attention cost grows with the square of the token count. Removing it reduces the cost per sample from $1.94\times10^{12}$ to $0.56\times10^{12}$ FLOPs, a factor of 3.5, and increases throughput on 128 GPUs from 6,600 to 12,300 samples per second. A compromise for this slowdown could be to oversample shorter buckets while keeping the 120-second bucket. As shown in Table~\ref{tab:ablations}, this oversampling approach does not produce better results with 41.9\% mean MAEB score.  We fix the policy that drops longer audio samples for the model-size grid. It leaves 366,006,046 of the 499,213,734 audio samples.

\textbf{Caption ratio:} Training on shorter or longer captions typically leads to performance trade-offs in downstream benchmarks. We tested training only on the short or long caption provided by our dataset. We observed that no caption policy is better than another within run-to-run variation. At 128M samples the 50/50 mix of short and long captions scores 39.1\%, short captions only 38.6\%, and long captions only 38.5\% on the MAEB benchmark (Table~\ref{tab:ablations}). Given these results and the fact that an even split offers the most caption diversity, we fix the default to an even split for further experiments. 

\textbf{Data domain:} Our training dataset has significantly more speech data than any other domain. To test whether this is positive or negative, we tried balancing the dataset evenly between speech, music and others. We do this by dropping samples during loading, with probabilities of 0\% for music, 40\% for speech, and 4\% for other classes. This keeps the data pool the same and oversamples to achieve a more even training data distribution. A model trained on 256 million rebalanced samples scores 41.7\% on MAEB-30, compared with 43.0\% without rebalancing. We therefore preserve the dataset's original data distribution for other experiments.

\section{HEAR and MSEB results}
\label{app:mseb}
Table~\ref{tab:hear} reports the HEAR scores of Figure~\ref{fig:size_vs_score} by domain and task type. Table~\ref{tab:mseb} reports the English tasks of the Massive Sound Embedding Benchmark (MSEB)~\citep{heigoldMassiveSoundEmbedding2026} for most models of Table~\ref{tab:scaling-final}.

\begingroup\small
\setlength{\tabcolsep}{2.5pt}\setlength{\LTcapwidth}{\textwidth}
\begin{longtable}{l r r rrr rrr}
\caption{HEAR benchmark results for the models of Table~\ref{tab:scaling-final} and the audio-only encoders of
Figure~\ref{fig:size_vs_score}, with the same embeddings as there. We report the mean over 17 tasks (all HEAR tasks except Beehive and Maestro 5h),
domain-specific means and the three
HEAR task types: clip classification, multi-label tagging (FSD50K) and timestamp event detection (DCASE 2016).
Audio-only models are evaluated by the models' authors or the official HEAR 2021 leaderboard. Audio-text models are evaluated by us. Within each group models are sorted by the mean, and only models with every value available are listed.
For audio-only models, we retain the best one or two variants of each method by mean score
All scores are in \%. The best listed value in each column is bolded, separately for the audio-only models and for the audio-text models together with ours.}
\label{tab:hear}\\
\toprule
& & & \multicolumn{3}{c}{\textbf{Domain}} & \multicolumn{3}{c}{\textbf{Task type}} \\
\cmidrule(lr){4-6}\cmidrule(lr){7-9}
\textbf{Model} & \textbf{Params} & \textbf{Mean} & \textbf{Sound} & \textbf{Speech} & \textbf{Music} & \textbf{Clf.} & \textbf{Tagging} & \textbf{Events} \\
\midrule
\endfirsthead
\multicolumn{9}{l}{\textit{Table~\ref{tab:hear}, continued}} \\
\toprule
& & & \multicolumn{3}{c}{\textbf{Domain}} & \multicolumn{3}{c}{\textbf{Task type}} \\
\cmidrule(lr){4-6}\cmidrule(lr){7-9}
\textbf{Model} & \textbf{Params} & \textbf{Mean} & \textbf{Sound} & \textbf{Speech} & \textbf{Music} & \textbf{Clf.} & \textbf{Tagging} & \textbf{Events} \\
\midrule
\endhead
\bottomrule
\endlastfoot
\multicolumn{9}{l}{\textit{Audio-only models}} \\
USAD 2.0-XXLarge+ \citep{changUSAD20Scaling2026} &  1.0B & \textbf{86.8} & \textbf{74.7} & \textbf{90.8} & 92.6 & \textbf{87.9} & 62.5 & 95.0 \\
USAD 2.0-XLarge+ \citep{changUSAD20Scaling2026} &  695M & \textbf{86.8} & \textbf{74.7} & 90.6 & \textbf{92.7} & 87.8 & 62.8 & \textbf{95.3} \\
AF3 Whisper encoder \citep{goelAudioFlamingo32025} &  637M & 85.6 & 74.6 & 89.2 & 90.9 & 86.6 & 61.4 & 93.8 \\
Dasheng-1.2B \citep{dinkelScalingMaskedAudio2024} &  1.2B & 83.7 & 70.4 & 87.0 & 90.8 & 85.0 & 54.2 & 94.2 \\
USAD-Large \citep{changUSADUniversalSpeech2025} &  330M & 81.8 & 69.3 & 85.7 & 87.9 & 82.9 & 53.0 & 93.9 \\
GURA Fuse Cat H+w+C \citep{wuEfficacySelfSupervisedSpeech2022} &  1.3B & 79.6 & 62.2 & 81.9 & 90.4 & 81.9 & 42.0 & 82.6 \\
RedRice ced\_base \citep{dinkelCEDConsistentEnsemble2024} &   86M & 79.0 & 73.3 & 70.4 & 89.2 & 79.0 & \textbf{65.5} & 92.2 \\
Hybrid BYOL-S/CvT \citep{elbannaBYOLSLearningSelfSupervised2022,elbannaHybridHandcraftedLearnable2022} &    5M & 77.9 & 67.5 & 76.3 & 86.5 & 78.9 & 52.0 & 88.9 \\
BEATs Iter3+ \citep{chenBEATsAudioPreTraining2023} &   90M & 75.7 & 62.3 & 71.2 & 88.6 & 79.0 & 63.6 & 39.6 \\
OpenL3 \citep{cramerLookListenLearn2019} &  4.7M & 70.8 & 61.2 & 59.3 & 86.0 & 71.7 & 44.7 & 83.3 \\
CP-JKU PaSST 2lvl+mel \citep{koutiniEfficientTrainingAudio2022} &   86M & 70.0 & 72.7 & 57.0 & 77.2 & 68.8 & 64.1 & 92.5 \\
wav2vec2 \citep{baevskiWav2vec20Framework2020} &  315M & 67.7 & 49.9 & 71.2 & 78.0 & 70.0 & 34.2 & 66.3 \\
CREPE \citep{kimCREPEConvolutionalRepresentation2018} & 22.2M & 54.6 & 37.6 & 28.3 & 85.6 & 57.5 & 15.9 & 50.4 \\
\midrule
\multicolumn{9}{l}{\textit{Audio-text models}} \\
e5-omni-7B \citep{chenE5OmniExplicitCrossModal2026} &  7.7B & 81.2 & 73.4 & 81.2 & 86.8 & 81.5 & 64.7 & \textbf{93.1} \\
jina-embeddings-v5-omni-nano \citep{honickeJinaEmbeddingsv5OmniGeometryPreservingEmbeddings2026} &  855M & 80.7 & 71.7 & 81.8 & 86.4 & 81.5 & 59.6 & 89.8 \\
M2D-CLAP \citep{niizumiM2DCLAPMaskedModeling2024} &  199M & 80.6 & \textbf{75.2} & 72.2 & \textbf{90.4} & 80.7 & 68.1 & 92.3 \\
BidirLM-Omni-2.5B \citep{boizardBidirLMTextOmnimodal2026} &  2.0B & 80.3 & 69.0 & 82.3 & 86.8 & 81.5 & 52.0 & 90.5 \\
LCO-Embedding-Omni-7B \citep{xiaoScalingLanguageCentricOmnimodal2025} &  7.7B & 79.8 & 70.3 & 80.1 & 86.3 & 80.1 & 65.9 & 88.4 \\
OmniEmbed-v0.1 \citep{maTevatron20Unified2025} &  7.8B & 79.7 & 71.6 & 80.4 & 85.1 & 79.9 & 65.1 & 91.8 \\
e5-omni-3B \citep{chenE5OmniExplicitCrossModal2026} &  3.7B & 79.7 & 71.2 & 78.4 & 86.7 & 80.1 & 63.6 & 90.1 \\
LCO-Embedding-Omni-3B \citep{xiaoScalingLanguageCentricOmnimodal2025} &  3.7B & 79.5 & 72.4 & 77.5 & 86.1 & 79.9 & 64.3 & 89.2 \\
Qwen2-Audio-7B \citep{chuQwen2AudioTechnicalReport2024} &  7.8B & 79.5 & 68.5 & 81.8 & 85.7 & 80.0 & 62.2 & 88.8 \\
jina-embeddings-v5-omni-small \citep{honickeJinaEmbeddingsv5OmniGeometryPreservingEmbeddings2026} &  1.3B & 78.1 & 69.9 & 78.9 & 83.5 & 78.6 & 58.4 & 90.3 \\
TP-CLAP \citep{liTextPromptedCLAPLearning2026} &  198M & 76.6 & 72.2 & 66.5 & 87.0 & 76.2 & 67.7 & 91.7 \\
Qwen2.5-Omni-7B \citep{xuQwen25OmniTechnicalReport2025} &  7.7B & 75.2 & 65.9 & 76.8 & 80.6 & 76.1 & 49.0 & 87.7 \\
MS-CLAP 2023 \citep{elizaldeNaturalLanguageSupervision2024} &  160M & 75.1 & 72.6 & 65.8 & 83.6 & 74.8 & 70.4 & 85.7 \\
GLAP \citep{dinkelGLAPGeneralContrastive2026} &  855M & 71.8 & 67.2 & 70.7 & 75.8 & 71.4 & 61.7 & 87.8 \\
LAION Larger CLAP \citep{wuLargeScaleContrastiveLanguageAudio2023} &  194M & 69.5 & 71.9 & 58.6 & 75.5 & 68.2 & \textbf{72.5} & 86.2 \\
LAION-CLAP HTSAT unfused \citep{wuLargeScaleContrastiveLanguageAudio2023} &  153M & 69.2 & 72.5 & 59.2 & 73.9 & 67.9 & 72.3 & 85.7 \\
EmbeddingGemma 2 \cite{EmbeddingGemma2Open2026} &  577M & 69.1 & 62.2 & 79.5 & 66.7 & 69.7 & 43.1 & 86.3 \\
CLAP-IPA small \cite{zhuTasteIPAOpenVocabulary2024} &  174M & 67.4 & 52.2 & 72.5 & 74.7 & 68.5 & 34.2 & 84.2 \\
SpeechT5 \citep{aoSpeechT5UnifiedModalEncoderDecoder2022} &  180M & 67.3 & 57.3 & 74.5 & 69.3 & 68.5 & 31.8 & 84.9 \\
LAION Larger CLAP (music+speech) \citep{wuLargeScaleContrastiveLanguageAudio2023} &  194M & 66.4 & 72.6 & 49.3 & 74.1 & 64.5 & 70.7 & 89.5 \\
MS-CLAP 2022 \citep{elizaldeCLAPLearningAudio2023} &  196M & 63.2 & 65.6 & 51.4 & 69.9 & 62.1 & 59.1 & 83.9 \\
wav2clip \citep{wuWav2CLIPLearningRobust2022} &   75M & 56.8 & 52.6 & 38.8 & 72.7 & 57.3 & 35.1 & 71.1 \\
\midrule
\textit{CLAP2-XXS}                    &   27M & 79.1 & 71.2 & 81.6 & 82.9 & 79.8 & 59.3 & 87.5 \\
\textit{CLAP2-XS}                     &   48M & 80.6 & 73.5 & 83.5 & 83.6 & 81.4 & 60.1 & 88.8 \\
\textit{CLAP2-S}                      &  104M & 81.8 & 73.4 & 84.5 & 85.9 & 82.7 & 60.9 & 89.9 \\
\textit{CLAP2-M}                      &  153M & 81.4 & 73.5 & 84.6 & 84.7 & 82.1 & 61.5 & 89.5 \\
\textit{CLAP2-L}                      &  340M & \textbf{81.9} & 72.7 & \textbf{85.2} & 86.1 & \textbf{82.8} & 61.1 & 88.8 \\
\end{longtable}
\endgroup

\begin{table}[!ht]
\centering\small
\caption{MSEB benchmark results on the English tasks, for most models of Table~\ref{tab:scaling-final} (same
parameter counts and checkpoints). We report the mean over 62 higher-is-better scores from 52 tasks (speaker clustering scores age,
gender and speaker identity separately), domain-specific means (40 speech, 2 environmental sound, 20 bird sound scores)
and task-specific means: classification (16), clustering (26), reranking (5),
reasoning (5), and retrieval (10). Stability (Stab., 5 tasks) is the continuous edit distance between the
embeddings of clean and noise-augmented speech; it is lower-is-better and not part of the mean. Speech transcription
and salient-term segmentation are not reported, since they score text output that an embedding model does not
produce. Multilingual tasks, the image--speech retrieval tasks, and the full-index document retrieval (we use the
small index) are not run. All scores are in \%. The best value in each column is bolded.}
\label{tab:mseb}
\setlength{\tabcolsep}{4pt}
\resizebox{\textwidth}{!}{%
\begin{tabular}{l r r rrr rrrrr r}
\toprule
& & & \multicolumn{3}{c}{\textbf{Domain}} & \multicolumn{6}{c}{\textbf{Task type}} \\
\cmidrule(lr){4-6}\cmidrule(lr){7-12}
\textbf{Model} & \textbf{Params} & \textbf{Mean} & \textbf{Speech} & \textbf{Sound} & \textbf{Bird} & \textbf{Clf.} & \textbf{Clust.} & \textbf{Rerank.} & \textbf{Reason.} & \textbf{Retr.} & \textbf{Stab.}$\downarrow$ \\
\midrule
M2D-CLAP \citep{niizumiM2DCLAPMaskedModeling2024}& 199M & \textbf{30.7} & 36.4 & 52.5 & \textbf{17.1} & \textbf{34.6} & 50.6 &  1.4 &  5.5 &  0.2 &  9.4 \\
LAION-CLAP HTSAT unfused \citep{wuLargeScaleContrastiveLanguageAudio2023}& 153M & 30.0 & 36.0 & 58.6 & 15.3 & 33.9 & 49.3 &  1.6 &  6.1 &  0.0 & 11.8 \\
MS-CLAP 2023 \citep{elizaldeNaturalLanguageSupervision2024}& 160M & 28.5 & 33.6 & 55.1 & 15.7 & 25.5 & \textbf{50.9} &  1.6 &  5.4 &  0.0 & 10.2 \\
LAION Larger CLAP \citep{wuLargeScaleContrastiveLanguageAudio2023}& 194M & 28.2 & 32.9 & \textbf{58.8} & 16.0 & 33.7 & 45.3 &  1.4 &  5.7 &  0.0 & 21.5 \\
LCO-Embedding-Omni-7B \citep{xiaoScalingLanguageCentricOmnimodal2025}& 7.7B & 27.9 & 33.4 & 55.7 & 14.2 & 23.4 & 12.7 & 41.7 & 15.1 & \textbf{74.3} & 13.5 \\
LCO-Embedding-Omni-3B \citep{xiaoScalingLanguageCentricOmnimodal2025}& 3.7B & 27.2 & 32.3 & 53.7 & 14.4 & 20.2 & 13.0 & 55.5 & 13.6 & 68.0 & 13.8 \\
e5-omni-7B \citep{chenE5OmniExplicitCrossModal2026}& 7.7B & 26.7 & 32.4 & 33.9 & 14.5 & 21.1 & 16.5 & \textbf{62.8} & 14.7 & 50.1 & 10.7 \\
BidirLM-Omni-2.5B \citep{boizardBidirLMTextOmnimodal2026}& 2.0B & 26.7 & 32.1 & 34.1 & 15.0 & 33.4 & 15.0 & 32.5 & 14.6 & 49.2 & 13.7 \\
e5-omni-3B \citep{chenE5OmniExplicitCrossModal2026}& 3.7B & 25.8 & 31.2 & 32.4 & 14.4 & 18.7 & 16.1 & 53.7 & \textbf{16.0} & 53.4 &  9.3 \\
EmbeddingGemma 2 \cite{EmbeddingGemma2Open2026}& 577M & 25.0 & 30.2 & 28.7 & 14.0 & 17.0 & 14.3 & 43.8 & 11.2 & 62.8 & 10.6 \\
jina-embeddings-v5-omni-nano \citep{honickeJinaEmbeddingsv5OmniGeometryPreservingEmbeddings2026}& 855M & 22.5 & 25.2 & 43.5 & 14.9 & 25.0 & 14.8 & 41.7 & 14.7 & 32.8 & 15.3 \\
jina-embeddings-v5-omni-small \citep{honickeJinaEmbeddingsv5OmniGeometryPreservingEmbeddings2026}& 1.3B & 22.1 & 25.3 & 41.9 & 13.8 & 25.9 & 12.4 & 42.5 & 14.4 & 35.2 & 16.5 \\
SpeechT5 \citep{aoSpeechT5UnifiedModalEncoderDecoder2022}& 180M & 16.7 & 19.2 & 16.6 & 11.8 & 18.7 & 26.1 &  1.7 & 10.0 &  0.2 &  9.5 \\
GLAP \citep{dinkelGLAPGeneralContrastive2026}& 855M & 15.7 & 16.1 & 35.5 & 12.8 & 20.0 & 16.5 & 34.6 &  8.9 &  0.6 & 12.6 \\
\midrule
\textit{CLAP2-XXS}              &  27M & 28.0 & 33.3 & 36.2 & 16.4 & 32.9 & 39.0 & 30.2 &  6.2 &  1.2 & \textbf{8.1} \\
\textit{CLAP2-XS}               &  48M & 27.6 & 32.6 & 36.6 & 16.8 & 34.3 & 37.6 & 28.6 &  6.0 &  1.4 &  8.4 \\
\textit{CLAP2-S}                & 104M & 29.8 & 36.0 & 37.5 & 16.7 & 33.6 & 39.7 & 39.9 &  6.1 &  4.9 &  8.7 \\
\textit{CLAP2-M}                & 153M & 29.4 & 35.4 & 37.1 & 16.6 & 33.5 & 37.8 & 44.3 &  6.9 &  4.7 &  8.8 \\
\textit{CLAP2-L}                & 340M & 30.3 & \textbf{37.0} & 38.6 & 16.1 & 33.8 & 39.9 & 41.2 &  7.1 &  6.0 &  9.4 \\
\bottomrule
\end{tabular}
}
\end{table}

\clearpage
\section{Dataset catalogue}
\label{app:catalog}
The catalogue lists each of the 425 sources with at least 10 LAION-Audio-XL audio samples at three stages: as stored audio in the archive, as selected audio in the 500M selection, and as audio samples of at most 30 seconds in LAION-Audio-XL. The archive counts come from the collection of stored audio samples with their measured durations. The archive lists 1.43B of the 1.48B stored samples. Rows that link to the same source page are combined into one row, and so are copies of one or similar datasets stored under different names.

Sources are grouped as academic datasets, internet datasets, web crawl and web collections, or YouTube-derived data. A source is academic when a citable publication was found for it: an arXiv preprint, a DOI that resolves to a paper, or a citation found on the source or derivative page. A source is an internet dataset when the same search did not find such publication, which is the usual case for a Hugging Face upload released without a paper. Websites such as Kaggle uploads count as web-crawled data unless the Kaggle page cites a paper that introduces the data, and so do uploads whose page says nothing about the data. The distinction of academic dataset, internet dataset, etc. describes the evidence we could find, and a later publication can move a source between the two groups. A paper that only uses the data does not count. Only the YouTube entry forms the YouTube-derived group, as other datasets that built from YouTube but are their own publication, such as AudioSet~\citep{gemmekeAudioSetOntology2017} and VGGSound~\citep{chenVGGSoundLargescaleAudiovisual2020}, are academic.

In total, about 94.4\% (345.4 of 366.0 million) of all LAION-Audio-XL audio samples have a Creative Commons license. The 14.1 million Common Crawl audio samples, whose individual licenses are not known, are counted as not Creative Commons licensed. When counting relatively per data source, about 69\% have a Creative Commons declaration: 26\% CC BY, 11\% CC BY-SA, 23\% variants with non-commercial or no-derivatives restrictions, 5\% CC0, and 4\% mixed or unspecified Creative Commons terms. About 6\% list software licenses such as Apache, MIT or GPL on the data, 2\% public-domain declarations, and 13\% other or source-specific terms. The remaining 10\% do not have a verified license declaration. These percentages count each source equally, without weighting by audio samples or hours. They summarize recorded declarations, which can include audio, annotations, metadata, or code.  
\begin{table}[ht]
\centering\scriptsize
\caption{Sources, stored audio samples and indexed audio hours by group, at three stages: the archive as collected, the 500M selection, and LAION-Audio-XL (audio samples at most 30 seconds long); the last two count their training split.  A clip re-cut from a recording above 120 seconds is charged once to its source in the last two stages. The archive counts the recording as collected. Rows count the sources listed in Table~\ref{tab:catalog}. The deduplicated pool between the first two stages is given in Figure~\ref{fig:steps}.}\label{tab:catalog-totals}
\setlength{\tabcolsep}{2.6pt}
\begin{tabular}{lrrrrrrr}\toprule
& & \multicolumn{2}{c}{Archive} & \multicolumn{2}{c}{500M selection} & \multicolumn{2}{c}{LAION-Audio-XL} \\
\cmidrule(lr){3-4}\cmidrule(lr){5-6}\cmidrule(lr){7-8}
Source group & Sources & Audio & Hours & Audio & Hours & Audio & Hours \\\midrule
Academic datasets & 316 & 83,689,086 & 231,011 & 32,815,997 & 82,820 & 31,283,585 & 57,789 \\
Internet datasets & 54 & 9,614,142 & 22,287 & 2,131,152 & 5,204 & 2,112,543 & 4,911 \\
Web crawl and web collections & 54 & 158,932,602 & 1,276,651 & 33,547,324 & 394,100 & 17,050,765 & 98,007 \\
YouTube-derived data & 1 & 1,179,581,779 & 9,420,695 & 430,719,261 & 2,865,365 & 315,559,153 & 1,228,570 \\
\midrule Total & 425 & 1,431,817,609 & 10,950,644 & 499,213,734 & 3,347,490 & 366,006,046 & 1,389,277 \\
\bottomrule\end{tabular}
\end{table}

\begingroup
\fontsize{6.2}{7.2}\selectfont
\setlength{\tabcolsep}{1.6pt}
\renewcommand{\arraystretch}{1.02}
\setlength{\LTpre}{3pt}\setlength{\LTpost}{3pt}
\begin{longtable}{@{}>{\raggedright\arraybackslash}p{.310\linewidth}>{\raggedleft\arraybackslash}p{.120\linewidth}>{\raggedleft\arraybackslash}p{.090\linewidth}>{\raggedleft\arraybackslash}p{.120\linewidth}>{\raggedleft\arraybackslash}p{.090\linewidth}>{\raggedleft\arraybackslash}p{.120\linewidth}>{\raggedleft\arraybackslash}p{.090\linewidth}@{}}
\caption{Complete dataset catalogue, alphabetized inside each group. \emph{Archive} counts every clip stored in the collection before deduplication. \emph{500M selection} counts the selected audio samples. \emph{LAION-Audio-XL} counts the subset of audio samples of at most 30 seconds. Both count their training split. A source counts as academic if it is linked to a publication, in the broadest sense. Otherwise it counts as an internet source. Sources with fewer than 10 LAION-Audio-XL audio samples were removed from both corpora unless they are part of a listed source, whose row counts them, so every source of the two corpora is listed. Names link to source pages where the dataset is found, or in some cases the paper itself. }\label{tab:catalog}\\
\toprule & \multicolumn{2}{c}{Archive} & \multicolumn{2}{c}{500M selection} & \multicolumn{2}{c}{LAION-Audio-XL} \\
\cmidrule(lr){2-3}\cmidrule(lr){4-5}\cmidrule(lr){6-7}
Dataset & Audio & Hours & Audio & Hours & Audio & Hours \\\midrule
\endfirsthead
\multicolumn{7}{@{}l}{\textit{Table \thetable{} continued.}}\\
\toprule & \multicolumn{2}{c}{Archive} & \multicolumn{2}{c}{500M selection} & \multicolumn{2}{c}{LAION-Audio-XL} \\
\cmidrule(lr){2-3}\cmidrule(lr){4-5}\cmidrule(lr){6-7}
Dataset & Audio & Hours & Audio & Hours & Audio & Hours \\\midrule
\endhead
\bottomrule\endfoot
\multicolumn{7}{@{}l}{\rule{0pt}{11pt}\textbf{Academic datasets} --- 316 sources}\\*
\href{http://mir.dei.uc.pt/downloads.html}{4Q Audio Emotion}~\citep{pandaNovelAudioFeatures2020} & 2,878 & 23.8 & 150 & 1.2 & 60 & 0.5\\
\href{https://zenodo.org/records/5794629}{AAM}~\citep{ostermannAAMDatasetArtificial2023} & 14,741 & 464.6 & 9,247 & 274.7 & 340 & 1.9\\
\href{https://github.com/KeisukeImoto/ACalt4}{ACalt4}~\citep{niizumiM2DCLAPMaskedModeling2024} & 120,464 & 330.0 & 1,433 & 3.9 & 1,433 & 3.9\\
\href{https://web.archive.org/web/20200811090336/http://www.marsyas.info/tempo/}{ACM MIRUM}~\citep{peetersPerceptualTempoEstimation2012} & 1,312 & 13.5 & 1,261 & 12.9 & 728 & 6.1\\
\href{https://zenodo.org/records/3383118}{AdoVoc Pro}~\citep{gomezgutierrezCanteFlamencoTech2018} & 2,111 & 1.4 & 20 & $<0.1$ & 20 & $<0.1$\\
\href{https://data.vision.ee.ethz.ch/cvl/ae\_dataset/}{AED}~\citep{takahashiDeepConvolutionalNeural2016} & 10,083 & 26.3 & 3,811 & 9.9 & 3,811 & 9.9\\
\href{https://m3c.web.auth.gr/research/aesdd-speech-emotion-recognition/}{AESDD Greek}~\citep{vryzasSpeechEmotionRecognition2018} & 604 & 0.7 & 223 & 0.3 & 223 & 0.3\\
\end{longtable}
\endgroup

\end{document}